\documentclass[twocolumn,prb,superscriptaddress,preprintnumbers,amsmath,amssymb,floatfix,longbibliography]{revtex4-2}

\usepackage{graphicx,xcolor,hyperref,ulem}
\usepackage{amsmath}
\usepackage{amsfonts}
\usepackage{amssymb}
\usepackage{subfigure}
\usepackage{dsfont}

\begin{document}

\title{
Probing the topological Anderson transition with quantum walks
}
\author{Dmitry Bagrets}
\affiliation{Institut f\"ur Theoretische Physik, Universit\"at zu K\"oln, Z\"ulpicher Stra\ss e 77, 50937 K\"oln, Germany}

\author{Kun Woo Kim}
\affiliation{Institut f\"ur Theoretische Physik, Universit\"at zu K\"oln, Z\"ulpicher Stra\ss e 77, 50937 K\"oln, Germany}
\affiliation{Department of Physics, Chung-Ang University, 06974 Seoul, Korea}

\author{Sonja Barkhofen}
\affiliation{Integrated Quantum Optics Group, Institute for Photonic Quantum Systems (PhoQS), Paderborn University, Warburger Stra\ss{}e 100, 33098 Paderborn, Germany}

\author{Syamsundar De }
\affiliation{Integrated Quantum Optics Group, Institute for Photonic Quantum Systems (PhoQS), Paderborn University, Warburger Stra\ss{}e 100, 33098 Paderborn, Germany}

\author{Jan Sperling}
\affiliation{Integrated Quantum Optics Group, Institute for Photonic Quantum Systems (PhoQS), Paderborn University, Warburger Stra\ss{}e 100, 33098 Paderborn, Germany}

\author{Christine Silberhorn}
\affiliation{Integrated Quantum Optics Group, Institute for Photonic Quantum Systems (PhoQS), Paderborn University, Warburger Stra\ss{}e 100, 33098 Paderborn, Germany}

\author{Alexander Altland}
\affiliation{Institut f\"ur Theoretische Physik, Universit\"at zu K\"oln, Z\"ulpicher Stra\ss e 77, 50937 K\"oln, Germany}

\author{Tobias Micklitz}
\affiliation{Centro Brasileiro de Pesquisas F\'isicas, Rua Xavier Sigaud 150, 22290-180, Rio de Janeiro, Brazil }

\date{\today}

\begin{abstract}
	We consider one-dimensional quantum walks in optical linear networks with synthetically introduced disorder and tunable system parameters allowing for the engineered realization of distinct topological phases. The option to directly monitor the walker's probability distribution makes this optical platform ideally suited for the experimental observation of the unique signatures of the one-dimensional topological Anderson transition. 
	We analytically calculate the  probability distribution describing the quantum critical walk in terms of a (time staggered) spin polarization signal and propose a concrete experimental protocol for its measurement. Numerical simulations  back the realizability of our blueprint with current date experimental hardware. 
\end{abstract}

\maketitle

\section{Introduction}

	Low-dimensional disordered quantum systems can escape the common fate of Anderson localization once topology comes into play, as first witnessed at the integer quantum Hall plateau transitions~\cite{Klitzing:1980, Khmelnitskii:1983}.
	The advent of topological insulators has brought a systematic understanding of topological Anderson insulators and their phase transitions~\cite{Mirlin:2010}. The hallmark of Anderson insulating phase with non-trivial topology 
	(coined 'topological Anderson insulator') is the presence of topologically protected 
	chiral edge states~\cite{obuse2011topological,rakovszky2015localization}.  
The Anderson localization transition by itself is characterized by the critical states which typically show unusual spectral- and  wave-function statistics~\cite{Evers:2008}, as well as  anomalous diffusive dynamics. 
	From the single parameter scaling theory of localization one e.g. 
	expects a scaling  $\langle \bold{q}^2(t)\rangle \sim t^{2/d}$ 
	of the mean square displacement at a 
	conventional $d$-dimensional 
	localization transition~\cite{Ohtsuki:1997}. Topological 
	localization transitions, on the other hand, 
	follow a two-parameter scaling and the situation is more complex~\cite{Pruisken:1984, Fu:2012, Altland:2015}.
	The controlled experimental study of a critical state at the Anderson localization transition presents an intriguing challenge.
	It has been first accomplished within a cold-atom realization of the quantum kicked rotor for the three-dimensional Anderson localization transition in the orthogonal class~\cite{Chabe:2008}.
	A corresponding study of a {\it topological} localization transition requires the control over additional internal degrees of freedom and has, to our knowledge, not been realized so far~\footnote{See Ref.~\cite{meier2018observation} for the recent realization of a $1d$ 
	 wire with chiral symmetry, where evidence for the topological Anderson insulator phase was given.}.
\begin{figure}[t]
	\includegraphics[width=.4\textwidth]{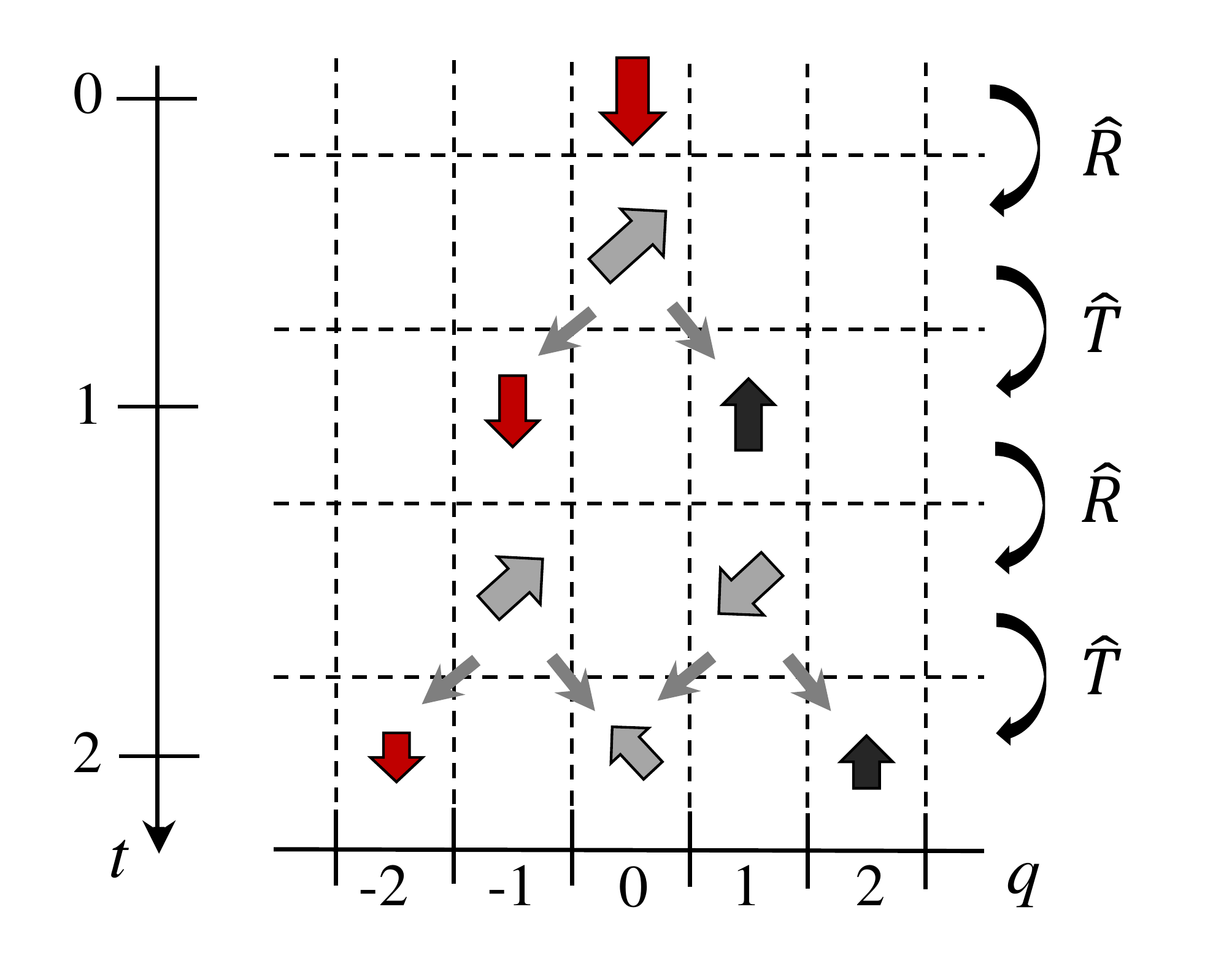}
	\caption
		{A schematic evolution of a discrete-time quantum walk of a spin $1/2$ particle over two steps in time.
		Alternating application of coin $\hat{R}$ and step $\hat{T}$ operators describe the dynamics on discrete lattice sites $q \in \mathbb{Z}$.
		In step two at position $q=0$, the first interference takes place.
	}\label{fig1}
\end{figure} 

	 The physics of the topological Anderson localization transition is
	 particularly intriguing in one-dimensional ($1d$) systems, where disorder is
	 exceptionally efficient in inducing quantum localization on short length scales.
	 Topological quantum criticality then reflects a competition between two powerful
	 principles: strong localization vs the enforced change of an integer topological
	 invariant. Topology trumps localization and forces long range correlations
	 across the system. In practical terms, this implies a divergent localization
	 length, and finite conduction. However, the reluctance of the system to conduct
	 shows in an extremely (logarithmically) slow spreading of quantum states  at
	 criticality~\cite{Balents:1997,Bagrets:2016}, strikingly different to the
	 diffusive dynamics conventionally observed at quantum phase transitions between
	 disordered phases. In this paper, we connect the physics of one-dimensional
	 topological quantum criticality to the unique  opportunities offered by quantum
	 optics experimentation. We  present a concrete and experimentally realistic blueprint for a
	 tunable $1d$ quantum walk in which the unique signatures of topological quantum
	 criticality show via a (time staggered) spin polarization signal.  
 
Quantum walks \cite{Aharonov:1993} have
	been implemented on various experimental platforms, such as
	photons 	\cite{bouwmeester1999optical, perets_realization_2008, peruzzo2010quantum, broome_discrete_2010, schreiber_photons_2010, Schreiber:2012, sansoni_two-particle_2012, crespi2013anderson, cardano_quantum_2015, xue2015localized}, ions \cite{schmitz2009quantum, zahringer2010realization}, atoms \cite{genske2013electric, karski2009quantum, preiss2015strongly} and nuclear magnetic resonance \cite{du2003experimental}.
	A detailed introduction to experimental implementations of quantum walks can be found in Ref.~\cite{wang2013physical}.
Quantum walks allow for a large tunability of the system parameters and have been used experimentally to observe Anderson localization \cite{Schreiber:2011, crespi2013anderson, xue2015localized}, dynamical localization \cite{genske2013electric} and topological effects \cite{Kitagawa:2012,  rechtsman_photonic_2013,  zeuner_observation_2015, cardano_statistical_2015, xiao2017observation, cardano2017detection, barkhofen2017measuring, wang2018detecting}. 
Quantum walk systems thus open the perspective of a low-dimensional system in which disorder and nontrivial topology can be introduced in controlled manners.
	Direct experimental access to the probability distribution allows, moreover, for a full characterization of the walker's dynamics.
	
	A prototypical quantum walk is depicted in Fig.~\ref{fig1}. 
	It is generated by the single time-step evolution $\hat U=\hat R\, \hat T$, iteratively acting on a 
	walker with a two dimensional internal degree of freedom, refered to as `spin' in the following. 
	Here $\hat T$ translates the particle on a discrete one-dimensional lattice. 
	Depending on its internal spin-state, the walker propagates to the left or right, and $\hat R$ is a rotation in spin-space. 
Using linear optical elements, discrete time quantum walks have been used to measure probability distributions of walkers exposed to tunable disorder and decoherence~\cite{Schreiber:2011, geraldi2020subdiffusion}. 
Specifically, conditions for ballistic, Anderson-localized and classically diffusive dynamics were prepared, and the corresponding walkers probability distributions [see also Eq.~\eqref{msq}] were observed.
	That is, the following scenarios apply: 
	(i) $P_{\sigma'\sigma}(t, q)\sim \delta(t- \sigma q)\delta_{\sigma\sigma'}$ for a translational invariant quantum system (here $\sigma=\pm$ denotes the spin state);
	(ii) $P_{\sigma'\sigma} (t,q)\sim \exp(-|q|/\xi_{\rm loc})$ for a disordered quantum system;
	and (iii) $P_{\sigma'\sigma} (t,q)\sim \exp(-q^2/Dt)$ for a disordered classical system, where $\xi_{\rm loc}$ and $D$ are localization length and diffusion constant. 
In the photonic implementation, the internal `spin'-states correspond to horizontal, $|H\rangle$,  and vertical, $|V\rangle$, polarization directions, 
	and disorder is controlled by local variations of  
	polarization plates~\cite{Schreiber:2011}.  Rotations that do not explore all ${\rm SU}(2)$-angles independently leave symmetries, which can place the walk into one of the five nonstandard symmetry classes hosting topologically interesting phases~\cite{Schnyder:2008, Tarasinski:2014, Cedzich:2018}.

In this paper, we explore a quantum walk operating at a topological Anderson localization transition.  
We derive the walker's critical probability distribution at the topological transition and find a (time-staggered) spin polarization as a smoking-gun evidence for the critical dynamics.
		We indicate a protocol which allows for an observation of the discussed 
		features within existing experimental platforms, and 
compare results of our effective field theory approach to numerical simulations. 
	The remainder of the paper is organized as follows. 
	In Sec.~\ref{ChiralWalk}, we introduce a quantum walk with chiral symmetry that can be tuned to a quantum critical point separating two topologically different Anderson insulators. 
	In Sec.~\ref{CriticalDistribution}, we analyze the probability distribution of the critical walker and propose, in Sec.~\ref{Experiment}, an experimental protocol that allows one to study the predicted effect.
	We  conclude in Sec.~\ref{Discussion} with a discussion and an outlook.
	Several technical discussions are relegated to appendices. 

\section{Chiral quantum walk}
\label{ChiralWalk}

\begin{figure}[t]
	\begin{center}
		\includegraphics[width=6.8cm]{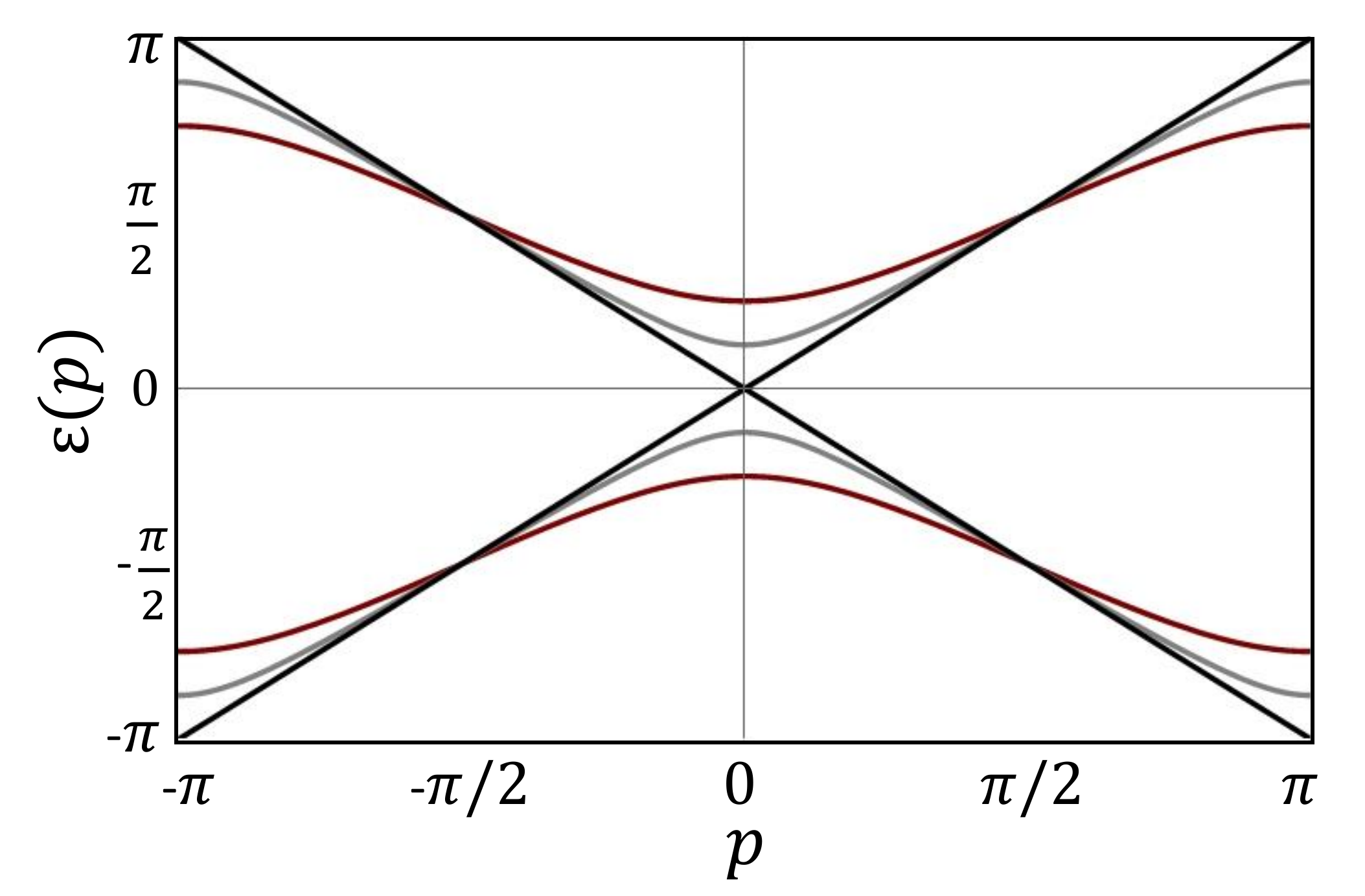}
	\end{center}
	\vspace{-15pt}
	\caption{
		Dispersion-relation $\epsilon(p)$ of Floquet bands shown for angles $\varphi=0$ and $\theta=0$ (black line), $\pi/8$ (grey), and $\pi/4$~(red). 
	}\label{fig2_0}
\end{figure}

	We start our discussion with a general one-dimensional quantum walk of a spin-$1/2$ particle, encoded in the single time-step evolution~\cite{Cedzich:2018}
	\begin{align}
		\hat U(\vartheta,\varphi,\theta) 
		&= 
		\hat R(\vartheta,\varphi,\theta)\hat T.
	\end{align}
	The spin-dependent `shift' operator $\hat T$ here is diagonal in the $\hat s_3$-eigenbasis, 
	\begin{align}
		\hat T 
		&= \sum_q \left( 
		|q+1, \uparrow \rangle  \langle \uparrow, q|
		+
		|q-1, \downarrow \rangle  \langle  \downarrow, q |
		\right),
	\end{align}
	where $q \in \mathbb{Z}$ are the lattice sites with unit spacing, and `spin' states 
	$\{ |\uparrow \rangle, |\downarrow\rangle \}$ parametrize the walker's two-dimensional 
	internal degrees of freedom, 
	see also Fig.~\ref{fig1}. 
	Local `coin' rotations, 
	\begin{align}
		\hat R
		&=
		\sum_{q, \sigma\sigma'} 
		|q, \sigma \rangle  R_q^{\sigma \sigma' }\langle q,\sigma'|,
	\end{align}
	are conveniently parametrized by (site-dependent) 
	Euler angles $R_q(\vartheta,\varphi,\theta)=\exp(\tfrac i2 \vartheta_q \sigma_1)\exp(i \varphi_q \sigma_3)\exp(\tfrac i2 \theta_q \sigma_1)$, where Pauli matrices $\sigma_i$ operate in spin-space.
	Employing a symmetrized time-splitting, we can write
	\begin{align}
	\label{sqw}
		\hat U(\vartheta,\varphi,\theta)
		&=
		R_z(\tfrac{\hat \varphi}{2}) R_x(\tfrac{\hat \theta}{2})
		\,\hat T \,
		R_x(\tfrac{\hat \vartheta}{2}) R_z(\tfrac{\hat \varphi}{2}),
	\end{align}
	with $\hat \vartheta$, $\hat \varphi$, $\hat \theta$ being site-diagonal matrices of angles and $R_i(\hat\alpha) = \exp(i\hat\alpha \sigma_i)$ defines a rotation along $i$-th direction.
	From Eq.~\eqref{sqw}, one readily verifies that quantum walks subject to the constraint $\hat \theta= \hat \vartheta$ exhibit a chiral symmetry~\cite{Asboth:2012}, 
	\begin{align}
	\label{chS}
		\sigma_2 \hat{U}\sigma_2 
		&=
		\hat{U}^\dagger.
	\end{align}
	That is, the latter are members of the chiral symmetry class ${\rm AIII}$, which may host $\mathbb{Z} \times \mathbb{Z}$ topological insulating phases for one-dimensional quantum walks~\cite{Tarasinski:2014}. 
	Throughout this paper we only focus on walks with $\hat\theta=\hat\vartheta$, i.e. 
	with the chiral symmetry denoted in Eq.~\eqref{chS}. 

	Chiral symmetry of the time-evolution operator reflects in a spectrum which is symmetric around zero in the $2\pi$-periodic `Brillouin zone' of quasi-energies $\epsilon_p^\pm=\pm\epsilon_p$.
	For spatially constant rotations, Floquet eigenstates are plane-waves and the two energy  bands are defined by the relation 
	\begin{equation}
	\begin{aligned}
		\cos(\epsilon_p(\theta,\varphi))
		=&
		\cos(\varphi + p)\cos^2(\tfrac{\theta}{ 2})
		-
		\cos(\varphi - p)\sin^2(\tfrac{\theta}{2}),
	\end{aligned}
	\end{equation}
	with $p$ being the momentum of the plane wave. 
	Finite angles $\varphi$ and $\theta$ shift the momentum and tune the bandwidth of Bloch-bands; 
	see Fig.~\ref{fig2_0}.
	Specifically, linearly dispersing bands that extend over the entire Brillouine zone exist at values $\theta=0,\pi$, with
	\begin{align}
		\epsilon_p(\theta,\varphi)
		&=
		\theta + p + e^{i\theta} \varphi,
	\end{align}
	and $\varphi= \pm \tfrac{\pi}{2}$, with 
	\begin{align}
	\epsilon_p(\theta,\varphi) 
	=\tfrac{\pi}{2} \pm {\rm sgn}(\varphi) p.{}
	\end{align}
	For any other values of the angles, the spectrum is gapped around $\epsilon=0$ and $\pi$. 
	Disorder can be introduced in a controlled manner by randomizing angles.
	Assuming short range site-to-site correlations, rotations are then characterized by average angles $\bar{\theta}$, $\bar{\varphi}$ and their deviations $\gamma_{\theta}$, $\gamma_{\varphi}$, which we assume to be identical for all lattice sites.
	In one dimension, even weak disorder $\gamma_{\theta,\varphi}\ll 1$ strongly affects the dynamics of the walker, turning its ballistic propagation into exponential Anderson-localization on the scale of the mean free path~\cite{Anderson:1958}, which is set by the spatial scale on which the random rotation angles fluctuate. 
	The presence of the chiral symmetry, Eq.~\eqref{chS}, on the other hand, allows the quantum walker to escape the common fate of Anderson localization.
	This happens when the system is fine tuned to the quantum critical point, separating two topologically different Anderson insulating phases, as we discuss next. 

\section{Critical distribution}
\label{CriticalDistribution}

	To elaborate on the last mentioned point, we consider the probability distribution,
	\begin{align}
	\label{msq}
		P_{\sigma'\sigma} (t,q) 
		&=
		\langle 
		|\langle q,\sigma' |\hat U^t |0,\sigma\rangle |^2
		\rangle_{\theta,\varphi},
	\end{align}
	for a walker who is initially in eigenstate $|\sigma\rangle=|\leftarrow\rangle,|\rightarrow\rangle$ of the chiral operator $\hat \sigma_2$. 
	This distribution yields the probability of the walker to be found after $t$ time-steps at a distance $q$ in eigenstate $|\sigma'\rangle$. 
	Here and in the following, $\langle \dots \rangle_{\theta,\varphi}$ denotes averages over distributions of angles.
Since particles conserve their (quasi-)energies, it is convenient to Fourier transform Eq.~\eqref{msq} to a 
	spectral representation 
	 	\begin{align}
	\label{msq_ft}
		P_{\sigma'\sigma} (\omega,q) 
		&= \int d\epsilon\, 
		\langle 
		\langle q,\sigma' |\hat G^R_{\epsilon+\frac{\omega}{2}} |0,\sigma\rangle 
		\langle 0,\sigma |\hat G^A_{\epsilon-\frac{\omega}{2}} |q,\sigma'\rangle 
		\rangle_{\theta,\varphi}.
		\end{align}
		Here we   
 	introduced the retarded (particle) and advanced (hole) propagators $\hat G^R_{\epsilon}= [1-e^{i\epsilon-0}\hat U]^{-1}$ and $\hat G^A_{\epsilon}=[\hat G^R_{\epsilon}]^\dagger$, respectively.
	 	
	The chiral symmetry relates particle and hole dynamics for states in the vicinity of particle-hole symmetric points  $\epsilon\simeq0, \pi$ in the $2\pi$-periodic spectrum. 
 More specifically, the chiral symmetry Eq.~\eqref{chS} translates into the relation, 
	\begin{equation}
 	\label{GRA}
 		\hat G^A_{-\epsilon}
 		=
 		\sigma_2 \hat G^R_{\epsilon}\sigma_2,
 	\end{equation}
indicating that the dynamics of particles and holes at a fixed energy 
is only related for 
  $\epsilon\simeq -\epsilon$. That is, for 
  states in the vicinity of particle-hole symmetric points $\epsilon\simeq 0, \pi$.

To account for the (breaking of) symmetry between particle and hole propagators in different ranges of the quasi-energy spectrum, we change to an energy representation of Eq.~\eqref{msq}, and separate Fourier components into two contributions~\cite{Altland:2015}, 
	\begin{align}
		P_{\sigma'\sigma}(\omega,q)
		&\simeq
		P^{\rm reg}
		(\omega,q)
		+
		P^{\rm chiral}_{\sigma'\sigma}
		(\omega,q).
	\end{align}
	Herein, the first (spin-independent) contribution results from single-particle states, with $|\epsilon|, |\epsilon-\pi |\gtrsim\omega$, for which energy detuning is large enough to break the chiral symmetry between propagators 
	that compose the probability distribution 
	as $P\sim G^RG^A$.
	Consequently, the probability distribution for states breaking the 
	chiral symmetry 
	is (on long time and length scales) identical 
	to that of conventional Anderson insulators.  
	That is, upon Fourier transform the first contribution is given by a static
probability distribution,
	\begin{align}
		P^{\rm reg} (t,q)  
		&\sim 
		\theta(t) e^{-|q|/\xi_{\rm loc}}.
	\end{align} 
	By contrast, the second contribution results from states with quasi-energies $|\epsilon|,|\epsilon-\pi |\lesssim |\omega|$, for which both propagators are related by chiral symmetry.
	That is,
	\begin{equation}
	\label{Pchiral}
		P^{\rm chiral}_{\sigma'\sigma}(\omega,q) \simeq
		|\omega| 
		\langle
		\langle q,\sigma| \hat G^R_{{\omega\over2}}|0,\sigma\rangle
		\langle 0,\sigma|  \hat G^A_{{- {\omega \over2}}}|\sigma,q\rangle
		\rangle_{\theta,\varphi}
	\end{equation} 
	encodes the critical dynamics of the walker, and a non-trivial time resolved behavior can be expected.
	We next apply field-theory methods to identify quantum critical points of the chiral walk and calculate the critical distribution in Eq.~\eqref{Pchiral}.

\begin{figure}[t]
	\begin{center}
		\includegraphics[width=6.0cm]{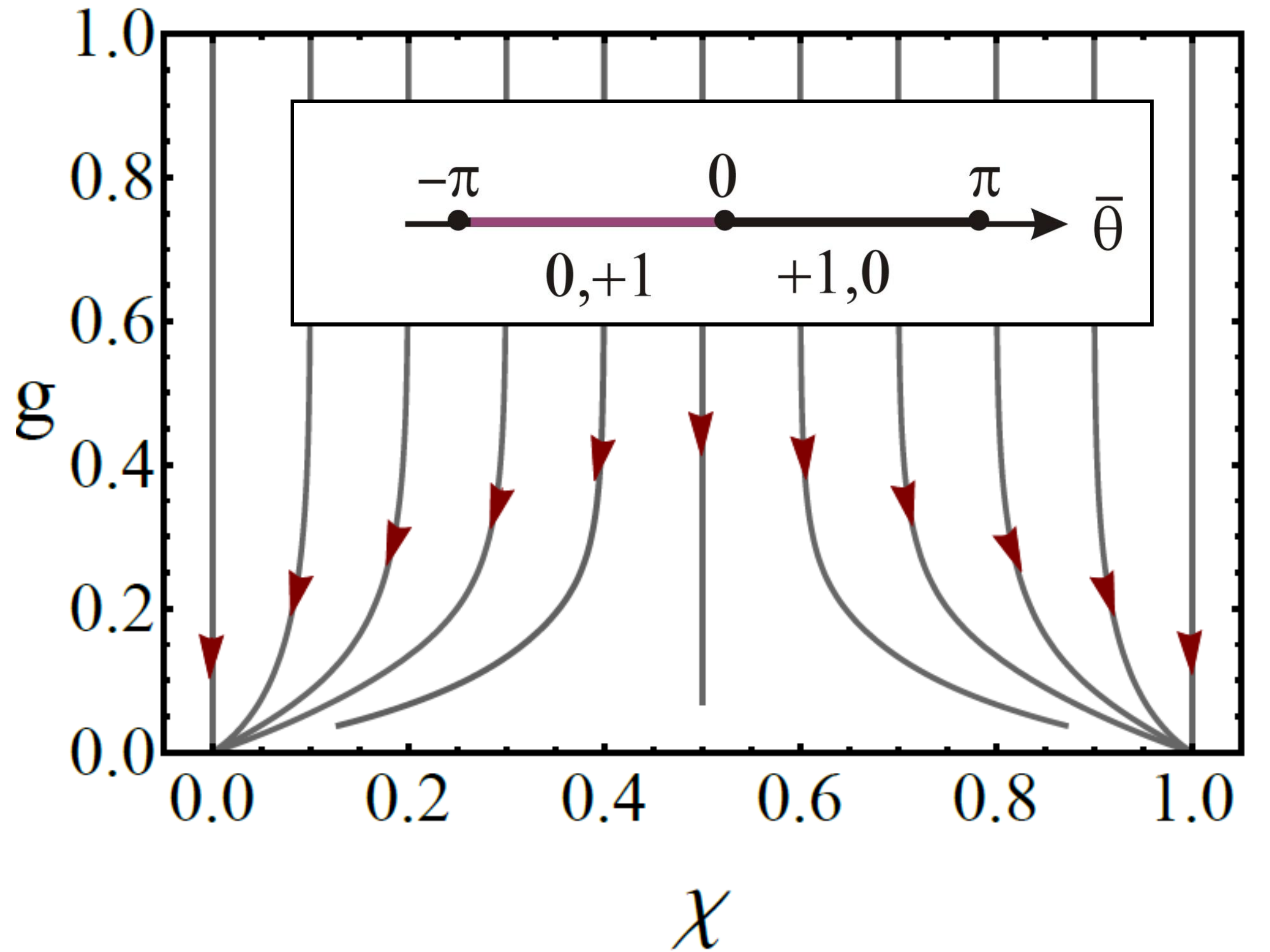}
	\end{center}
	\vspace{-15pt}
	\caption{
		Two-parameter flow of conductance $g(L)$ and  average topological index $\chi(L)$ 
		for class AIII nonlinear $\sigma$-model with bare values $\mathfrak{g}(1) \sim 1$ and $\chi(1) = \bar\chi$. 
		Inset: the phase diagram of the quantum walk. 
		Assuming $\bar \varphi \neq \pm \pi/2$ the system is at criticality provided $\bar \theta = 0$ or $\pi$. 
		Away from criticality, the pair of topological indices $(\chi^0, \chi^\pi)$ flow to either $(0,1)$ or $(1,0)$ which
		defines two distinct Anderson localized topological phases. For $\bar \varphi = \pm \pi/2$ the system is always critical. 
		}
	\label{fig2}
\end{figure} 

\subsection{Sinai diffusion}

		Following standard approach to disordered systems~\cite{wegner1979,efetov1980zh,pruisken1982anderson, efetov1983kinetics, Pruisken:1984, Efetov-book} 
		we describe the physics of the critical states 
		around $\epsilon\simeq 0,\pi$ 
		by a Ginzburg Landau type effective theory. 
		More specifically, we derive in Appendix~\ref{FieldTheory} a nonlinear $\sigma$-model action
		 which encodes the full quantum dynamics of soft diffusion modes and interference processes which eventually drive strong Anderson localization. What changes this conventional behavior in our case is a topological contribution to the effective action. 
		
		 The $\sigma$-model is parametrized by the frequency $\omega$ (which however does not `flow' in an renormalization group sense), and two coupling constants, viz. the 
		 conductance $\mathfrak{g}$ and topological angle $\chi$, see Eq.~\eqref{app_eq:S_0}. 
In a conventional disordered system $\mathfrak{g}$ follows a single parameter scaling~\cite{Abrahams:1979}, 
which in $1d$ predicts a single, Anderson insulating phase. 
The angle $\chi$, on the other hand, allows for a characterization of 
topologically different Anderson insulating phases. This opens the possibility to escape Anderson localization when fined tuned to specific, critical values, separating two topologically distinct Anderson insulators.
	
	The bare topological angle for the chiral quantum walk reads (see Appendix~\ref{FieldTheory} for details) 
	\begin{align}
		\bar\chi^\epsilon ={1\over2}\left(1 - e^{i\epsilon}\langle\sin(\theta)\cos(\varphi)\rangle_{\theta,\varphi}\right),
	\end{align} 
	with $\epsilon=0,\pi$ indicating the critical states described by the effective action.
	The presence of two coupling constants in the $\sigma$-model action --- the `conductance' and `topological angle'~\cite{Fulga:2011,Altland:2015,Altland:2014} ---
	leads to a two-parameter flow and resulting phase-diagram shown in Fig.~\ref{fig2}~\cite{Altland:2015}. 
	For generic bare values, the average topological angle flows  to the closest integer value $\chi=0$ 
	 and $1$, characterizing the two Anderson insulating phases realized by the chiral quantum walk, while $\chi=1/2$ defines the critical fixed-point line. 
	For general $\bar\varphi$, a critical line corresponds to $\bar\theta=0,\pm\pi$.
	The same configuration of angles in the clean limit gives rise to gapless linearly dispersing bands, as expected from analogy to the corresponding Hamiltonian system. 
	For generic $\bar\theta$, on the other hand, criticality is achieved at $\bar\varphi=\pm\pi/2$. 
	Finally, we remark that, in the strong disorder limit where angels are randomly drawn from the entire unit circle, $\gamma_{\theta,\varphi}=2\pi$, the quantum walk is always critical.
	The same effective action also describes disordered quantum wires with chiral symmetry and bare localization length $\xi_{\rm loc}=1/2$~\cite{Altland:2015}.
	Criticality in the strong disorder limit is, however, a peculiarity of the Floquet system.

	Concentrating then on the vicinity of a critical point, we can calculate the walker's critical distribution Eq.~\eqref{Pchiral}.
	The rather technical calculation is detailed in Appendix~\ref{TransferMatrix} and indicates the scaling form
	\begin{align}
	\label{Pscal}
		P^{\rm chiral}_{\sigma'\sigma}
		&={\cal N}(t){\cal F}_{\sigma'\sigma}(|q|\xi^{-1}_t),
		\qquad
		\xi_t\equiv \frac{2}{\pi^2}\ln^2t,
	\end{align} 
	with a time-dependent normalization factor 
	 ${\cal N}(t)\propto \ln^{-5}\!t$ and the explicit form of ${\cal F}(x)$ to be given given below.~\footnote{Notice that this implies that $\sum_{q, \sigma'\sigma} P^{\rm chiral}_{\sigma'\sigma}(t,q) = 1/(4\ln^2 t)$ as shown in Appendix~\ref{TransferMatrix}.}
	The scaling of  $\xi_t$ implies anomalously retarded `Sinai diffusion' of critical states, characterized by a mean displacement~\cite{Sinai:1982,Bouchaud:1990,Comtet:1998}  
	\begin{align}
		\langle |q| \rangle 
		&\sim \log^2 t.
	\end{align}
	Another feature of the critical distribution is the dependence on spin orientations $\sigma \sigma' = \pm 1$ with reference to the chiral symmetry $\sigma_2$. 
	This can be seen from the scaling functions in the long time and distance limits, $t,q \gg 1$,  
	\begin{align}
	\label{chPt}
		{P}^{\rm chiral}_{\sigma'\sigma}(t,q) &\propto
		\frac{1}{\ln^5 t}\sum_{n=1}^\infty (\sigma \sigma')^{(n+1)} n^2 e^{-n^2 |q|/\xi_t}.
	\end{align} 
	Focusing on the tails $|q|\gg \xi_t$, one finds from Eq.~\eqref{chPt} the exponential profiles   
	\begin{align}
		{\cal F}_{\pm \sigma, \sigma}(x) = e^{-x} \pm 4e^{-4 x} + \ldots ,
	\end{align} 
	where the leading spin-independent contribution reminds us of conventional Anderson insulators. 
	The  directly following terms indicate, however, dependence of the critical distribution on the spin orientation of the final state $\pm\sigma$,  with interesting implications, being discussed in the next subsection. 
	The full distributions [cf. Eq.~\eqref{chPt}] are shown in Fig.~\ref{fig3}, and we refer the interested reader to Appendix~\ref{TransferMatrix} for more detailed analytical expression.
	We next discuss how the characteristic features of the critical distribution, viz.
	(i) slowly increasing width $\xi_t$ in time
	and (ii) dependence on spin-orientation with respect to the basis of the chiral symmetry operator, can be observed in experiments.

\subsection{``Time-staggered" spin polarization}

\begin{figure}[t]
	\centering
	\includegraphics[width=.46\textwidth]{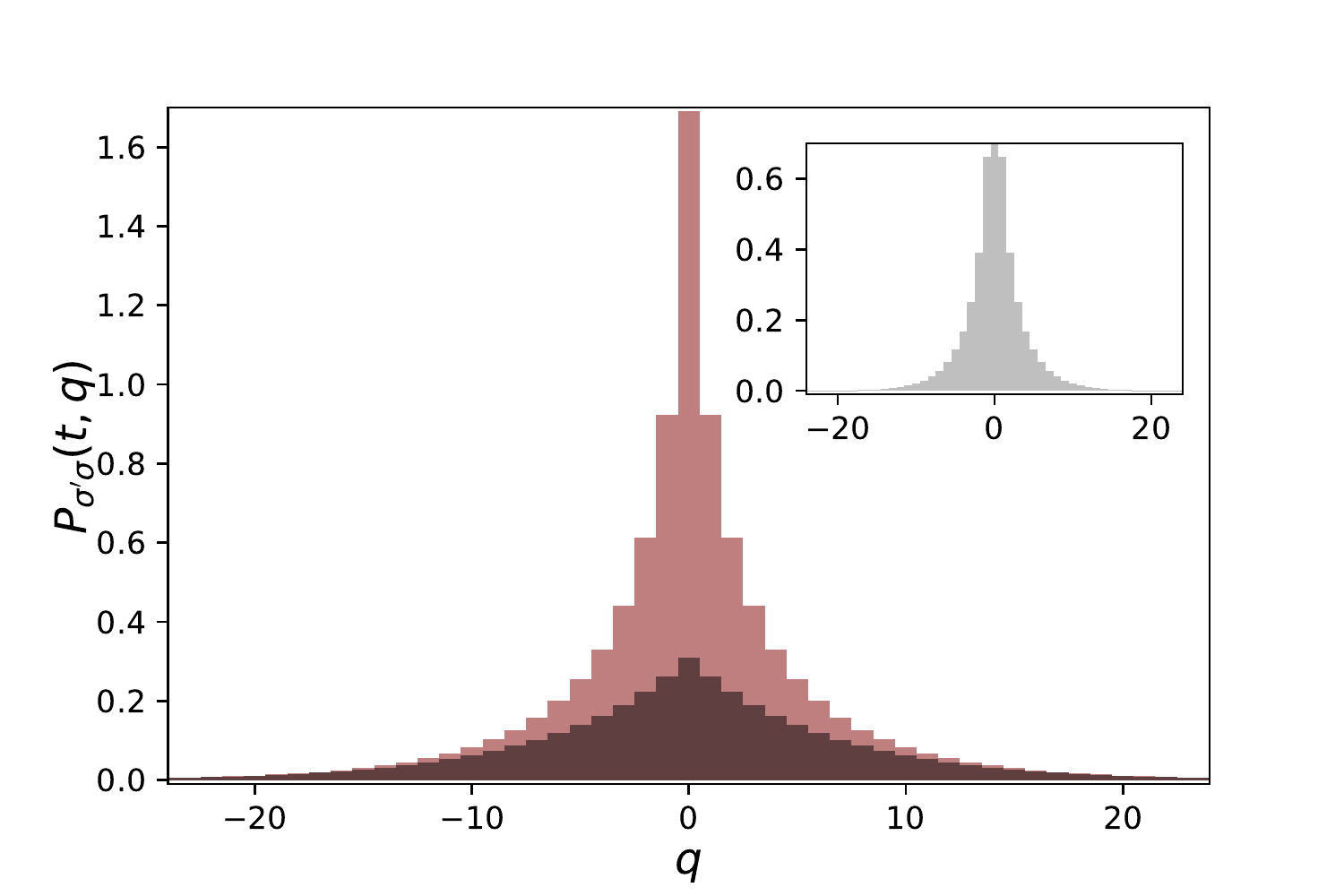}
	\vspace{-15pt}
	\caption{Walker's critical probability distribution 
			$P^{\rm chiral}_{\sigma'\sigma}(t,q)$, Eq.~\eqref{chPt}, for $t=10$. 
			Spin-configurations $(\sigma',\sigma)$ of final and initial states are aligned
			$(\rightarrow,\rightarrow)$ (light red) and anti-aligned $(\leftarrow,\rightarrow)$ (dark red), 
			and distributions are
			normalized by the average return probability per spin. Inset: Spin polarization $\Delta P(t,q)$ (peak at origin is not fully shown). }
	\label{fig3}
\end{figure}

	Sinai diffusion has previously been predicted for disordered one-dimensional systems with particle-hole symmetry~\cite{Balents:1997,Bagrets:2016}, and our above result for a system with chiral symmetry indicates  that it is a universal dynamical feature at one-dimensional topological Anderson localization transitions.
	Arguably, observation of the weak logarithmic time-dependence presents an experimentally intriguing challenge~\footnote{Sinai diffusion has recently been suggested to 
	leave traces of unconventional heat propagation in the form of non-monotonically propagating thermal current pulses in quasi-one-dimensional topological superconducting wires near criticality~\cite{Bagrets:2016}.}. 
	Recalling, moreover, the contribution from non-critical states to the total probability distribution implies that Sinai diffusion is generally masked by conventionally Anderson localization.
	Complicating this matter even further, the number of critical states resolved in time $t$ reduces as $|\omega|\sim 1/t$.
	This generates additional time dependencies in the distributions of non-critical and critical states, as summarized in the normalization ${\cal N}(t)$ of Eq.~\eqref{Pscal}. 
	The optical linear network realization of a quantum walk  discussed in the introduction, on the other hand, allows for a  direct observation of spin-resolved probability distributions. 
	This opens an interesting opportunity to observe the second feature, i.e., the peculiar spin-dependence of the critical walk.
	Specifically, this suggests to measure the difference  
	\begin{align}
		\Delta P(t,q)\equiv P^{\rm chiral}_{\rightarrow\rightarrow}(t,q)-P^{\rm chiral}_{\leftarrow\rightarrow}(t,q),
	\end{align} 
	which only depends on the critical contribution to the total probability. 
	A finite spin polarization of the critical walker may be viewed as a precursor of spin polarized boundary states emerging in the topologically non trivial phase~\cite{Asboth:2013,Mondragon-Shem:2014}~\footnote{Topological boundary modes can be introduced, for example, by connecting two 1-dimensional quantum walk systems characterized by $\hat U(\varphi=0,\theta)$ and $\hat U(\varphi=0,-\theta)$, respectively. The spin expectation of boundary modes are eigen values of chiral operator $\sigma_2$. }.

	The $q$-dependence of the difference $\Delta P(t,q)$ is shown in the inset of Fig.~\ref{fig3}. 
	The corresponding long time probability distributions for non-critical states 
	and conventional Anderson insulators (with spin orbit interaction) do not keep the memory 
	of the  initial spin-configuration. The observation of $\Delta P(t,q)$ 
 would thus provide clear evidence for the critical walk at an Anderson localization transition. 

	Further smoking-gun evidence for the critical distribution is then obtained from an additional symmetry of the Floquet operator, not discussed so far.
	The discrete lattice structure motivates the introduction of the sublattice operator~\cite{zhao2015disordered}
	\begin{align}
	\label{SubLat}
		\hat S
		&\equiv 
		\sum_q |q\rangle(-1)^q\langle q|,
	\end{align} 
	anti-commuting with the Floquet operator $\hat U$.
	From Eq.~\eqref{SubLat}, one can construct a chiral-sublattice operator
	\begin{align}
		\hat {\cal C}_{\rm sl}\equiv\sigma_2 \otimes  \hat S,
	\end{align} 
	satisfying $\hat {\cal C}_{\rm sl} \, i \hat U \, \hat {\cal C}_{\rm sl}  =  (i \hat U)^\dagger$, and consequently resulting in
	\begin{align}
		\hat G^A_{ \epsilon_0-\epsilon}
		&=
		\hat {\cal C}_{\rm sl} \, \hat G^R_{ \epsilon_0 + \epsilon} \, \hat {\cal C}_{\rm sl},
	\end{align}
	whenever $\epsilon_0=\pm \tfrac{\pi}{2}$. 
	That is,  $\hat {\cal C}_{\rm sl}$ is an additional chiral symmetry that applies to critical states $\epsilon\simeq\pm \tfrac{\pi}{2}$.
	This symmmetry is also visible in the density of states as we show in Appendix~\ref{DoS}.  
	Interestingly, $\hat {\cal C}_{\rm sl}$ has different implications for time evolution when extending over an even or odd number of time steps $t$.  
	We show in Appendix~\ref{TimeStaggering} that, for critical states induced by the chiral-sublattice symmetry $\hat {\cal C}_{\rm sl}$, the spin-polarization alternates in between time-steps; that is 
	\begin{align}
	\label{stagg}
		\Delta P(t,q)
		&= 
		(-1)^t |\Delta P(t,q)|
	\end{align}
	holds true.
	The two main obervations leading to Eq.~\eqref{stagg} are the following (for a more rigorous explanation,  see Appendix~\ref{TimeStaggering}).
	(i) The Floquet operator induces transitions between subspaces of opposite site-parity.
	That is, walkers positioned at an even site propagate in the following time step to an odd site, and vice versa. 
	(ii) Eigenstates of $\hat {\cal C}_{\rm sl}$ have alternating spin-polarization on even and odd sites;  e.g., $|q,\rightarrow\rangle$ are eigenstates of $\hat {\cal C}_{\rm sl}$ with eigenvalues $(-1)^q$, and 
	analogously for $|q,\leftarrow\rangle$. 
	Combining both observation, we notice that walkers propagating for an {\it even} number of time steps have spins aligned if their initial and final states are eigenstates of $\hat {\cal C}_{\rm sl}$ to the same eigenvalue. 
	(The same applies for the chiral operator $\sigma_2$.)
	In contrast, for an {\it odd} number of time steps, the walker has spins anti-aligned if initial and final states are eigenstates of $\hat {\cal C}_{\rm sl}$ to the same eigenvalue.
	This difference simply follows from the observation that, for an odd number of time steps, the walk starts and ends in opposite parity sectors.   
	Finally, we also remark that the probability $P_{\sigma'\sigma}$ has to be read as the transition within ($\sigma'=\sigma$) or between ($\sigma'=-\sigma$) eigenspaces of the chiral operator.
	Combining the above, it follows that for odd numbers of time steps spin polarization reverses its sign. 
	We thus conclude that, for critical states induced by the chiral-sublattice symmetry $\hat {\cal C}_{\rm sl}$,  the spin-polarization alternates in between time steps, as indicated in Eq.~\eqref{stagg}.

\section{Experimental proposal} 
\label{Experiment}

	We now devise an experimental protocol which allows us to observe the discussed characteristic features of the quantum critical walk.

\subsection{Experimental protocols}
\label{Exp_protocols}

\begin{figure}[t]
	\centering
	\includegraphics[width=.45\textwidth]{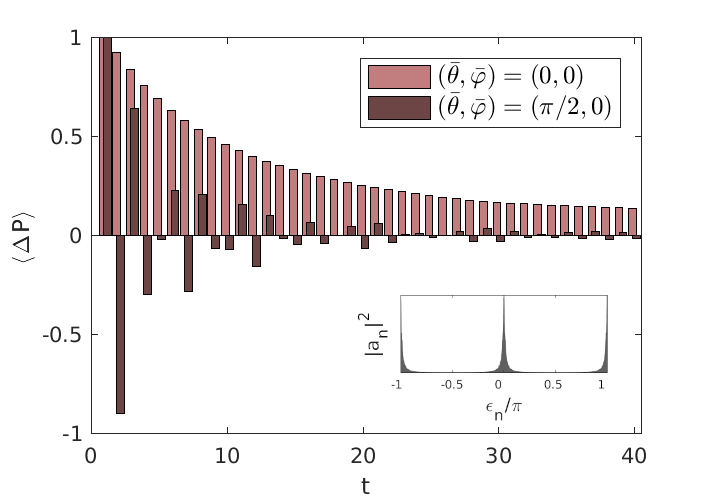}
	\caption{
		Spin-polarization $\Delta P(t)$ as a function of time steps $t$ for the initial states $|\psi_M^{0}\rangle $ with 
		$M = 10^2$ and different choices of mean angles $(\bar\theta, \bar\varphi)$.
		$\Delta P(t)$ remains finite for a large number, $t \sim {\cal O}(10^2)$, of time steps (light red) if the walker probes critical states $\epsilon \simeq 0$ corresponding 
		to $(\bar\theta, \bar\varphi) = (0,0)$ (cf. a spectral decomposition of the initial state as shown in the inset, with $a_n= |\langle \epsilon_n|\psi_M^0 \rangle|^2$).
		On other hand, $\Delta P(t)$ is scrambled (dark red) if 
		non-critical states at $(\bar\theta, \bar\varphi) = (\pi/2,0)$ are probed.
	}\label{fig4}
\end{figure}

\begin{figure}[t]
	\centering
	\includegraphics[width=.45\textwidth]{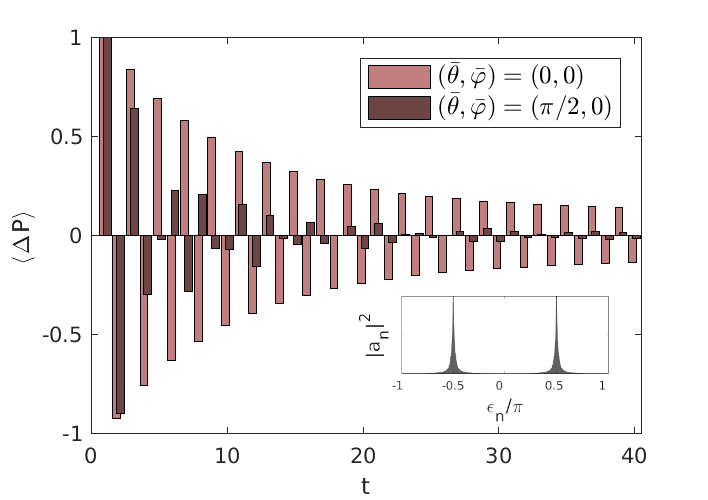}
	\caption{
		The spin-polarization $\Delta P(t)$ shows the predicted time-staggered behavior with a long-lived finite amplitude for the initial state $|\psi_M^{\pi/2}\rangle$ with $M=10^2$
		which is chosen to filter critical energies $\epsilon \simeq \pm \tfrac \pi 2$ 
		related to the chiral-sublattice symmetry $\hat C_{\rm sl}$.
		The criticality implies $(\bar\theta, \bar\varphi) = (0,0)$  (light red).
		Similar to Fig.~\ref{fig4}, $\Delta P(t)$ is scrambled 
		when non-critical states at $(\bar\theta, \bar\varphi) = (\pi/2,0)$ are probed (dark red).
	}\label{fig5}
\end{figure}

	Quantum walks are typically initialised at a localised position and thus involve eigenmodes from the entire quasi-energy domain.
	The experimental challenge then is to restrict the dynamics to states that approximately preserve chiral symmetry. 
	As discussed in the previous section, the Floquet evolution operator of the quantum walk induces the transitions between the sites
	of different parity only.
	We thus suggest to prepare an initial state of the single photon as a coherent superposition described by
	\begin{equation} 
	\label{eq:psi_i}
		|\psi_M^{p_0}\rangle=\frac{1}{\sqrt{M+1}}\sum\limits_{|q|\leq M/2}  |(2 q)\rangle \otimes |\rightarrow\rangle \,e^{ 2 i q p_0 },
	\end{equation}
	which occupies ($M+1$) even lattice sites, 
	where $M\gg 1$ can be used to enhance the population of states at energies $\pm\epsilon_{p_0}$
	(see also below discussion of initial states in Figs.~\ref{fig4} and~\ref{fig5}).

	Alternatively, one can make use of the equivalence of coherent light and a single quantum particle when propagating in a linear optical network and directly apply a train of coherent laser pulses.
	With $|\psi_M^{p_0}\rangle$ as the delocalised initial state, time dependence of the localization length $\xi_t$ cannot be captured, but the spatially integrated spin-polarization $\Delta P(t)$ can serve as a key measure for critical phases.  
	Specifically, we define the latter as 
	\begin{align}
		\Delta P(t) 
		&\equiv
		\sum_q \Delta P_\psi(t,q),
	\end{align}
	where $\Delta P_\psi(t,q) \equiv P_{\rightarrow\psi}(t,q) - P_{\leftarrow\psi}(t,q)$ and the spin-dependent local probabilities are
	\begin{equation}
	\label{eq:P_sigma_psi}	
		P_{\sigma \psi}(t,q) \equiv \langle |\langle q,\sigma |\hat U^t |\psi_M^{p_0}\rangle |^2 \rangle_{\theta,\varphi}.
	\end{equation}

	Figures~\ref{fig4} and~\ref{fig5} show our numerical simulations for the spin polarization $\Delta P(t)$ using the initial state from Eq.~\eqref{eq:psi_i} for $M=10^2$. We here did not assume periodic boundary conditions, i.e. the signal could propagate without restrictions to the left and right.	In these plots, light red histograms simulate the critical walk at the topological Anderson localization transition, corresponding to critical energies $\epsilon \simeq 0,\pi$  
	(Fig.~\ref{fig4}) and $\epsilon \simeq \pi/2$ (Fig.~\ref{fig5}), respectively.
	It can be clearly seen that a finite (spin-staggered) polarization is maintained up to over $t=40$ time-steps, which is in the reach of current experiments, indeed \cite{boutari2016large,xu2018measuring, nitsche2018probing}. 
	By contrast, the dark red curve is a simulation of the quantum walk in a conventional Anderson insulating phase 
	(see also  phase diagram shown in inset of Fig.~\ref{fig2}). 
	In this case, the spin polarization of the initial state $|\psi_M^{p_0}\rangle$ quickly scrambles and decays. 
	In all simulations static uncorrelated angles $\theta_q$ and $\varphi_q$ were randomly drawn from intervals 
	$(\bar \theta -\delta, \bar \theta + \delta)$ and $(\bar \varphi-\delta, \bar \varphi + \delta)$
	of size $2\delta=\pi/4$ with $\bar \theta$ and $\bar \varphi$ referring to their mean values, and we performed ensemble average 
	over $5 \cdot 10^3$ realizations.
	Overall, we find clear evidence of the discussed features in different variants of the suggested protocol, starting at a number of $t\sim {\cal O}(10)$ time steps. As we have also checked, the results demonstrated in Figs.~\ref{fig4} and \ref{fig5} remain qualitatively unchanged 
	provided only one angle, $\varphi_q$, is random but $\theta_q = \bar\theta $ does not fluctuate, which potentially is easier to realize in practice as discussed in the next subsection.
	
\subsection{Experimental setup}

\begin{figure}[t]
	\includegraphics[width=.45\textwidth]{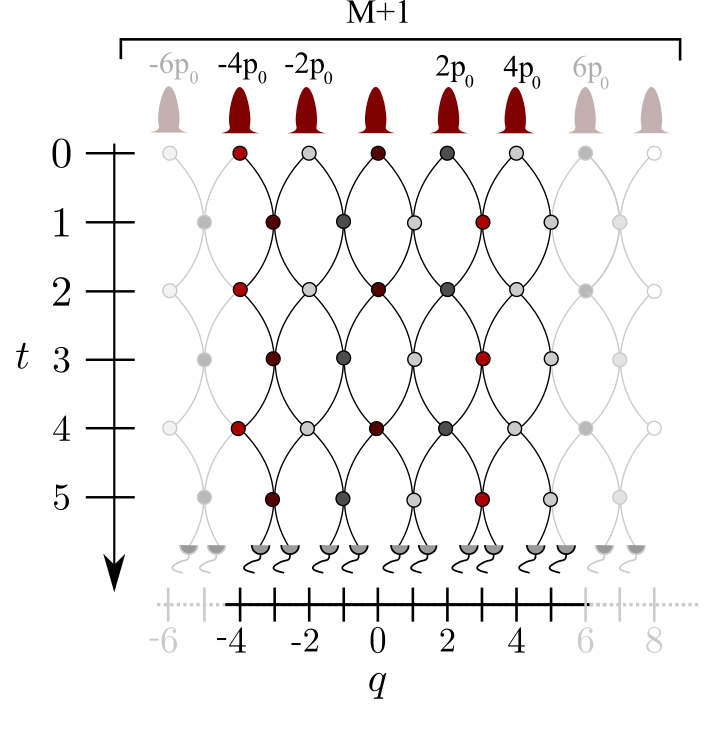}
	\caption{
		A prototype of a linear optical network to realize a quantum critical walk discussed in details in
		the main text.
		A train of $M+1$ coherent pulses with a fixed relative phase difference, $2 p_0$, between adjacent pulses enters neighbouring ports of the network.
		Each vertex illustrated the coupling of two spatial modes and is implemented by a series of waveplates and a polarizing beam splitters, realising the operators according to Eq.~(\ref{sqw}). 
		The different colors of the vertices denote the (static) randomness in the phase $\varphi_q$. 
		Eventually, the distribution for each disorder realization is measured polarization-resolved in the circular basis. 
	}\label{fig:network}
\end{figure}

	A schematic drawing of such quantum critical walk is shown in Fig.~\ref{fig:network}. 
	Pulses with a fixed phase relation are entering neighbouring input modes of the network.
	Over the course of the evolution, they undergo polarisation rotations with particular angles (indicated by the different colours of vertices) and a polarisation dependent routing.
	Finally, the detectors resolve internal (i.e., polarization) as well as external degree of freedom for extracting spin-resolved probability distributions.
	Ensemble averages over a few thousand realisations of disorder are necessary since a single realisation only shows very little signatures of the critical states because of the impact of all the non-critical states.
	Only through this averaging procedure, the staggering behaviour of the critical states becomes visible and can be reliably extracted.

    In addition to the precise control of all local coin rotations and the easy reconfigurability of the experiment to programme the high number of realisations, the generation of the delocalized input state is one of the main experimental challenges.
	When considering a spatial implementation of the quantum walk network \cite{do2005experimental,broome_discrete_2010, xue_experimental_2015}, a spatial-multiplexing techniques, as in \cite{wang2019boson}, can be adopted to produce the input state to be fed into the network ports.
	Analogously, time-multiplexing networks \cite{schreiber_photons_2010,he2017time} can be adapted in the (temporal) position spacing to be directly compatible with the pulse train produced by a coherent cavity laser source.
	For the advanced control of the phase between the pulses, we envision the usage of directly modulated light source~\cite{Yuan:2016}.
	Alternatively, an external time-multiplexing loop, as suggested for driven quantum walks in \cite{hamilton2016driven} and which controls timings and phases of the initial pulse train, can be connected to the setup.
	Standard optical waveplates take care of the desired circular input polarization resulting in the state $|\psi_M^{p_0}\rangle$---i.e., a state in the form of Eq.~(\ref{eq:psi_i}).

	The non-random rotations $R_x(\bar\theta/2)$ and the random rotation $R_z(\bar\varphi_q/2)$, forming the coin operator, are implemented by (spatially varying) half (HWP) and quarter waveplates (QWPs), being positioned independently at each node of the network.
	In time-multiplexing networks, fast switching electro-optic modulators can be utilized to introduce the random phases $\varphi_q$  \cite{Schreiber:2011, Schreiber:2012} in a controlled fashion.
	Since the evolution in the network typically takes place in horizontal and vertical polarisation, the measurement basis has to be rotated again to circular states, such as by using QWPs at $45^\circ$ angle in front of the detectors.
	Crucially, to extract the spin-polarization, $ \Delta P(t)$, 
	both polarization modes must be measured separately for every step.
	In Appendix~\ref{Details}, we provide estimates for time and spacial scales which validate a feasibility of our proposal within existing experimental techniques.

\section{Discussion}
\label{Discussion}

We have studied the one-dimensional quantum walk of a spin-1/2 particle
with chiral symmetry and tunable disorder. 
The quantum walk allows to realize topologically different Anderson insulating phases, and 
can be tuned to a quantum critical point separating two 
such phases. 
Building on a low energy effective field theory, we have derived the walker's phase-diagram 
 as a function of the average values of the coin operators rotation angles and their
 variances.  
Notice that in a periodically driven system all eigenmodes in the quasienergy domain typically contribute
to physical responses. In the presence of delocalized bulk modes, Floquet systems therefore  
cannot be called topological insulators (regardless of the presence of topological boundary modes 
and a finite gap in the quasi energy spectrum). 
This is different in the presence of disordered, where quantum interference drives 
the system into Anderson localized phases (as summarized in 
the two-parameter flow shown Fig.~\ref{fig2}). 
We found that the critical point between two topologically distinct Anderson insulators 
is characterized by the spin dependent probability distribution.

 The critical dynamics reflects the 
 competition between strong localization in a $1d$ system on the scale of the mean free path, 
	 and the formation of a nontrivial topological invariant which forces long range correlation through the system.
We calculated the critical probability distribution of the walker, and verified that 
  the powerful, opposing strong localization in $1d$ manifests in extremely slow critical dynamics. That is, 
  in Sinai diffusion as previously also found for quasi one-dimensional 
 disordered topological superconductors~\cite{Bagrets:2016}. 
We identified a (time-staggered) spin polarization as a promissing observable signature witnessing the quantum 
	  critical correlations. More specifically, we noticed that the walker's critical distribution keeps
 memory of the initial spin configuration, when prepared in an eigenstate of the chiral operator. 
Moreover, we noted that the combination of chiral hopping on a discrete lattice and the 
chiral symmetry leads to a second, `chiral lattice symmetry'.  
The spin-polarization then becomes staggered in time, 
when critical states protected by this second symmetry are probed. 
The underlying mechanism suggests that, quite generally,  
in systems with chiral hopping on a discrete lattice one may
expect time-staggering of observables which are senstitive to the eigenvalues of 
the chiral symmetry operator.

Taking advantage of the versatile opportunities offered by optics, 
we have proposed a protocol that should allow for the observation of 
the spin polarization within existing experimental set up. 
One experimental challenge is 
to minimize contributions from uncritical states that suffer from conventional Anderson localization and 
which may mask the spin polarization. 
We proposed to filter critical states in the quasi-energy spectrum by preparing 
the walker initially in a spatially extended state. 
We confirmed the viability of our proposal by
numerically simulations of the protocol for experimentally realistic system parameters, and 
verfied a strong suppression of the spin polarization 
by either activating uncritical states, breaking chiral symmetry or introducing dephasing. 
For an experimental platform we e.g. 
 indicate an optical linear network similar to that used in Ref.~\cite{Schreiber:2011}. 
 The preparation of an extended initial state with a stable fixed phase relation may 
 still be challenging. We are, however, optimistic that some variant of the protocol is 
 in experimental reach in one of the discussed platforms. 
The experimental observation of the time staggered spin polarization 
would provide intriguing evidence for the quantum critical dynamics
manifesting as a competition of strong localization and nontrivial topology 
in disordered quantum systems.

\begin{acknowledgments}
	We wish to thank  Benjamin Brecht for discussions. 
	TM acknowledges financial support by Brazilian agencies CNPq and FAPERJ. 
	AA, DB, and KWK were funded by the Deutsche Forschungsgemeinschaft (DFG) Project No. 277101999 TRR~183 (project A01/A03).
	 The Integrated Quantum Optics group acknowledges financial support through the Gottfried Wilhelm Leibniz-Preis (Grant No. SI1115/3-1) and the European Commission through the ERC project QuPoPCoRN (Grant No. 725366).
\end{acknowledgments}

\appendix

\section{Effective field theory}
\label{FieldTheory}

	In this section, we discuss how the evaluation of the probability distributions $P_{\sigma'\sigma}(t,q)$ at large distance scales $q \gg 1$ can be brought in the framework of the effective SUSY field theory, as discussed in the main text.  
	We start by introducing a fictitious local gauge transformation of the basis, defined by the unitary operator 
	\begin{align}
		\hat G_\psi 
		&=   \sum_{q=0}^{N-1}  |q , \sigma  \rangle e^{i\psi_q}  \langle q , \sigma|,
	\end{align}
	with $\psi_q = q \phi_0 + \psi_0$. 
	Here $\phi_0$ and $\psi_0$ are some constants; the phase $\psi_q$ is a linear ramp; and, to comply with the periodic boundary conditions, we require that $\phi_0 = 2\pi n/N$ with $n \in \mathbb{Z}$.
	Then, by definition, $ \hat G_\psi   |q , \sigma  \rangle =   |q , \sigma  \rangle e^{i\psi_q} $ holds true, and this enables one to represent Eq.~(4) in the equivalent form 
	\begin{align}
		P_{\sigma'\sigma}(t,q)
		&=
		\langle 
		|\langle q,\sigma' |G_\psi^\dagger \hat U^t  G_\psi |0,\sigma\rangle |^2
		\rangle_{\theta,\varphi}.
	\end{align}
	This is only nominally $\psi$-dependent, and the usefulness of such artificial representation is going to be evident momentarily.
	As one can see, the operator $\hat G_\psi$ commutes with the (local) coin operator $\hat R$.
	On the other hand, $\hat G^\dagger_\psi \hat T  \hat G_\psi = e^{i\phi_0 \sigma_3 } \hat T$ applies.
	With this observation, we can write the probability distribution as
	\begin{align}
		P_{\sigma'\sigma}(t,q) 
		&= 
		\langle | \langle q , \sigma' | [\hat U_{\!\phi_0}]^{t}  |0 , \sigma \rangle |^2\rangle_{\theta,\varphi}, 
		\\
		\text{ where }
		\hat U_{\phi_0} 
		&= 
		e^{i\frac{\hat \varphi}{2} \sigma_3 }  e^{i \frac{\hat\theta}{2}\sigma_1} e^{i\phi_0 \sigma_3 }  \hat T 
		e^{\frac{\hat\theta}{2}\sigma_1} e^{i\frac{\hat \varphi}{2}\sigma_3},
	\end{align}
	and average the latter over the auxiliary angle $\phi_0$. 

	For that, let us consider the (disorder specific) SUSY action 
	\begin{align}
		S_0[\bar{\psi},\psi] 
		&= 
		\int dq' \, \bar \psi_{q'} (1 -  e^{i\frac\omega 2 - 0 }\hat U_{\phi_0}) \psi_{q'}. 
	\end{align}
	Here the fields $\bar{\psi}_q=\{\bar{\psi}^{\alpha,\sigma}_q\}$ are four-component supervectors, consisting of (anti-)commuting components $\alpha=({\rm f}){\rm b}$ and carrying spin index $\sigma = \pm$. 
	The latter denote the eigenvalue of a spin operator $\hat s_i$, which we here keep rather general, i.e. $i=1,2,3$, although our focus was on the chiral operator $i=2$ in the main text.
	The notation $\int\,dq$ above is symbolic in the sense that operator $\hat U_{\phi_0}$ in fact maps the spinor $\psi_q$ onto $\psi_{q \pm 1}$.
	On taking into account the chiral symmetry, $\hat G^A_{-\epsilon} = \sigma_2 G^R_{\epsilon} \sigma_2$, the probability in Eq.~\eqref{Pchiral}  
	can be then obtained via a Gaussian functional average
	\begin{align}
	\label{eq:P_chiral_SM}
		&P_{\sigma'\sigma}^{\rm chiral}(\omega,q)  \\
		=& |\omega|\bigl \langle 
		\int {\cal D}(\bar{\psi},\psi) \,
		\psi_q^{{\rm b}\sigma'} \bar \psi_0^{{\rm b}\sigma} 
		[\sigma_2 \psi_0^{{\rm f}}]^\sigma [\bar \psi_q^{{\rm f}} \sigma_2 ]^{\sigma'}
		e^{-S_0[\bar{\psi},\psi]}
		\bigr\rangle_{\theta,\varphi} \nonumber
	\end{align} 
	(here we used $(1 -  e^{i\frac\omega 2 - 0 }\hat U_{\phi_0}) \equiv [G_{\frac{\omega}{2}}^R]^{-1}$).
	To facilitate the subsequent derivation, it is advantageous to
	augment the action by a source term, 
	\begin{align}
		S_J[\bar{\psi},\psi] 
		&= 
		S_0[\bar{\psi},\psi] - \int dq' \, \bar \psi_{q'} ( j_{q'}  \sigma_2 )\psi_{q'}.
	\end{align}
	Here the current $j_{q'} = \alpha\delta_{q'0}\pi^{\rm bf}\otimes \pi^\sigma_i + \beta\delta_{q'q}\pi^{\rm fb}\otimes \pi^{\sigma'}_i $ involves projection matrices in spin- and graded-space, 
	$\pi^\sigma_i=\tfrac{1}{2}(1+\sigma\sigma_i)$, 
	$\pi^{\rm bf}=(\begin{smallmatrix}0&1\\0&0\end{smallmatrix})$, 
	$\pi^{\rm fb}=(\begin{smallmatrix}0&0\\1&0\end{smallmatrix})$,
	and affords a calculation of the probability distribution according to 
	\begin{align}
		P_{\sigma'\sigma}^{\rm chiral}(\omega,q) 
		&=
		|\omega|\langle \partial^2_{\alpha\beta}{\cal Z}_J\rangle_{\theta,\varphi}|_{\alpha=\beta=0},
	\end{align}
	where ${\cal Z}_J$ is a partition sum of the action $S_J[\bar{\psi},\psi]$.
	This identity can be checked by a straightforward computation which invokes the Wick's theorem for the Gaussian action $S_0$. 

	The probability in Eq.~\eqref{eq:P_chiral_SM} is $\phi_0$--independent by construction.
	Thus, one can average the generating function over all equivalent gauge configurations, $\langle {\cal Z}_J \rangle_{\phi_0} =   \int_0^{2\pi} \tfrac{d\phi_0}{2\pi} {\cal Z}_J $.
	This integral can be done via the color-flavor transformation~\cite{Zirnbauer:1996} by trading the `microscopic' degrees of freedom $\bar{\psi}_q$, $\psi_q$ for bi-local matrix fields $\bar{Z}_{qq'}$, $Z_{qq'}$, representing the Goldstone, viz. diffusion modes, of the disordered single-particle system.
	To see its working principle, let us decompose the Floquet evolution operator as 
	\begin{align}
		U_{\phi_0} 
		&= 
		V_1(\theta,\varphi)T_+ e^{i\phi_0 \sigma_3} T_- V_2(\theta,\varphi),
	\end{align}
	where partial `coin' rotations are $V_1(\varphi,\theta) = e^{i\frac{\varphi_q}{2} \sigma_3 } e^{i\frac{\theta_q}{2} \sigma_1}$ and $V_2(\varphi,\theta) = e^{i\frac{\theta_q}{2} \sigma_1} e^{i\frac{\varphi_q}{2}\sigma_3}$, while $T_\sigma = T \pi_3^\sigma + \pi_3^{-\sigma}$ are `shift' operators describing individual hopping of the spin up and down particle to the left and right, respectively.
	If one further introduces auxiliary spinors
	\begin{equation}
	\label{eq:spinors}
	\begin{aligned}
		(\psi_1^T, \psi_2^T) =&  \bar \psi  e^{i \frac\omega 2}\,  V_1(\theta,\varphi)T_+,  \\
		(\psi_2' , \psi_1')^T =&  T_- V_2(\theta,\varphi) \,
		\psi,
	\end{aligned}
	\end{equation}
	where the two component structure refers to the spin subspace, then the free action $S_0$ can be cast into the equivalent form 
	\begin{align}
	\begin{split}
		&S_0[\psi,\bar\psi] 
		\\
		=&  
		\int dq \, \left( \bar \psi_q \, \psi_q - \psi_{1,q}^T e^{i\phi_0} \psi_{2,q}' - \psi_{2,q}^T e^{-i\phi_0} \psi_{1,q}' \right). 
	\end{split}
	\end{align}

	At the heart of the color-flavor transformation lies the identity~\cite{Zirnbauer:1996, Altland:2015a}
	\begin{align}
	\label{cft}
		&\int_0^{2\pi} \frac{d\phi_0}{2\pi}\, 
		e^{ \sum_q \, \left( \psi_{1,q}^T e^{i\phi_0} \psi_{2,q}'  + \psi_{2,q}^T e^{-i\phi_0} \psi_{1,q}' \right)} 
		\\
		=& 
		\int dZd\bar Z\, {\rm sdet}(1 - \bar Z Z)\,
		e^{\sum_{qq'}\,  \left(  \psi_{1,q}^T  Z_{qq'} \psi_{1,q'}'  +  \psi_{2,q}^T\bar Z_{qq'} \psi_{2,q'}'  \right)}, \nonumber
	\end{align}
	and `${\rm sdet}$' referes to the graded determinant. 
	Here $\bar{Z}=\{ \bar{Z}^{\alpha\alpha'} \}$ and $Z=\{ Z^{\alpha\alpha'} \}$ are the (graded) matrix-fields mentioned above, with components $\alpha,\alpha'\in\{{\rm b}, {\rm f}\}$,  $\bar{Z}^{\rm bb}=-[Z^{\rm bb}]^\dagger$ and $\bar{Z}^{\rm ff}=[Z^{\rm ff}]^\dagger$ to guarantee convergence of Eq.~\eqref{cft}, and the additional matrix structure of $Z$ is in spin-space.
	The anti-commuting blocks, $Z^{\alpha \alpha'}$ and $\bar Z^{\alpha \alpha'}$ with $\alpha \neq \alpha'$ are independent varibales.
	On applying this identity to the partition sum $\langle {\cal Z}_J \rangle_{\phi_0}$ and then integrating over fields $(\bar\psi_q, \psi_q)$, we can reduce the former to the path integral over collective matrix fields $(\bar Z,Z)$ with an action
	\begin{align}
	\label{eq:S_Z_sym}
		S[\bar Z, Z] 
		=& -{\rm str} \ln(1 - \bar Z Z) 
		\\\nonumber
		&
		+ {\rm str}\ln \left(1- e^{i\frac \omega 2} V_1   T_+  
		\left(
		\begin{smallmatrix}
		0 &Z \\
		\bar Z & 0
		\end{smallmatrix}
		\right)  T_-   V_2 - j \sigma_2\right ).
	\end{align}
	Recalling the chiral symmetry, $\sigma_2 T_-   V_2 \sigma_2 = (T_+)^\dagger  V^\dagger_1$, one identifies two Goldstone modes of this action, $Z = - \bar Z$, whenever $\omega \to 0$ or $2\pi$ are satisfied (this corresponds to particle/hole energies $\pm\epsilon$ being close to $0$ or $\pm \pi$, respectively). 
	Indeed, if $Z = - \bar Z$ are constant in space and $j=0$, then action~\eqref{eq:S_Z_sym} vanishes identically. 
	Physically, the field $Z_{qq'}^{\alpha\alpha'} \sim \psi_q^\alpha \bar\psi_{q'}^{\alpha'}$ describes a pairwise propagation of a retarded and an advanced single-particle amplitude at a slight (${\rm mod}\, 2\pi$) difference in frequency $\omega$.
	At long spatial scales, off-diagonal components (with $q\neq q'$) relax quickly due to accumulation of random phases, and Goldstone modes assume the form $Z_{qq'} = Z_q \delta_{qq'}$.
	Assuming $Z_q$ to vary slowly in space, we further expand~\eqref{eq:S_Z_sym} in small spatial gradients and frequency.

\subsubsection{Topological and 2nd order gradient terms}
Let us first discuss terms with spatial derivatives and set $j=\omega=0$~---~this corresponds to the 1st Goldstone 
mode~---~and we comment on the 2nd one (with $\omega\to 2\pi$) in the end of this subsection. 
By defining ${\cal Z} = (1- i Z \sigma_2)$, one can rewrite the action~(\ref{eq:S_Z_sym}) as 
\begin{align}
S[Z] 
&= 
- {\rm str} \ln({\cal Z})  + {\rm str} \ln ({\cal Z} + \delta {\cal Z})
\end{align}
with 
\begin{align}
\delta {\cal Z} 
&= 
- V_1 T_+ [ Z ,T_+^\dagger ]  V_1^\dagger\, i\sigma_2.
\end{align}
To second order, 
\begin{align}
S[Z]
&\simeq
{\rm Str}\left( {\cal Z}^{-1}\delta{\cal Z}
 \right)
-
\tfrac{1}{2}
{\rm Str}\left( 
{\cal Z}^{-1}\delta{\cal Z}{\cal Z}^{-1}\delta{\cal Z}
 \right)
 +\dots
 \nonumber\\
 & = S^{(1)}[Z] + S^{(2)}[Z] + \dots,
\end{align} 
while 
$\langle q| [T_+[Z,T_+^\dagger] |q\rangle = 
( Z'_q  + \frac 12 Z''_q ) P_ + \dots$.
The topological and so-called Gade terms originate from the 1st-order terms in these series.
Using that ${\cal Z}^{-1} = (1 + i \sigma_2 Z)/(1+Z^2)$ and evaluating traces in the spin subspace, one arrives at
\begin{align}
\label{eq:S_11_SM}
\langle S_1^{(1)}\rangle_{\theta,\varphi} 
&=  \bar\chi^0 \int dq\,  {\rm str} \left( g^{-1} \partial_q g\right) 
- 
\int dq \, \partial_q \,{\rm str} \ln (1+g) 
\nonumber\\
&\equiv 
S_{\rm top} + S_{\rm r}.
\end{align}
Here 
\begin{align}
\bar\chi^0 
&= 
\frac 12 (1 - \langle\sin \theta_q\cos\varphi_q \rangle_{\theta,\varphi}), 
\end{align}
and we introduced $g=(1+iZ)/(1-iZ)$. 
Geometrically, the unconstrained pair $(\bar Z,Z)$ defines a set of stereographic coordinates 
parametrizing a two-dimensional sphere in the `fermionic' ${\rm ff}$-sector, respectively,
hyperboloid in the `bosonic'  ${\rm bb}$-sector.
This is readily verified recalling that $\bar{Z}^{\rm ff/bb}=\pm[Z^{\rm ff/bb}]^*$ and stereographic coordinates 
\begin{equation}
(x_1,x_2,x_3) =\frac{1}{1\pm \bar z z}\left( \pm2\mathrm{Re}z,\mp2\mathrm{Im}(z),1\mp\bar z z  \right)
\end{equation}
for the two-sphere/hyperboloid, respectively. The Goldstone-mode restriction $\bar Z=-Z$ 
defines one-dimensional submanifolds which result from their intersection with two-dimensional planes, 
viz. a circle, respectively, hyperbola. The latter identifies $g \in {\rm Gl}(1|1)$ as a supersymmetric group manifold.   

For a system with periodic boundary conditions we can omit the 2nd (residual) term $S_r$
and keep only the 1st (topological) one. In fact, both terms are full derivatives since 
${\rm str} (g^{-1}\partial_q g) = \partial_q \ln {\rm det } (g)$. 
However, $S_{\rm top}$ is non-trivial. Consider a configuration
$g = \left(
\begin{smallmatrix}
e^{x} & 0\\
0 &1
\end{smallmatrix}
\right)_{\rm bf}$,
where $x$ is a compact fermion angle.  Assuming periodic boundary conditions, mappings $x_q: S_1 \to S_1$ 
may have windings, i.e. $x_L = x_0 + 2\pi W$, where $W \in \mathbb{Z}$. 
Then action $S_{\rm top}$ on such configuration becomes non-zero, $S_{\rm top} =  2i \pi n \chi$. 
For the residual term one finds 
\begin{align}
S_r 
&=    
\int_0^{2\pi n}   \frac { i e^{i x} }{1 + e^{i x}} \, dx  = n \oint_{|w| = 1} \frac{dw}{(1+w)}. 
\end{align}
If one regularizes this integral by slightly shifting the pole $w=1$ outside the unit circle $|w|=1$, then $S_r$ vanishes. 

The Gade term is obtained if one keeps the 2nd order cumulant expansion when averaging over disorder, 
\begin{align}
S_{\rm G}[g] 
&=  
- \frac  12 c \int dq\,  {\rm str}^2 (g^{-1}\partial_q g),
\end{align}
where
$c = 
\langle\langle \chi^2(\theta,\varphi) \rangle\rangle = \langle \chi^2(\theta,\varphi)\rangle_{\theta,\varphi}  - \bar\chi^2$. 
This term is exactly zero at criticality (where $\chi = \tfrac 12 $ and does not fluctuate), 
and is known to give inessential modifications away from it~\cite{Lamacraft:2004}.

Since in this paper we are interested in critical quantum walks only, we can derive the diffusive action $S_0[g]$ 
by setting $V_1 = V_2= \openone$ (this corresponds to $\theta=0,\pi$).  The latter simplifies the variation,  
$\delta {\cal Z} = - (Z'_q  + \tfrac 12 Z''_q) P_+ i  \sigma_2 \dots$, and action $S_0[g]$ originates from two pieces.
The 1st piece is 
\begin{align}
S^{(2)}[Z] 
&=  
-\frac 12 {\rm str} \bigl( ({1+Z^2})^{-1} Z Z'  \bigr)^2,
\end{align}
while the 2nd piece stems from the 2nd order gradient term ($ \propto Z''$) in $S^{(1)}[Z]$. It evaluates to
\begin{align}
S_2^{(1)}[Z] 
&=  
\frac 12 {\rm str} \bigl(   ({1+Z^2})^{-1} Z Z''\bigr). 
\end{align}
By adding these two contributions and integrating by parts one finds the diffusive action of the class AIII $\sigma$-model
\begin{equation}
S_{0}[g] = -  \frac 12  {\rm str} \left(    \frac {1}{1+Z^2}  Z'\right)^2  = 
- \frac 18 \int dq\, {\rm str} (\partial_q g^{-1} \partial_q g).
\end{equation}

Let us now comment on the 2nd Goldstone mode with $\omega \to 2\pi$. 
If we change $Z \to -Z$ in the prototype action~(\ref{eq:S_Z_sym}), it is reduced to the one 
with the 1st Goldstone ($\omega \to 0$) and at the same time $g \to g^{-1}$.  The latter does not change $S_0[g]$, but
transforms the topological angle, $\bar \chi^0 \to \bar\chi^\pi = 1-\bar \chi^0$, in the action~(\ref{eq:S_11_SM}). 
At criticality both Goldstone modes are described by the same action with $\bar\chi = 1/2$.

To summarize, 
		using supersymmetric techniques for disordered systems~\cite{Efetov-book} and the colour-flavour transformation~\cite{Zirnbauer:1996,Altland:2015a}, we arrive at the effective action,
	$S^\epsilon = S_{\rm 0} + S^\epsilon_{\rm top}$, where
	\begin{align}
	\label{app_eq:S_0}
		&S_{\rm 0} =
		\frac{1}{2}\int dq  \left[- \mathfrak{g}_0\, {\rm str} \left(\partial_q g^{-1} \partial_q g \right) +i\omega\, {\rm str} (g+g^{-1} ) \right], 
		\nonumber
		\\
	&S_{\rm top}^\epsilon 
		= 
		\bar\chi^\epsilon \int dq\,  {\rm str}( g^{-1} \partial_q g).
	\end{align}
Here $g$ denotes a group-valued matrix field that describes the critical fluctuations in the system, 
$\mathfrak{g}_0=1/4$ is the `bare 'conductance, 
and 
	\begin{align}
		\bar\chi^\epsilon ={1\over2}\left(1 - e^{i\epsilon}\langle\sin(\theta)\cos(\varphi)\rangle_{\theta,\varphi}\right) 
	\end{align} 
	is the bare topological angle 
	with $\epsilon=0,\pi$ indicating the critical states described by the effective action.
	Its is also worth mentioning here that the action $S^\epsilon$ is identical to the one describing disordered quantum 
	wires of a symmetry class AIII~\cite{Altland:2015}.

\subsection{Sources}

Finally let us turn to source contributions $S_J$ to the action. 
Relevant contributions result from an expansion of the action~(\ref{eq:S_Z_sym}) to linear order in $j$,
\begin{align}
\label{eq:S_Z_source_lin}
S_J
&= -
{\rm str}\left(
j\sigma_2 {\cal Z}^{-1}
\right)
\nonumber\\
&=
-\alpha \left[
{iZ_0+\sigma\delta_{i2}\over 1+Z_0Z_0}
\right]^{\rm fb}
-
\beta \left[
{iZ_q+\sigma'\delta_{i2}\over 1+Z_qZ_q}
\right]^{\rm bf},
\end{align}
or 
\begin{equation}
\label{eq:S_g_source_lin}
S_J[g] = - 
\begin{cases}
{\alpha\over 4} \left[
g_0-g_0^{-1}
\right]^{\rm fb}
+
{\beta\over 4} 
\left[
g_q-g_q^{-1}
\right]^{\rm bf},
\, i=1,3
\\
{\sigma \alpha\over 2} \left[
g_0^{\sigma}
\right]^{\rm fb}
+
{\sigma'\beta\over 2} 
[g_q^{\sigma'}
]^{\rm bf},
\,
i=2,
\end{cases}
\end{equation}
resulting in
\begin{align}
P^{\rm chiral}_{\sigma'\sigma}
\label{eq:SM_P1}
&=
\sum_{\sigma,\sigma'=\pm}
\tfrac{\sigma\sigma'}{16}|\omega|\
 \langle \langle
[g^\sigma_0]^{\rm fb}
[g_q^{\sigma'}]^{\rm bf}
 \rangle \rangle,
 \qquad 
 i=1,3,
 \\
\label{eq:SM_P2}
P^{\rm chiral}_{\sigma'\sigma}
&=
\tfrac{\sigma\sigma'}{4}|\omega|
 \langle \langle
[g^\sigma_0]^{\rm fb}
[g_q^{\sigma'}]^{\rm bf}
 \rangle \rangle,
 \qquad 
 i=2,
\end{align}
We evaluate these propagators in the next section.

\section{Transfer matrix method}
\label{TransferMatrix}

When evaluating the probability distributions~\eqref{eq:SM_P1} and \eqref{eq:SM_P2}, 
the non-perturbative nature of Anderson localization requires the functional integration over the entire group-manifold, 
which usually is a highly complicated task. We are here, however, in a better situation  
since powerful alternative non-perturbative methods are available for the one-dimensional
$\sigma$-model~\cite{Efetov-book, Altland:2001}. The latter is based on the interpretation of the action $S_0[g]$ as 
the action of a quantum mechanical particle with coordinate $g$ moving in the potential 
$V(g)=\eta\,{\rm str}(g+g^{-1})$ where $\eta = - i\omega$. Changing then from the path-integral- to  the  
Schr\"odinger-description, one expresses the probability distribution in a spectral decomposition with respect to the corresponding 
Hamilton-operator   
\begin{align}
\hat H 
&= 
\Delta_g+V(g)
\end{align}
where $\Delta_g=-J^{-1}\partial_i G^{ij} J \partial_j$ 
is the Beltrami-Laplace operator on the ${\rm AIII}$-manifold, with metric tensor
 $G^{ij}$ and Jacobian $J=\sqrt{{\rm sdet}G}$.

In what follows we sketch the details of such program at criticality when $\bar\chi =1/2$ and derive a propagator
of the quantum Sinai diffusion. We start by parameterizing the field $g$ in terms of 4 coordinates $z=(x,y,\bar\xi, \xi)$ such that
\begin{equation}
g = {\cal U} \left(
\begin{array}{cc}
e^{x} & 0\\
0 & e^{iy}
\end{array}
\right)_{\rm bf}
{\cal U}^{-1}, \quad
{\cal U}=\exp \left(\begin{array}{cc} 0 & \xi \crcr   \bar\xi & 0  \end{array}\right)_{\rm bf}
\end{equation}
with $x,y \in \mathbb{R}$ being commutative while $\bar\xi,\xi$ being Grassmann anti-commutative fields,
which results in the following metric 
\begin{eqnarray}
dl^2 &=& - {\rm str}(dgdg^{-1})= G_{ij} dz^i dz^j \\
\label{eq:metric_g}
&=& dx^2 + dy^2 + 8 \sinh^2(\tfrac{x-iy}{2}) d\bar\xi d\xi
\end{eqnarray} 
on the $\mathrm{GL}(1|1)$ manifold. The Eq.~(\ref{eq:metric_g}) above defines non-zero elements of the tensor $G_{ij}$.
With the Jacobian $J(z) = \tfrac 14 \sinh^{-2} (\tfrac{x-iy}{2})$ and the vector potential
$A = \bar \chi(i,1,0,0)$ this metric defines the transfer matrix Hamiltonian
\begin{align}
{\cal H} 
&= 
- J^{-1}(z)(\partial_\mu -  i A_\mu) G^{\mu\nu} J(z)(\partial_\nu - i A_\nu)  + V(z),  
\end{align}
where 
$V(x,y) = \eta(\cosh x - \cos y)$ is the potential energy due to frequency term in the action and $\eta = - i \omega$.
Then the Sutherland transformation, 
\begin{align}
H  
&= 
e^{  \bar \chi (x - i y)}  J^{1/2}  {\cal H} J^{-1/2} e^{ - \bar \chi (x - i y) },
\end{align}
complemented by the 'gauge' transform eliminating the vector potential brings the Hamiltonian to a simpler form 
\begin{equation}
\hat H = -\partial_x^2 - \partial_y^2 - \frac{1}{2}\sinh^{-2}\left(\tfrac{x- i y}{2}\right) \partial_{\bar\xi} \partial_\xi + V(x,y) .
\end{equation}  
The ground state $|0\rangle \equiv \Phi_0(x,y)$ of $\hat H$ --- it obeys $\hat H|0\rangle = 0$ due to supersymmetry --- depends only on bosonic angles $(x,y)$ and can be approximated 
by 
\begin{align}
\Phi_0(x,y) 
&= 
- \coth\left(\frac{x - i y}{2}\right) K_0\left( \sqrt{2\eta} e^{|x|/2}\right)/{\ln \eta}. 
\end{align}
If $\eta \ll 1$ then the latter correctly interpolates between the two analytically known  expressions for the ground state $|0\rangle$ 
in the limit $x \sim 1$ and $|x| \gg 1$, resp.~\cite{Altland:2015a}. 
The excited states $|k\rangle \equiv \Phi_k(z)$ of $\hat H$ with energies $E_k >0 $ can be labeled by a set of quantum numbers 
$k=(n,l,\bar\lambda,\lambda)$, where $n$ and $l$ are integers and $\bar\lambda,\lambda$ are Grassmanns.
Specifically,  
\begin{align}
\Phi_k(z) 
&= 
{\cal R}_k(x,y) \times e^{\bar\xi \lambda + \xi \bar \lambda}
\end{align}
can be split into radial
and angular parts where ${\cal R}_k(x,y)$ satisfies to the radial Schr\"odinger equation
\begin{align}
\label{eq:H_R_Phi}
&\left( -\partial_x^2 - \partial_y^2  + V(x,y) - 
\frac{1}{2}\sinh^{-2}\left(\tfrac{x- i y}{2}\right) \bar\lambda \lambda \right)  {\cal R}_k \Phi_k 
\nonumber\\
&= E_k {\cal R}_k .  
\end{align}
Since $\bar\lambda \lambda$ is the nilpotent of the Grassmann algebra,
the spectrum and eigenstates of the above radial equation should have the following form:  
$E_k = \epsilon_{n,l} + \bar\lambda \lambda\, \epsilon_{n,l}' $ and 
\begin{align}
{\cal R}_k(x,y) 
&= 
R_{n,l}(x,y) + \bar\lambda \lambda R_{n,l}'(x,y), 
\end{align}
where $n=1,2,\dots$, and $l \in \mathbb{Z}$ are radial quantum numbers.
It turns out (see Sec.\ref{sec:SDP} below) that only the 0th order terms in bilinear $\bar\lambda \lambda$ are required to evaluate the propagator $P_{\sigma'\sigma}^{\rm chiral}(\eta, q)$ of the quantum Sinai diffusion. 
We proceed by constructing an asymptotic form of the radial wave function
$R_{n,l}(x,y)$ at $\eta \ll 1$ (or $t \gg 1$)  in the next section and then find $P_{\sigma'\sigma}$ in Sec.\ref{sec:SDP}.

\subsubsection{Radial wave function}
We now concentrate on the spectrum $\epsilon_{n,l}$ and eigenstates  $R_{n,l}(x,y)$ of the 0th order 
Hamiltonian 
\begin{align}
\hat H_0 
&=
-\partial_x^2 - \partial_y^2  +\eta(\cosh x  - \cos y).
\end{align}
It will be seen in Sec.\ref{sec:SDP} that
in the limit $\eta \ll 1$ which we are going to explore essential $x$'s satisfy $\eta e^{|x|} \sim 1$ and hence
$\cos y$ term in $\hat H_0$ can be neglected. We thus approximate $R_{n,l}(x,y) \approx R_n(x) e^{i l y}$,
which leads to $\epsilon_{n,l} = \epsilon_n + l^2$ together with a simple radial equation
\begin{equation}
\label{eq:Rn}
[-\partial_x^2 + \eta \cosh x] R_n(x) = \epsilon_n R_n(x).
\end{equation}
To solve it we introduce momenta $k_n = \sqrt{\epsilon_n}$ and
divide the $x$-axis in three intervals: (I) 'small' angles with $|x| < 1 $; (II) 'intermediate' ones,  such that
$ 1 < |x| < \ln (1/{\eta})$ and (III) 'large' angles, where $|x| > \ln (1/{\eta})$. In the following, it will be sufficient to consider
the domain $x>0$ since the potential $\cosh x $ is symmetric.  In the intervals II \& III one can approximate~(\ref{eq:Rn}) by
\begin{align}
\left[ -\partial_x^2 +  \tfrac 12 \eta e^{x}\right] R_{n}(x) 
&= 
k_{n}^2 R_{n}(x).
\end{align}
Up to a normalization factor which is found below, the solution of this equation is a modified  
Bessel function $R_{n}(x) \propto K_{2 ik_{n}}(\sqrt{2\eta} e^{x/2})$. 
Taking a limit of $K_\nu(z)$ at small argument, the wave function $R_{n}(x)$ in the interval II 
is reduced to the plane wave
\begin{align}
\label{eq:R_n_region_II_A}
R_{n}(x) 
&\propto 
A(k_{n}) e^{i  k_{n} x } + A^*(k_{n}) e^{-i k_{n} x }, 
\nonumber\\
A(k) 
&= 
\Gamma(-2 ik) \left(\eta/ 2\right)^{ik}.
\end{align}
As to interval I, one can neglect $\eta$-dependent potential whatsoever, and therefore by continuity the 
plane wave~(\ref{eq:R_n_region_II_A}) is also a solution in the interval I. We can thus introduce a scattering matrix and a phase shift from the
right potential barrier, $S(k) = A(-k)/A(k) = e^{- i \phi(k)}$, which finally gives us a quantization condition $\phi(k_n) = \pi n$.
Here $n = 0,1, 2, \dots$ with even/odd $n$ corresponding to even/odd wave functions $R_n(x)$, resp., i.e.
$R_n(-x) = (-1)^n R_n(x)$. For small momenta, $k_n \ll 1$, we get with log-accuracy $\eta^{2ik} \simeq e^{i \pi (n+1)}$ which
leads to the spectrum
\begin{equation}
\epsilon_{n,l} = \frac {\pi^2}{4\ln^2 \eta} ( n +1)^2 + l^2, \quad n=0,1,2,\dots, \,\, l \in \mathbb{Z}.
\end{equation}

We now proceed to find a normalization factor for the radial wave function. 
For that let's note that the main contribution to its norm 
$\int_{-\infty }^{+\infty } R^2_{n}(x) dx=1$ comes from the intervals I and II.
The wave function in these regions is a plane wave,
\begin{align}
R_{n}(x) 
&\propto 
|A(k_{n})| \cos\left( k_{n} x + \tfrac 12 \pi n \right).
\end{align}
It can be matched to the one found within the semiclassical approximation, 
\begin{align}
R_{n}(x) 
&= 
({C_{n}}/{\sqrt{k_n}}) \cos\left( k_n x + \tfrac 12 \pi n \right), 
\end{align}
where the normalization constant is fixed by 
$C_{n}^2 = (2 {k_{n}}/{\pi})({\partial k_{n}}/{\partial n})$. 
The comparison of these two representations leads to the following normalized radial wave function
\begin{align}
\label{eq:Rnl_xy_norm}
R_{n,l}(x,y) 
&= e^{i l y }\left(\frac {2}{\pi} \frac{\partial k_{n}}{\partial n} \right)^{1/2} 
|A(k_{n})|^{-1} K_{2 ik_{n}}(\sqrt{2 \eta} e^{x/2}),
\nonumber \\
|A(k)|^{-1} 
&= 
\left(\frac{2 k \sinh 2 \pi k}{\pi}\right)^{1/2}, \qquad x>0.
\end{align}
We use this intermediate result in the next subsection to evaluate the series expansion of the propagator
$P^{\rm chiral}_{\sigma' \sigma}$.

\subsubsection{Propagator of Sinai diffusion}
\label{sec:SDP}

Employing a spectral decomposition, the propagator~(7) can be written as a sum over
excited eigenstates $|k\rangle$, 
\begin{equation}
P^{\rm chiral}_{\sigma' \sigma}(\eta,q) 
= 
\eta \sum_{n,l}\int d\lambda d\bar\lambda \,
\Gamma^{\sigma'}_k \bar \Gamma^{\sigma}_k e^{- 2 E_k |q|}, \quad
\end{equation}
where
\begin{eqnarray}
\label{eq:Gamma_k}
\Gamma^{\sigma'}_k 
&=& 
\langle 0 | [g^{\sigma'}]^{\rm bf}|k\rangle \\
&=& 
\frac{\sigma'}{4\pi}\int\limits_{-\infty}^{+\infty}dx\!\int\limits_{0}^{2\pi}  dy\! 
\int  d\bar\xi d\xi\,
\Phi_0(x,y)  [g^{\sigma'}]^{\rm bf} \Phi_k(z). 
\nonumber
\end{eqnarray}
is a matrix element of the field $[g^{\sigma'}]^{\rm bf} =  (e^{i \sigma' y}-e^{\sigma' x})\xi$ between
the ground and excited states and a similar expression is valid for a conjugated matrix element $\bar\Gamma_k$ of the field
$[g^{\sigma}]^{\rm fb}=([g^{\sigma}]^{\rm bf})^*$.
Using the explicit form of the excited state, 
\begin{align}
\Phi_k(z) 
&= 
{\cal R}_k(x,y)  \times e^{\bar\xi \lambda + \xi \bar \lambda},
\end{align}
one can first perform the integral over Grassmanns  $(\bar\xi,\xi)$ in Eq.~(\ref{eq:Gamma_k})
and verify that the nilpotent part $\sim R_{n,l}'(x,y)$ of the radial wave function does not contribute to the matrix elements. 
The latter are then simplified to $\Gamma^{\sigma'}_k = -\lambda \Gamma^{\sigma'}_{n,l}$ and 
$\bar\Gamma^{\sigma}_k = \bar\lambda \Gamma^\sigma_{n,l}$ with
\begin{equation}
\Gamma^{\sigma}_{n,l} = \frac{\sigma}{4\pi} \int\limits_{-\infty}^{+\infty} dx\!\int\limits_{0}^{2\pi}  dy \,
\Phi_0(x,y) (e^{\sigma x}-e^{i\sigma y}) R_{n,l}(x,y).
\end{equation}
Separating  here $y$-dependent parts of the wave functions, the integration over the compact angle $y$ yields
\begin{align}
\frac{\sigma}{2\pi} \int_{-\pi}^{\pi} dy \coth\left(\frac{x - i y}{2}\right) ( e^{\sigma x} - e^{ i \sigma y} ) e^{i l y} 
&= 
e^{\sigma x} \delta_{l} + \delta_{l+\sigma}. 
\end{align}
Here the $l=\pm 1$ terms induce a gap in the spectrum ($\epsilon_{n,\pm} = \epsilon_n + 1$) 
thus we keep $l=0$ contribution only, the latter reads
\begin{align}
\Gamma^{\sigma}_{n,0} 
&= 
- \tfrac{1}{2} \int e^{\sigma x} K_0(\sqrt{2\eta} e^{|x|/2}) R_{n,0}(x) dx / \ln \eta.
\end{align}
It is worth mentioning that $R_{n,0}(x)$ is either even or odd depending on a parity of $n$, 
thus $\Gamma^{+}_{n,0} = (-1)^n \Gamma^-_{n,0}$. On changing the integration variable to
$z=\sqrt{2\eta} e^{x/2}$, the remaining integral for $\Gamma^{\sigma}_{n,0}$ is reduced to a table one,
\begin{align}
\int_0^{+\infty} z K_0(z) K_{2ik}(z) dz 
&= 
(k^2 \pi^2/2) \sinh^{-2}(\pi k).
\end{align} 
Finally, taking into account
proper normalization factors given in Eq.~(\ref{eq:Rnl_xy_norm}), one finds the following matrix elements
\begin{align}
M_n^{\sigma'\sigma} 
&=\Gamma^{\sigma'}_{n,0} \Gamma^{\sigma}_{n,0} 
\nonumber\\
& =
(\sigma'\sigma)^n \frac{\pi^2}{2}\frac{k_n^2}{\eta^2 \ln^2 \eta} 
\left(\frac{\partial k_{n}}{\partial n}\right)\times \frac{k_{n}^3 \cosh (\pi k_{n})}
{\sinh^3 (\pi k_{n})}
\nonumber\\
&\overset{k_{n} \ll 1}{\longrightarrow } \,\,
\frac{ (\sigma'\sigma)^n (n+1)^2}{\eta^2 \ln^5(1/\eta)}.
\end{align}
From here the propagator of Sinai diffusion is constructed as
\begin{align}
\label{eq:P_chiral_M}
P^{\rm chiral}_{\sigma'\sigma}(\eta, q) 
&= \eta\sum_{n,l=0} \int d\lambda d\bar\lambda\, 
\Gamma^{\sigma'}_k \bar\Gamma^{\sigma}_k
e^{- 2 |q| E_k} 
\nonumber\\
&= 
\eta \sum_{n=0}^{+\infty} M_n^{\sigma' \sigma} e^{- 2 |q| \epsilon_{n,0}}.
\end{align}
When evaluating the above integral over Grassmann variables one may notice that the nilpotent correction to the spectrum,
$\bar \lambda \lambda \epsilon'_{n,l}$, does not contribute to the net result. 
At large distances, $q \gg 1$, essential momenta are small, $k_{n} \ll 1$, 
and the Laplace transform of~(\ref{eq:P_chiral_M}) from $\eta$ to the time domain yields the result Eq.~\eqref{chPt} 
in the main text.

As a final remark let us evaluate the integrated probability 
\begin{equation}{}
P^{\rm chiral}(\eta) = \sum_{\sigma'\sigma}\int dq\, P^{\rm chiral}_{\sigma'\sigma}(\eta,q) = 4\eta \sum_{k=0}^{+\infty}{}
\frac{M^{++}_{2k}}{\epsilon_{2k,0}}.
\end{equation}
This series is convergent owing to the exponential decay of $M^{\sigma'\sigma}_n$ at large momenta $k_n >1$.
In the limit $\eta \ll 1$ one may substitute the sum by an integral to obtain, 
\begin{equation}
P^{\rm chiral}(\eta) = 
\frac{\pi^2}{\eta \ln^2 \eta}
\int_0^{+\infty} dk {}
\,
\frac{k^3 \cosh \pi k}{\sinh^3 \pi k} 
=\frac{1}{ 4\,\eta \ln^2 \eta}.
\end{equation}
Hence the overall contribution of the critical states to the walker's probability decreases in time as $P^{\rm chiral}(t) = 1/(4\ln^2 t)$.

\section{Density of states}
\label{DoS}

\begin{figure*}[t!]
\begin{center}
\includegraphics[width=0.85\columnwidth]{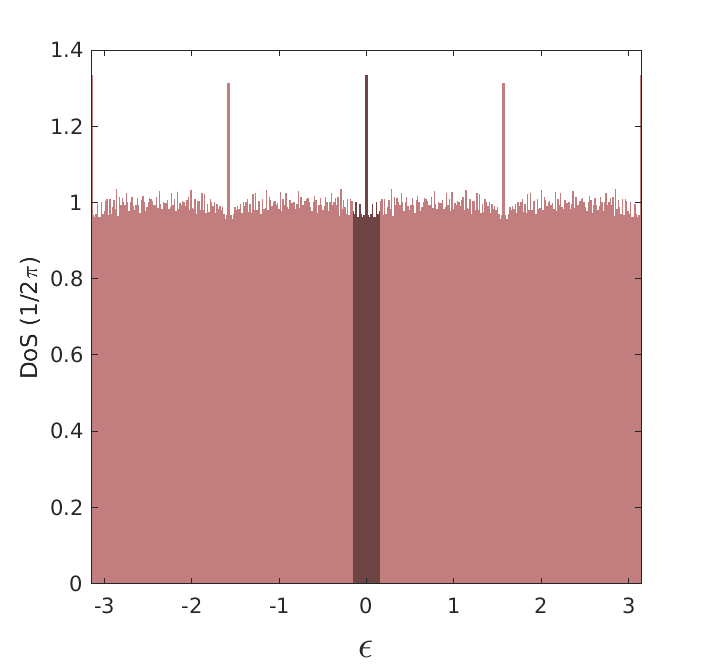}
\includegraphics[width=0.85\columnwidth]{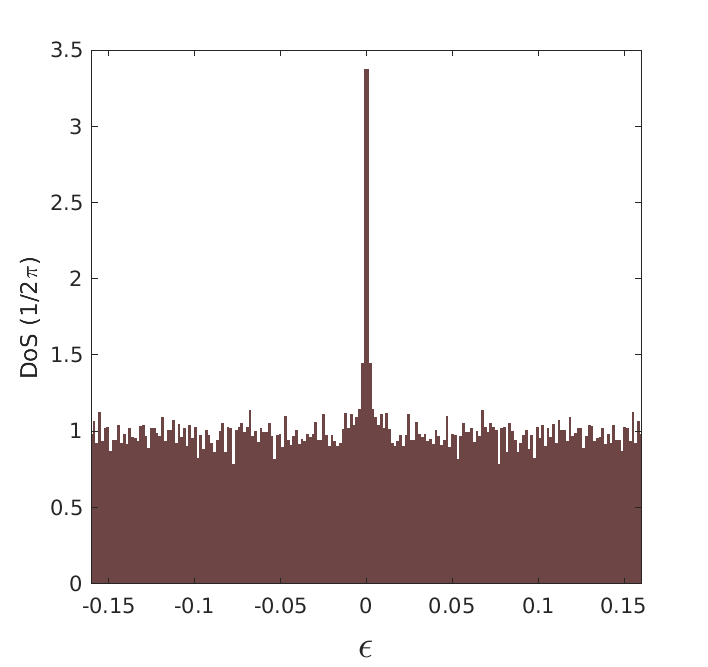}
\caption{
Density of states (DoS) of the quantum walk model with averaged angles $(\bar\theta,\bar\varphi)=(0,0)$
and disorder strengths $\gamma_\theta = \gamma_\varphi=\pi/4$ 
for the quasi-energy domain $\epsilon\in[-\pi,\pi]$~(upper panel) and $\epsilon \in [-0.16,0.16]$~(lower pannel) 
using different energy resolution. 
In the calculations we used averages over $10^3$ disorder configurations. 
Peaks of the DoS at chiral symmetric energy $\epsilon=0,\pm \pi/2, \pi$ are clearly visible. 
}
\label{sm_fig:1}
\end{center}
\end{figure*}

In the main text we focused on the walker's critical dynamics at the topological Anderson localization 
transition. As discussed there, the critical dynamics describes quasi-energy states centered around 
the chiral symmetric energies $\epsilon=0,\pm\pi/2,\pi$. To substantiate this statement we provide 
here the numerical results for the density of state (DoS) of the quantum walker with periodic
boundary conditions. Fig.~\ref{sm_fig:1} (left) shows the disorder averaged 
DoS in the entire quasi-energy domain for a system of $N_x=400$ sites. 
As expected, sharp peaks are visible at the chiral symmetric energies $\epsilon=0,\pm \pi/2, \pi$. 
Fig.~\ref{sm_fig:1} (right) shows a magnified view of the region colored in red of Fig.~\ref{sm_fig:1}. 
With a smaller energy scale for the histogram, we can see further structures of the DoS, which 
is known to diverge as $\sim 1/(\epsilon \ln^3 \epsilon)$~\cite{Balents:1997}. 

For our discussion it is important to notice that   
the number of eigenstates within these energy domains is   
not the dominant contribution to the total density of states.  
This indicates that a walker initially localized on a single site is not 
the optimal choice for a protocol aiming to test the walker's critical dynamics, 
as it involves quasi energy states from the entire energy band approximately with equal weight. 
That is why in the main text we propose to use the plane wave with momentum $p_0=0,\pi/2$ as an initial state, 
which can be used to select only states within quasi-energy regions centered around the chiral symmetric energies.

\section{Time-staggered spin polarization}
\label{TimeStaggering}

In this Appendix we discuss the time-staggered spin polarization, 
observable in a quantum critical walk at a topological Anderson localization transition.
As stated in the main text, the time-staggered spin polarization    
involves critical states $\epsilon\simeq \pm \pi/2$, related to the chiral sublattice symmetry  
$\hat {\cal C}_{\rm sl}\equiv \sigma_2\otimes \hat{S}$, where
$\hat{S}\equiv \sum_q|q\rangle (-1)^q \langle q|$ the sublattice operator. 
Before discussing the relation between $\hat {\cal C}_{\rm sl}$ and 
a time-staggered signal, it is instructive to reformulate our discussion in the main text on the
chiral symmetry $\hat{\cal C}_0\equiv\hat s_2$, and related 
spin polarization $\Delta P$, in a more formal way which readily allows for 
an extension to the chiral sublattice symmetry of interest. 

{\it Chiral symmetry:---}In the main text we introduced the probability distribution,
\begin{align}
\label{app_msq}
P_{\sigma'\sigma} (t,q) &=
\langle 
|\langle q,\sigma' |\hat U^t |0,\sigma\rangle |^2
\rangle_{\theta,\varphi},
\end{align}
for a walker initially prepared in eigenstate $|\sigma\rangle=|\leftarrow\rangle,|\rightarrow\rangle$ of the chiral operator $\hat{\cal C}_0$   
to be found after $t$ time-steps at a distance $q$ in eigenstate $|\sigma'\rangle$. 
More formally, we can separate the  walker's Hilbert space into the direct sum of subspaces characterized 
by the quantum numbers $s=\pm $ of the chiral operator $\hat{\cal C}_0$,
${\cal H}={\cal H}^0_+\oplus {\cal H}^0_-$, and spanned by 
\begin{align}
{\cal H}^0_+ 
&= {\rm span}\{ |q,\leftarrow\rangle\},
\\
{\cal H}^0_- 
&= {\rm span}\{ |q,\rightarrow\rangle\}.
\end{align}
The statement on the positive `spin polarization' discussed in the main text, can then be restated as follows: 
for critical states related to the chiral symmetry $\hat{\cal C}_0$ the probability distributions
$P_{s's}\hspace{-.1cm}: {\cal H}_s \overset{\hat{U}^t}{\longrightarrow} {\cal H}_{s'}$
for initial and final states belonging to the same and different subspaces, $s'=s$ respectively
$s'=-s$, differ and their difference is strictly positive  
\begin{align}
\label{app_pd}
P_{ss}(t,q)-P_{-ss}(t,q)>0.
\end{align} 
Formulated in terms of quantum numbers of the chiral operator, the statement on the positivity~\eqref{app_pd} 
holds for critical states related to the chiral symmetry, independently of its specific form.   
For the specific chiral symmetry $\hat{\cal C}_0=\hat s_2$ quantum numbers are simply spin-orientations, and the positive difference
is indeed equivalent to the positive `spin polarization distribution',
$P^{\rm chiral}_{\rightarrow\rightarrow}(t,q)-P^{\rm chiral}_{\leftarrow\rightarrow}(t,q) \equiv \Delta P(t,q)>0$, 
discussed in the main text. Statement~\eqref{app_pd} can now be applied to the chiral sublattice symmetry $\hat{\cal C}_{\rm sl}$,   
where it shows more interesting consequences.

{\it Chiral sublattice symmetry:---}Separating the walker's Hilbert space  
into the direct sum of subspaces, ${\cal H}={\cal H}^{\rm sl}_+\oplus {\cal H}^{\rm sl}_-$, 
characterized by the quantum numbers $s=\pm $ of the chiral sublattice 
symmetry $\hat{\cal C}_{\rm sl}\equiv \sigma_2\otimes \hat{S}$,
we notice that quantum numbers differ from the spin orientations, and subspaces are now spanned by
\begin{align}
\label{app_ssp_sl_p}
{\cal H}^{\rm sl}_+ 
&= 
{\rm span}\{ |2q,\leftarrow\rangle, |2q-1,\rightarrow \rangle \},
\\
\label{app_ssp_sl_m}
{\cal H}^{\rm sl}_- 
&= {\rm span}\{ |2q,\rightarrow\rangle, |2q-1,\leftarrow\rangle\}.
\end{align}
Positivity~\eqref{app_pd} holds for critical states related to the chiral operator independently of its specific form,  
and we next have to relate this statement to the spin polarization. The relation  
is more involved for $\hat {\cal C}_{\rm sl}$ than for $\hat {\cal C}_0$,   
since the spin structure of eigenstates of the former alternates between even and odd sites. 
More specifically, this implies that the spin structure of $P_{s's}$
depends on the (parity of the) propagated distance $q$, i.e. 
\begin{align}
\label{app_msq_sl}
P_{ss} (t,q) 
&=
\begin{cases}
P_{\sigma\sigma} (t,q),
\qquad \text{$q$ even},
\\
P_{-\sigma\sigma} (t,q),
\qquad \text{$q$ odd},
\end{cases}
\nonumber 
\\
P_{-ss} (t,q) 
&=
\begin{cases}
P_{-\sigma\sigma}(t,q),
\qquad \text{$q$ even},
\\
P_{\sigma\sigma}(t,q),
\qquad \text{$q$ odd},
\end{cases}
\end{align}
where $s,s'$ are the eigenvalues of $\hat{\cal C}_{\rm sl}$ and 
$\sigma,\sigma'$ those of $\sigma_2$.

To structure then above probabilities~\eqref{app_msq_sl} according to the parity of propagated time steps $t$,
we notice that the single time-step evolution $\hat U$ propagates states by exactly one lattice site. 
Starting e.g. from the even site $q=0$ and propagating for an even number of time steps $t$
one, therefore, ends again on an even site. For an odd number of time steps $t$, on the other hand, 
one ends on an odd site. That is, 
\begin{align}
{\rm span}\{ |2q,\sigma\rangle\rangle \}
&\overset{\hat U^{2t} }{\longrightarrow}
{\rm span}\{ |2q,\sigma\rangle\rangle \},
\\ 
{\rm span}\{ |2q,\sigma\rangle\rangle \}
&\overset{ \hat{U}^{2t+1} }{\longrightarrow}
{\rm span}\{ |2q+1,\sigma\rangle\rangle \}. 
\end{align}
and we can relate probabilities Eqs.~\eqref{app_msq_sl} 
to the parity of propagated steps $t$ as follows. For even numbers of time steps probabilities 
$P_{s's}\hspace{-.1cm}:{\cal H}^{\rm sl}_s\overset{\hat U^{2t}}{\longrightarrow} {\cal H}^{\rm sl}_{s'}$, 
conserving (changing) the quantum number of the chiral sublattice operator coincides with 
probabilities preserving (changing) spin orientation, 
$P_{s's}=P_{\sigma'\sigma}$. 
The difference~\eqref{app_pd} is again the spin polarization,
$P_{ss}(t,q)-P_{-ss}(t,q)=
P^{\rm chiral}_{\rightarrow\rightarrow}(t,q)-P^{\rm chiral}_{\leftarrow\rightarrow}(t,q)$. 
For odd numbers of time steps, on the other hand, probabilities 
$P_{s's}\hspace{-.1cm}:{\cal H}^{\rm sl}_s\overset{\hat U^{2t+1}}{\longrightarrow} {\cal H}^{\rm sl}_{s'}$ 
conserving (changing) the quantum number of the chiral sublattice operator
correspond to probabilities changing (preserving) spin orientation. In this case 
$P_{ss}(t,q)-P_{-ss}(t,q)=
P^{\rm chiral}_{\leftarrow\rightarrow}(t,q)- P^{\rm chiral}_{\rightarrow\rightarrow}(t,q)$ 
is the negative spin polarization distribution. Summarizing, we find that 
for critical states $\epsilon \simeq \pm \pi/2$ of the chiral sublattice symmetry $\hat {\cal C}_{\rm sl}$ 
positivity~\eqref{app_pd} translates into a time-staggered spin polarization distribution,
\begin{align}
 P^{\rm chiral}_{\rightarrow\rightarrow}(t,q)-P^{\rm chiral}_{\leftarrow\rightarrow}(t,q)
&=
(-1)^t|\Delta P(t,q)|,
\end{align}
as stated in the main text. 

\section{Details of experimental proposal}

\begin{figure}[tt]
	\includegraphics[width=.49\textwidth]{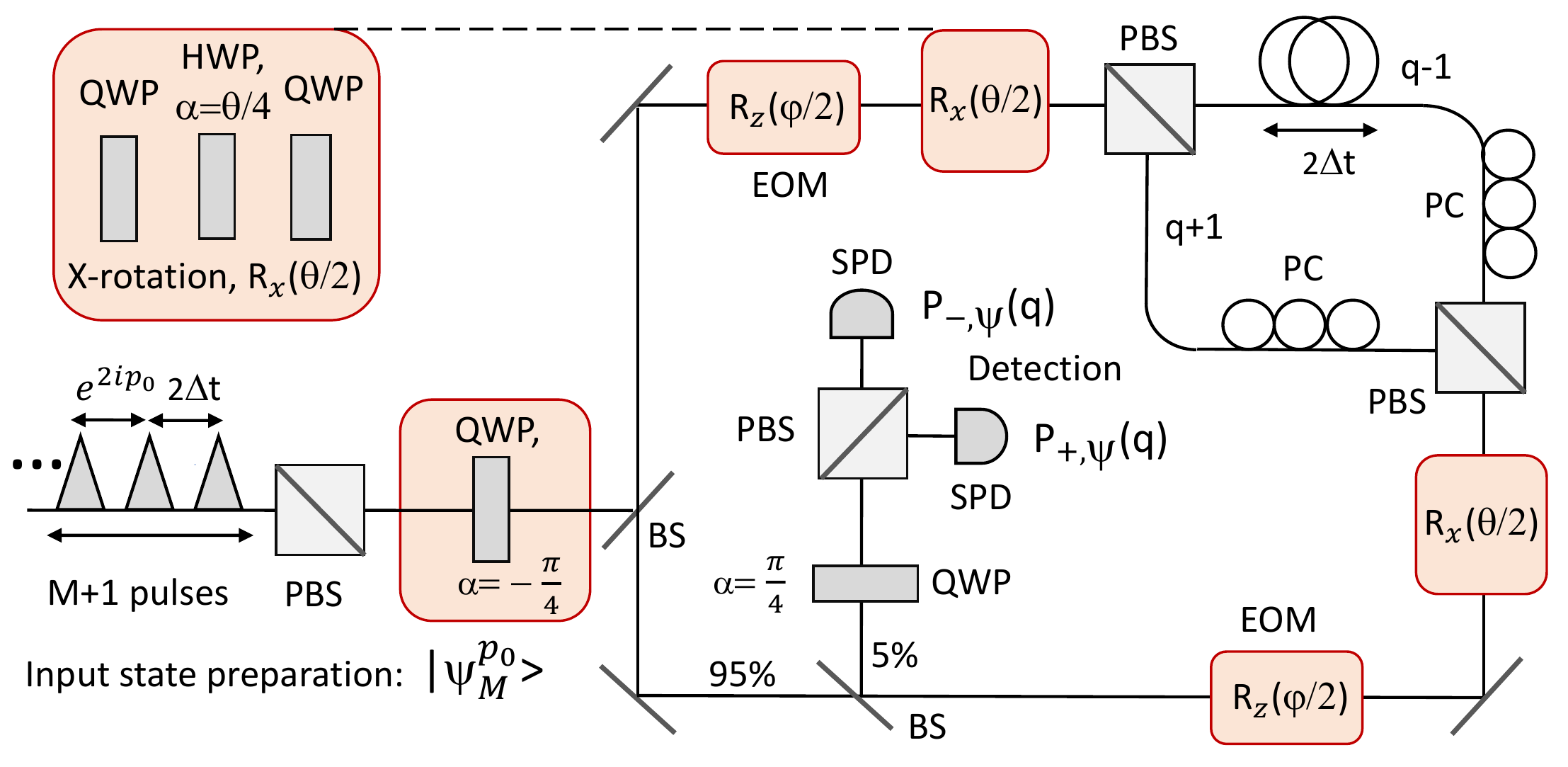}
	\caption{
		A prototype of a linear optical network to realize a quantum walk discussed in details in
		the main text (along the lines of Refs.~\cite{Schreiber:2011, Schreiber:2012}). 
		A phase modulated laser source (not shown) generates a train of
		pulses with a fixed time interval, $\Delta t$, and relative phase difference, $2p_0$, between adjacent pulses.  
		HWP: half-wave plate rotated by angle $\alpha$. QWP: quarter wave plate; PBS: polarizing beam splitter; BS: beam sampler;
		EOM: fast switching electro-optic modulator; SPD: single-photon detector; PC: polarization controller. Fibers of different lengths ensure a time
		delay $2\Delta t$ between $|H\rangle$ and $| V\rangle$ states thereby realizing 'shift' operator $T$.
	}
	\label{Opt_Net}
\end{figure}

\label{Details}

	Here we discuss few technical details related to the time-multiplexing experimental proposal mentioned in the main text, see Fig.~\ref{Opt_Net}. 
	One envisions a train of equidistant pulses with controlled phase relation to be produced by a coherent laser source. 
	The half-wave (HWP) and quarter-wave (QWP) plates are used for the initialization of input state in the form~(\ref{eq:psi_i}), implementation of the rotation $R_x(\theta)$ 	as well as in the detection. 
	With a fast axis aligned horizontally, the plates in the basis of linearly polarized states, $\{|H\rangle, |V\rangle\}$, 
	are characterized by the diagonal Jones matrices 
    $M_{1/2} = {\rm diag}(1,-1)$ and $M_{1/4} = {\rm diag}(1,i)$.
	Then, for instance, the Jones matrix of the HWP rotated at $\alpha$ degrees becomes
	\begin{equation}
	M_{1/2}(\alpha) = \left(\begin{array}{cc}
	\cos 2\alpha & \sin 2\alpha \\ \sin 2\alpha & - \cos 2\alpha
	\end{array}\right),
	\end{equation}
	and, on other hand, $M_{1/4}(-\pi/4) \sim R_x(\pi/4)$ where the last equality holds up to inessential phase factor.  
	
	Consider now left/right circular polarized states, 
	$|L/R\rangle = \frac{1}{\sqrt{2}} (|H\rangle \pm i |V\rangle)$, which
	are eigenstates of the operator $\hat\sigma_2$  and thus can be identified with spin states $|\rightarrow\rangle$ and 
	$|\leftarrow\rangle$ discussed in the main text. Assuming that a light from a laser source is linearly polarized along $|H\rangle$, 
	one checks that $ M_{1/4}(-\pi/4)|H\rangle = |L\rangle$, which generates incoming state $|\psi_M^{p_0} \rangle$,
	cf. Eq.~(\ref{eq:psi_i}) in Sec.~\ref{Exp_protocols}. The same is true for the detection. Owing to polarizing beam splitters (PBS), two 
	single-photon detectors (SPDs) detect linearly polarized states.
	Because of identity $\langle R(L)| = \langle H (V)| M_{1/4}(\pi/4)$ the later are transformed
	into circular polarized ones and thereby the measurement of spin-dependent probabilities $P_{\sigma\psi}(t,q)$ defined by Eq.~(\ref{eq:P_sigma_psi}) 
	in Sec.~\ref{Exp_protocols} can be achieved. Finally, the identity
	\begin{equation}
	R_x(\theta/2) = M_{1/4} \cdot M_{1/2}(\theta/4) \cdot M_{1/4}
	\end{equation}
	is a key to implement a (half)-rotation along $x$-axis using three plates as shown in Fig.~\ref{Opt_Net}.  
	
	Few remarks are now in order with regard to possible time and spatial scales of the experiment. 
	Following Refs.~\cite{nitsche2018probing,geraldi2020subdiffusion} 	we assume that a laser emits a train of pulses with interval $\Delta t \sim 106$~ns at the telecom wavelength $\lambda \sim 1550$~nm.
	This timing is chosen such that it is compatible with typical switching speeds of the electro-optic modulators and the deadtimes of the detectors.
	The intensity of such pulses in initial state should be attenuated to the single photon level, $\langle n \rangle_{\rm in} \sim 1$,
	to eliminate many photon contributions in the click detectors.
	The interval $\Delta t$ requires a fibre length mismatch $\Delta L  \sim 20$~m between the long and the short path in order to implement 'shift' operator $T$ of the quantum walk.  
	An initial wave packet of pulses with a total time span $ M \Delta t$ spreads after $N$ walk's steps (each corresponding to a single run along the interferometer loop) to  $ (M + N) \Delta t$. Thus for $M = 10^2$ and $N = 20$ the length of a loop should exceed $L \sim 2.5$~km, easily realisable with optical fibres in the telecom regime. 
	Assuming that losses in the optical setup stem mainly from the coupling mismatch between in and outcoupling of the fibres and, in sum,  are 20\% per run, we obtain the occupation number  of the order of
	$\langle n \rangle_{\rm f} \sim 5 \cdot 10^{-5}$ after $N = 20$ steps.
	 With a repetition rate of 1~kHz, this leads to $0.05$ clicks per second per time bin, which should be easily detectable by superconducting singlephoton nanowire detectors within realistic measurement times.
	The required step numbers of 20 were already  exceeded for a localised input state in \cite{nitsche2018probing} in which 36 steps were demonstrated.
	Also the ensemble averages over 5000 realizations are in the reach of the experiment, as e.g. in Ref.~\cite{geraldi2020subdiffusion} 2400 disorder realizations were already measured.
	In summary, we strongly believe that the proposed experimental realisation of the  topological Anderson localization transition is feasible with current technologies and is thus in the reach of near-future measurements.

\bibliography{Literature}	

\begin{thebibliography}{78}%
\makeatletter
\providecommand \@ifxundefined [1]{%
 \@ifx{#1\undefined}
}%
\providecommand \@ifnum [1]{%
 \ifnum #1\expandafter \@firstoftwo
 \else \expandafter \@secondoftwo
 \fi
}%
\providecommand \@ifx [1]{%
 \ifx #1\expandafter \@firstoftwo
 \else \expandafter \@secondoftwo
 \fi
}%
\providecommand \natexlab [1]{#1}%
\providecommand \enquote  [1]{``#1''}%
\providecommand \bibnamefont  [1]{#1}%
\providecommand \bibfnamefont [1]{#1}%
\providecommand \citenamefont [1]{#1}%
\providecommand \href@noop [0]{\@secondoftwo}%
\providecommand \href [0]{\begingroup \@sanitize@url \@href}%
\providecommand \@href[1]{\@@startlink{#1}\@@href}%
\providecommand \@@href[1]{\endgroup#1\@@endlink}%
\providecommand \@sanitize@url [0]{\catcode `\\12\catcode `\$12\catcode
  `\&12\catcode `\#12\catcode `\^12\catcode `\_12\catcode `\%12\relax}%
\providecommand \@@startlink[1]{}%
\providecommand \@@endlink[0]{}%
\providecommand \url  [0]{\begingroup\@sanitize@url \@url }%
\providecommand \@url [1]{\endgroup\@href {#1}{\urlprefix }}%
\providecommand \urlprefix  [0]{URL }%
\providecommand \Eprint [0]{\href }%
\providecommand \doibase [0]{https://doi.org/}%
\providecommand \selectlanguage [0]{\@gobble}%
\providecommand \bibinfo  [0]{\@secondoftwo}%
\providecommand \bibfield  [0]{\@secondoftwo}%
\providecommand \translation [1]{[#1]}%
\providecommand \BibitemOpen [0]{}%
\providecommand \bibitemStop [0]{}%
\providecommand \bibitemNoStop [0]{.\EOS\space}%
\providecommand \EOS [0]{\spacefactor3000\relax}%
\providecommand \BibitemShut  [1]{\csname bibitem#1\endcsname}%
\let\auto@bib@innerbib\@empty
\bibitem [{\citenamefont {Klitzing}\ \emph {et~al.}(1980)\citenamefont
  {Klitzing}, \citenamefont {Dorda},\ and\ \citenamefont
  {Pepper}}]{Klitzing:1980}%
  \BibitemOpen
  \bibfield  {author} {\bibinfo {author} {\bibfnamefont {K.~v.}\ \bibnamefont
  {Klitzing}}, \bibinfo {author} {\bibfnamefont {G.}~\bibnamefont {Dorda}},\
  and\ \bibinfo {author} {\bibfnamefont {M.}~\bibnamefont {Pepper}},\
  }\bibfield  {title} {\bibinfo {title} {New method for high-accuracy
  determination of the fine-structure constant based on quantized {H}all
  resistance},\ }\href {https://doi.org/10.1103/PhysRevLett.45.494} {\bibfield
  {journal} {\bibinfo  {journal} {Phys. Rev. Lett.}\ }\textbf {\bibinfo
  {volume} {45}},\ \bibinfo {pages} {494} (\bibinfo {year} {1980})}\BibitemShut
  {NoStop}%
\bibitem [{\citenamefont {Khmelnitskii}(1983)}]{Khmelnitskii:1983}%
  \BibitemOpen
  \bibfield  {author} {\bibinfo {author} {\bibfnamefont {D.~E.}\ \bibnamefont
  {Khmelnitskii}},\ }\bibfield  {title} {\bibinfo {title} {{Quantization of
  Hall conductivity}},\ }\href
  {http://www.jetpletters.ac.ru/ps/1485/article_22668.shtml} {\bibfield
  {journal} {\bibinfo  {journal} {JETP Lett.}\ }\textbf {\bibinfo {volume}
  {38}},\ \bibinfo {pages} {552} (\bibinfo {year} {1983})}\BibitemShut
  {NoStop}%
\bibitem [{\citenamefont {Mirlin}\ \emph {et~al.}(2010)\citenamefont {Mirlin},
  \citenamefont {Evers}, \citenamefont {Gornyi},\ and\ \citenamefont
  {Ostrovsky}}]{Mirlin:2010}%
  \BibitemOpen
  \bibfield  {author} {\bibinfo {author} {\bibfnamefont {A.~D.}\ \bibnamefont
  {Mirlin}}, \bibinfo {author} {\bibfnamefont {F.}~\bibnamefont {Evers}},
  \bibinfo {author} {\bibfnamefont {I.~V.}\ \bibnamefont {Gornyi}},\ and\
  \bibinfo {author} {\bibfnamefont {P.~M.}\ \bibnamefont {Ostrovsky}},\
  }\bibfield  {title} {\bibinfo {title} {Anderson transitions: Criticality,
  symmetries and topologies},\ }\href
  {https://doi.org/10.1142/S0217979210064526} {\bibfield  {journal} {\bibinfo
  {journal} {International Journal of Modern Physics B}\ }\textbf {\bibinfo
  {volume} {24}},\ \bibinfo {pages} {1577} (\bibinfo {year}
  {2010})}\BibitemShut {NoStop}%
\bibitem [{\citenamefont {Obuse}\ and\ \citenamefont
  {Kawakami}(2011)}]{obuse2011topological}%
  \BibitemOpen
  \bibfield  {author} {\bibinfo {author} {\bibfnamefont {H.}~\bibnamefont
  {Obuse}}\ and\ \bibinfo {author} {\bibfnamefont {N.}~\bibnamefont
  {Kawakami}},\ }\bibfield  {title} {\bibinfo {title} {Topological phases and
  delocalization of quantum walks in random environments},\ }\href
  {https://doi.org/10.1103/PhysRevB.84.195139} {\bibfield  {journal} {\bibinfo
  {journal} {Phys. Rev. B}\ }\textbf {\bibinfo {volume} {84}},\ \bibinfo
  {pages} {195139} (\bibinfo {year} {2011})}\BibitemShut {NoStop}%
\bibitem [{\citenamefont {Rakovszky}\ and\ \citenamefont
  {Asboth}(2015)}]{rakovszky2015localization}%
  \BibitemOpen
  \bibfield  {author} {\bibinfo {author} {\bibfnamefont {T.}~\bibnamefont
  {Rakovszky}}\ and\ \bibinfo {author} {\bibfnamefont {J.~K.}\ \bibnamefont
  {Asboth}},\ }\bibfield  {title} {\bibinfo {title} {Localization,
  delocalization, and topological phase transitions in the one-dimensional
  split-step quantum walk},\ }\href
  {https://doi.org/10.1103/PhysRevA.92.052311} {\bibfield  {journal} {\bibinfo
  {journal} {Phys. Rev. A}\ }\textbf {\bibinfo {volume} {92}},\ \bibinfo
  {pages} {052311} (\bibinfo {year} {2015})}\BibitemShut {NoStop}%
\bibitem [{\citenamefont {Evers}\ and\ \citenamefont
  {Mirlin}(2008)}]{Evers:2008}%
  \BibitemOpen
  \bibfield  {author} {\bibinfo {author} {\bibfnamefont {F.}~\bibnamefont
  {Evers}}\ and\ \bibinfo {author} {\bibfnamefont {A.~D.}\ \bibnamefont
  {Mirlin}},\ }\bibfield  {title} {\bibinfo {title} {Anderson transitions},\
  }\href {https://doi.org/10.1103/RevModPhys.80.1355} {\bibfield  {journal}
  {\bibinfo  {journal} {Rev. Mod. Phys.}\ }\textbf {\bibinfo {volume} {80}},\
  \bibinfo {pages} {1355} (\bibinfo {year} {2008})}\BibitemShut {NoStop}%
\bibitem [{\citenamefont {Ohtsuki}\ and\ \citenamefont
  {Kawarabayashi}(1997)}]{Ohtsuki:1997}%
  \BibitemOpen
  \bibfield  {author} {\bibinfo {author} {\bibfnamefont {T.}~\bibnamefont
  {Ohtsuki}}\ and\ \bibinfo {author} {\bibfnamefont {T.}~\bibnamefont
  {Kawarabayashi}},\ }\bibfield  {title} {\bibinfo {title} {{Anomalous
  Diffusion at the Anderson Transitions}},\ }\href
  {https://doi.org/10.1143/JPSJ.66.314} {\bibfield  {journal} {\bibinfo
  {journal} {Journal of the Physical Society of Japan}\ }\textbf {\bibinfo
  {volume} {66}},\ \bibinfo {pages} {314} (\bibinfo {year} {1997})}\BibitemShut
  {NoStop}%
\bibitem [{\citenamefont {Pruisken}(1984)}]{Pruisken:1984}%
  \BibitemOpen
  \bibfield  {author} {\bibinfo {author} {\bibfnamefont {A.}~\bibnamefont
  {Pruisken}},\ }\bibfield  {title} {\bibinfo {title} {On localization in the
  theory of the quantized {H}all effect: A two-dimensional realization of the
  $\theta$-vacuum},\ }\href
  {https://doi.org/https://doi.org/10.1016/0550-3213(84)90101-9} {\bibfield
  {journal} {\bibinfo  {journal} {Nuclear Physics B}\ }\textbf {\bibinfo
  {volume} {235}},\ \bibinfo {pages} {277 } (\bibinfo {year}
  {1984})}\BibitemShut {NoStop}%
\bibitem [{\citenamefont {Fu}\ and\ \citenamefont {Kane}(2012)}]{Fu:2012}%
  \BibitemOpen
  \bibfield  {author} {\bibinfo {author} {\bibfnamefont {L.}~\bibnamefont
  {Fu}}\ and\ \bibinfo {author} {\bibfnamefont {C.~L.}\ \bibnamefont {Kane}},\
  }\bibfield  {title} {\bibinfo {title} {Topology, delocalization via average
  symmetry and the symplectic anderson transition},\ }\href
  {https://doi.org/10.1103/PhysRevLett.109.246605} {\bibfield  {journal}
  {\bibinfo  {journal} {Phys. Rev. Lett.}\ }\textbf {\bibinfo {volume} {109}},\
  \bibinfo {pages} {246605} (\bibinfo {year} {2012})}\BibitemShut {NoStop}%
\bibitem [{\citenamefont {Altland}\ \emph
  {et~al.}(2015{\natexlab{a}})\citenamefont {Altland}, \citenamefont
  {Bagrets},\ and\ \citenamefont {Kamenev}}]{Altland:2015}%
  \BibitemOpen
  \bibfield  {author} {\bibinfo {author} {\bibfnamefont {A.}~\bibnamefont
  {Altland}}, \bibinfo {author} {\bibfnamefont {D.}~\bibnamefont {Bagrets}},\
  and\ \bibinfo {author} {\bibfnamefont {A.}~\bibnamefont {Kamenev}},\
  }\bibfield  {title} {\bibinfo {title} {Topology versus anderson localization:
  Nonperturbative solutions in one dimension},\ }\href
  {http://link.aps.org/doi/10.1103/PhysRevB.91.085429} {\bibfield  {journal}
  {\bibinfo  {journal} {Phys. Rev. B}\ }\textbf {\bibinfo {volume} {91}},\
  \bibinfo {pages} {085429} (\bibinfo {year} {2015}{\natexlab{a}})}\BibitemShut
  {NoStop}%
\bibitem [{\citenamefont {Chab\'e}\ \emph {et~al.}(2008)\citenamefont
  {Chab\'e}, \citenamefont {Lemari\'e}, \citenamefont {Gr\'emaud},
  \citenamefont {Delande}, \citenamefont {Szriftgiser},\ and\ \citenamefont
  {Garreau}}]{Chabe:2008}%
  \BibitemOpen
  \bibfield  {author} {\bibinfo {author} {\bibfnamefont {J.}~\bibnamefont
  {Chab\'e}}, \bibinfo {author} {\bibfnamefont {G.}~\bibnamefont {Lemari\'e}},
  \bibinfo {author} {\bibfnamefont {B.}~\bibnamefont {Gr\'emaud}}, \bibinfo
  {author} {\bibfnamefont {D.}~\bibnamefont {Delande}}, \bibinfo {author}
  {\bibfnamefont {P.}~\bibnamefont {Szriftgiser}},\ and\ \bibinfo {author}
  {\bibfnamefont {J.~C.}\ \bibnamefont {Garreau}},\ }\bibfield  {title}
  {\bibinfo {title} {{Experimental Observation of the Anderson Metal-Insulator
  Transition with Atomic Matter Waves}},\ }\href
  {https://doi.org/10.1103/PhysRevLett.101.255702} {\bibfield  {journal}
  {\bibinfo  {journal} {Phys. Rev. Lett.}\ }\textbf {\bibinfo {volume} {101}},\
  \bibinfo {pages} {255702} (\bibinfo {year} {2008})}\BibitemShut {NoStop}%
\bibitem [{Note1()}]{Note1}%
  \BibitemOpen
  \bibinfo {note} {See Ref.~\cite {meier2018observation} for the recent
  realization of a $1d$ wire with chiral symmetry, where evidence for the
  topological Anderson insulator phase was given.}\BibitemShut {Stop}%
\bibitem [{\citenamefont {Balents}\ and\ \citenamefont
  {Fisher}(1997)}]{Balents:1997}%
  \BibitemOpen
  \bibfield  {author} {\bibinfo {author} {\bibfnamefont {L.}~\bibnamefont
  {Balents}}\ and\ \bibinfo {author} {\bibfnamefont {M.~P.~A.}\ \bibnamefont
  {Fisher}},\ }\bibfield  {title} {\bibinfo {title} {Delocalization transition
  via supersymmetry in one dimension},\ }\href
  {https://doi.org/10.1103/PhysRevB.56.12970} {\bibfield  {journal} {\bibinfo
  {journal} {Phys. Rev. B}\ }\textbf {\bibinfo {volume} {56}},\ \bibinfo
  {pages} {12970} (\bibinfo {year} {1997})}\BibitemShut {NoStop}%
\bibitem [{\citenamefont {Bagrets}\ \emph {et~al.}(2016)\citenamefont
  {Bagrets}, \citenamefont {Altland},\ and\ \citenamefont
  {Kamenev}}]{Bagrets:2016}%
  \BibitemOpen
  \bibfield  {author} {\bibinfo {author} {\bibfnamefont {D.}~\bibnamefont
  {Bagrets}}, \bibinfo {author} {\bibfnamefont {A.}~\bibnamefont {Altland}},\
  and\ \bibinfo {author} {\bibfnamefont {A.}~\bibnamefont {Kamenev}},\
  }\bibfield  {title} {\bibinfo {title} {{Sinai diffusion at quasi-1D
  topological phase transitions}},\ }\href
  {https://doi.org/10.1103/PhysRevLett.117.196801} {\bibfield  {journal}
  {\bibinfo  {journal} {Phys. Rev. Lett.}\ }\textbf {\bibinfo {volume} {117}},\
  \bibinfo {pages} {196801} (\bibinfo {year} {2016})}\BibitemShut {NoStop}%
\bibitem [{\citenamefont {Aharonov}\ \emph {et~al.}(1993)\citenamefont
  {Aharonov}, \citenamefont {Davidovich},\ and\ \citenamefont
  {Zagury}}]{Aharonov:1993}%
  \BibitemOpen
  \bibfield  {author} {\bibinfo {author} {\bibfnamefont {Y.}~\bibnamefont
  {Aharonov}}, \bibinfo {author} {\bibfnamefont {L.}~\bibnamefont
  {Davidovich}},\ and\ \bibinfo {author} {\bibfnamefont {N.}~\bibnamefont
  {Zagury}},\ }\bibfield  {title} {\bibinfo {title} {Quantum random walks},\
  }\href {https://doi.org/10.1103/PhysRevA.48.1687} {\bibfield  {journal}
  {\bibinfo  {journal} {Phys. Rev. A}\ }\textbf {\bibinfo {volume} {48}},\
  \bibinfo {pages} {1687} (\bibinfo {year} {1993})}\BibitemShut {NoStop}%
\bibitem [{\citenamefont {Bouwmeester}\ \emph {et~al.}(1999)\citenamefont
  {Bouwmeester}, \citenamefont {Marzoli}, \citenamefont {Karman}, \citenamefont
  {Schleich},\ and\ \citenamefont {Woerdman}}]{bouwmeester1999optical}%
  \BibitemOpen
  \bibfield  {author} {\bibinfo {author} {\bibfnamefont {D.}~\bibnamefont
  {Bouwmeester}}, \bibinfo {author} {\bibfnamefont {I.}~\bibnamefont
  {Marzoli}}, \bibinfo {author} {\bibfnamefont {G.~P.}\ \bibnamefont {Karman}},
  \bibinfo {author} {\bibfnamefont {W.}~\bibnamefont {Schleich}},\ and\
  \bibinfo {author} {\bibfnamefont {J.~P.}\ \bibnamefont {Woerdman}},\
  }\bibfield  {title} {\bibinfo {title} {Optical galton board},\ }\href
  {https://doi.org/10.1103/PhysRevA.61.013410} {\bibfield  {journal} {\bibinfo
  {journal} {Phys. Rev. A}\ }\textbf {\bibinfo {volume} {61}},\ \bibinfo
  {pages} {013410} (\bibinfo {year} {1999})}\BibitemShut {NoStop}%
\bibitem [{\citenamefont {Perets}\ \emph {et~al.}(2008)\citenamefont {Perets},
  \citenamefont {Lahini}, \citenamefont {Pozzi}, \citenamefont {Sorel},
  \citenamefont {Morandotti},\ and\ \citenamefont
  {Silberberg}}]{perets_realization_2008}%
  \BibitemOpen
  \bibfield  {author} {\bibinfo {author} {\bibfnamefont {H.~B.}\ \bibnamefont
  {Perets}}, \bibinfo {author} {\bibfnamefont {Y.}~\bibnamefont {Lahini}},
  \bibinfo {author} {\bibfnamefont {F.}~\bibnamefont {Pozzi}}, \bibinfo
  {author} {\bibfnamefont {M.}~\bibnamefont {Sorel}}, \bibinfo {author}
  {\bibfnamefont {R.}~\bibnamefont {Morandotti}},\ and\ \bibinfo {author}
  {\bibfnamefont {Y.}~\bibnamefont {Silberberg}},\ }\bibfield  {title}
  {\bibinfo {title} {Realization of quantum walks with negligible decoherence
  in waveguide lattices},\ }\href
  {https://doi.org/10.1103/PhysRevLett.100.170506} {\bibfield  {journal}
  {\bibinfo  {journal} {Phys. Rev. Lett.}\ }\textbf {\bibinfo {volume} {100}},\
  \bibinfo {pages} {170506} (\bibinfo {year} {2008})}\BibitemShut {NoStop}%
\bibitem [{\citenamefont {Peruzzo}\ \emph {et~al.}(2010)\citenamefont
  {Peruzzo}, \citenamefont {Lobino}, \citenamefont {Matthews}, \citenamefont
  {Matsuda}, \citenamefont {Politi}, \citenamefont {Poulios}, \citenamefont
  {Zhou}, \citenamefont {Lahini}, \citenamefont {Ismail}, \citenamefont
  {W{\"o}rhoff}, \citenamefont {Bromberg}, \citenamefont {Silberberg},
  \citenamefont {Thompson},\ and\ \citenamefont {OBrien}}]{peruzzo2010quantum}%
  \BibitemOpen
  \bibfield  {author} {\bibinfo {author} {\bibfnamefont {A.}~\bibnamefont
  {Peruzzo}}, \bibinfo {author} {\bibfnamefont {M.}~\bibnamefont {Lobino}},
  \bibinfo {author} {\bibfnamefont {J.~C.~F.}\ \bibnamefont {Matthews}},
  \bibinfo {author} {\bibfnamefont {N.}~\bibnamefont {Matsuda}}, \bibinfo
  {author} {\bibfnamefont {A.}~\bibnamefont {Politi}}, \bibinfo {author}
  {\bibfnamefont {K.}~\bibnamefont {Poulios}}, \bibinfo {author} {\bibfnamefont
  {X.-Q.}\ \bibnamefont {Zhou}}, \bibinfo {author} {\bibfnamefont
  {Y.}~\bibnamefont {Lahini}}, \bibinfo {author} {\bibfnamefont
  {N.}~\bibnamefont {Ismail}}, \bibinfo {author} {\bibfnamefont
  {K.}~\bibnamefont {W{\"o}rhoff}}, \bibinfo {author} {\bibfnamefont
  {Y.}~\bibnamefont {Bromberg}}, \bibinfo {author} {\bibfnamefont
  {Y.}~\bibnamefont {Silberberg}}, \bibinfo {author} {\bibfnamefont {M.~G.}\
  \bibnamefont {Thompson}},\ and\ \bibinfo {author} {\bibfnamefont {J.~L.}\
  \bibnamefont {OBrien}},\ }\bibfield  {title} {\bibinfo {title} {Quantum walks
  of correlated photons},\ }\href {https://doi.org/10.1126/science.1193515}
  {\bibfield  {journal} {\bibinfo  {journal} {Science}\ }\textbf {\bibinfo
  {volume} {329}},\ \bibinfo {pages} {1500} (\bibinfo {year}
  {2010})}\BibitemShut {NoStop}%
\bibitem [{\citenamefont {Broome}\ \emph {et~al.}(2010)\citenamefont {Broome},
  \citenamefont {Fedrizzi}, \citenamefont {Lanyon}, \citenamefont {Kassal},
  \citenamefont {{Aspuru-Guzik}},\ and\ \citenamefont
  {White}}]{broome_discrete_2010}%
  \BibitemOpen
  \bibfield  {author} {\bibinfo {author} {\bibfnamefont {M.~A.}\ \bibnamefont
  {Broome}}, \bibinfo {author} {\bibfnamefont {A.}~\bibnamefont {Fedrizzi}},
  \bibinfo {author} {\bibfnamefont {B.~P.}\ \bibnamefont {Lanyon}}, \bibinfo
  {author} {\bibfnamefont {I.}~\bibnamefont {Kassal}}, \bibinfo {author}
  {\bibfnamefont {A.}~\bibnamefont {{Aspuru-Guzik}}},\ and\ \bibinfo {author}
  {\bibfnamefont {A.~G.}\ \bibnamefont {White}},\ }\bibfield  {title} {\bibinfo
  {title} {Discrete {{Single}}-{{Photon Quantum Walks}} with {{Tunable
  Decoherence}}},\ }\href {https://doi.org/10.1103/PhysRevLett.104.153602}
  {\bibfield  {journal} {\bibinfo  {journal} {Phys. Rev. Lett.}\ }\textbf
  {\bibinfo {volume} {104}},\ \bibinfo {pages} {153602} (\bibinfo {year}
  {2010})}\BibitemShut {NoStop}%
\bibitem [{\citenamefont {Schreiber}\ \emph {et~al.}(2010)\citenamefont
  {Schreiber}, \citenamefont {Cassemiro}, \citenamefont {Poto{\v c}ek},
  \citenamefont {G\'abris}, \citenamefont {Mosley}, \citenamefont {Andersson},
  \citenamefont {Jex},\ and\ \citenamefont
  {Silberhorn}}]{schreiber_photons_2010}%
  \BibitemOpen
  \bibfield  {author} {\bibinfo {author} {\bibfnamefont {A.}~\bibnamefont
  {Schreiber}}, \bibinfo {author} {\bibfnamefont {K.~N.}\ \bibnamefont
  {Cassemiro}}, \bibinfo {author} {\bibfnamefont {V.}~\bibnamefont {Poto{\v
  c}ek}}, \bibinfo {author} {\bibfnamefont {A.}~\bibnamefont {G\'abris}},
  \bibinfo {author} {\bibfnamefont {P.~J.}\ \bibnamefont {Mosley}}, \bibinfo
  {author} {\bibfnamefont {E.}~\bibnamefont {Andersson}}, \bibinfo {author}
  {\bibfnamefont {I.}~\bibnamefont {Jex}},\ and\ \bibinfo {author}
  {\bibfnamefont {C.}~\bibnamefont {Silberhorn}},\ }\bibfield  {title}
  {\bibinfo {title} {Photons {{Walking}} the {{Line}}: {{A Quantum Walk}} with
  {{Adjustable Coin Operations}}},\ }\href
  {https://doi.org/10.1103/PhysRevLett.104.050502} {\bibfield  {journal}
  {\bibinfo  {journal} {Phys. Rev. Lett.}\ }\textbf {\bibinfo {volume} {104}},\
  \bibinfo {pages} {050502} (\bibinfo {year} {2010})}\BibitemShut {NoStop}%
\bibitem [{\citenamefont {Schreiber}\ \emph {et~al.}(2012)\citenamefont
  {Schreiber}, \citenamefont {G{\'a}bris}, \citenamefont {Rohde}, \citenamefont
  {Laiho}, \citenamefont {{\v S}tefa{\v n}{\'a}k}, \citenamefont {Poto{\v
  c}ek}, \citenamefont {Hamilton}, \citenamefont {Jex},\ and\ \citenamefont
  {Silberhorn}}]{Schreiber:2012}%
  \BibitemOpen
  \bibfield  {author} {\bibinfo {author} {\bibfnamefont {A.}~\bibnamefont
  {Schreiber}}, \bibinfo {author} {\bibfnamefont {A.}~\bibnamefont
  {G{\'a}bris}}, \bibinfo {author} {\bibfnamefont {P.~P.}\ \bibnamefont
  {Rohde}}, \bibinfo {author} {\bibfnamefont {K.}~\bibnamefont {Laiho}},
  \bibinfo {author} {\bibfnamefont {M.}~\bibnamefont {{\v S}tefa{\v n}{\'a}k}},
  \bibinfo {author} {\bibfnamefont {V.}~\bibnamefont {Poto{\v c}ek}}, \bibinfo
  {author} {\bibfnamefont {C.}~\bibnamefont {Hamilton}}, \bibinfo {author}
  {\bibfnamefont {I.}~\bibnamefont {Jex}},\ and\ \bibinfo {author}
  {\bibfnamefont {C.}~\bibnamefont {Silberhorn}},\ }\bibfield  {title}
  {\bibinfo {title} {A 2d quantum walk simulation of two-particle dynamics},\
  }\href {https://doi.org/10.1126/science.1218448} {\bibfield  {journal}
  {\bibinfo  {journal} {Science}\ }\textbf {\bibinfo {volume} {336}},\ \bibinfo
  {pages} {55} (\bibinfo {year} {2012})}\BibitemShut {NoStop}%
\bibitem [{\citenamefont {Sansoni}\ \emph {et~al.}(2012)\citenamefont
  {Sansoni}, \citenamefont {Sciarrino}, \citenamefont {Vallone}, \citenamefont
  {Mataloni}, \citenamefont {Crespi}, \citenamefont {Ramponi},\ and\
  \citenamefont {Osellame}}]{sansoni_two-particle_2012}%
  \BibitemOpen
  \bibfield  {author} {\bibinfo {author} {\bibfnamefont {L.}~\bibnamefont
  {Sansoni}}, \bibinfo {author} {\bibfnamefont {F.}~\bibnamefont {Sciarrino}},
  \bibinfo {author} {\bibfnamefont {G.}~\bibnamefont {Vallone}}, \bibinfo
  {author} {\bibfnamefont {P.}~\bibnamefont {Mataloni}}, \bibinfo {author}
  {\bibfnamefont {A.}~\bibnamefont {Crespi}}, \bibinfo {author} {\bibfnamefont
  {R.}~\bibnamefont {Ramponi}},\ and\ \bibinfo {author} {\bibfnamefont
  {R.}~\bibnamefont {Osellame}},\ }\bibfield  {title} {\bibinfo {title}
  {Two-{Particle} {Bosonic}-{Fermionic} {Quantum} {Walk} via {Integrated}
  {Photonics}},\ }\href {https://doi.org/10.1103/PhysRevLett.108.010502}
  {\bibfield  {journal} {\bibinfo  {journal} {Physical Review Letters}\
  }\textbf {\bibinfo {volume} {108}},\ \bibinfo {pages} {010502} (\bibinfo
  {year} {2012})}\BibitemShut {NoStop}%
\bibitem [{\citenamefont {Crespi}\ \emph {et~al.}(2013)\citenamefont {Crespi},
  \citenamefont {Osellame}, \citenamefont {Ramponi}, \citenamefont
  {Giovannetti}, \citenamefont {Fazio}, \citenamefont {Sansoni}, \citenamefont
  {De~Nicola}, \citenamefont {Sciarrino},\ and\ \citenamefont
  {Mataloni}}]{crespi2013anderson}%
  \BibitemOpen
  \bibfield  {author} {\bibinfo {author} {\bibfnamefont {A.}~\bibnamefont
  {Crespi}}, \bibinfo {author} {\bibfnamefont {R.}~\bibnamefont {Osellame}},
  \bibinfo {author} {\bibfnamefont {R.}~\bibnamefont {Ramponi}}, \bibinfo
  {author} {\bibfnamefont {V.}~\bibnamefont {Giovannetti}}, \bibinfo {author}
  {\bibfnamefont {R.}~\bibnamefont {Fazio}}, \bibinfo {author} {\bibfnamefont
  {L.}~\bibnamefont {Sansoni}}, \bibinfo {author} {\bibfnamefont
  {F.}~\bibnamefont {De~Nicola}}, \bibinfo {author} {\bibfnamefont
  {F.}~\bibnamefont {Sciarrino}},\ and\ \bibinfo {author} {\bibfnamefont
  {P.}~\bibnamefont {Mataloni}},\ }\bibfield  {title} {\bibinfo {title}
  {Anderson localization of entangled photons in an integrated quantum walk},\
  }\href {https://doi.org/10.1038/nphoton.2013.26} {\bibfield  {journal}
  {\bibinfo  {journal} {Nature Photonics}\ }\textbf {\bibinfo {volume} {7}},\
  \bibinfo {pages} {322} (\bibinfo {year} {2013})}\BibitemShut {NoStop}%
\bibitem [{\citenamefont {Cardano}\ \emph {et~al.}(2015)\citenamefont
  {Cardano}, \citenamefont {Massa}, \citenamefont {Qassim}, \citenamefont
  {Karimi}, \citenamefont {Slussarenko}, \citenamefont {Paparo}, \citenamefont
  {de~Lisio}, \citenamefont {Sciarrino}, \citenamefont {Santamato},
  \citenamefont {Boyd},\ and\ \citenamefont {Marrucci}}]{cardano_quantum_2015}%
  \BibitemOpen
  \bibfield  {author} {\bibinfo {author} {\bibfnamefont {F.}~\bibnamefont
  {Cardano}}, \bibinfo {author} {\bibfnamefont {F.}~\bibnamefont {Massa}},
  \bibinfo {author} {\bibfnamefont {H.}~\bibnamefont {Qassim}}, \bibinfo
  {author} {\bibfnamefont {E.}~\bibnamefont {Karimi}}, \bibinfo {author}
  {\bibfnamefont {S.}~\bibnamefont {Slussarenko}}, \bibinfo {author}
  {\bibfnamefont {D.}~\bibnamefont {Paparo}}, \bibinfo {author} {\bibfnamefont
  {C.}~\bibnamefont {de~Lisio}}, \bibinfo {author} {\bibfnamefont
  {F.}~\bibnamefont {Sciarrino}}, \bibinfo {author} {\bibfnamefont
  {E.}~\bibnamefont {Santamato}}, \bibinfo {author} {\bibfnamefont {R.~W.}\
  \bibnamefont {Boyd}},\ and\ \bibinfo {author} {\bibfnamefont
  {L.}~\bibnamefont {Marrucci}},\ }\bibfield  {title} {\bibinfo {title}
  {Quantum walks and wavepacket dynamics on a lattice with twisted photons},\
  }\href {https://doi.org/10.1126/sciadv.1500087} {\bibfield  {journal}
  {\bibinfo  {journal} {Science Advances}\ }\textbf {\bibinfo {volume} {1}},\
  \bibinfo {pages} {e1500087} (\bibinfo {year} {2015})}\BibitemShut {NoStop}%
\bibitem [{\citenamefont {Xue}\ \emph {et~al.}(2015{\natexlab{a}})\citenamefont
  {Xue}, \citenamefont {Zhang}, \citenamefont {Bian}, \citenamefont {Zhan},
  \citenamefont {Qin},\ and\ \citenamefont {Sanders}}]{xue2015localized}%
  \BibitemOpen
  \bibfield  {author} {\bibinfo {author} {\bibfnamefont {P.}~\bibnamefont
  {Xue}}, \bibinfo {author} {\bibfnamefont {R.}~\bibnamefont {Zhang}}, \bibinfo
  {author} {\bibfnamefont {Z.}~\bibnamefont {Bian}}, \bibinfo {author}
  {\bibfnamefont {X.}~\bibnamefont {Zhan}}, \bibinfo {author} {\bibfnamefont
  {H.}~\bibnamefont {Qin}},\ and\ \bibinfo {author} {\bibfnamefont {B.~C.}\
  \bibnamefont {Sanders}},\ }\bibfield  {title} {\bibinfo {title} {Localized
  state in a two-dimensional quantum walk on a disordered lattice},\ }\href
  {https://doi.org/10.1103/PhysRevA.92.042316} {\bibfield  {journal} {\bibinfo
  {journal} {Phys. Rev. A}\ }\textbf {\bibinfo {volume} {92}},\ \bibinfo
  {pages} {042316} (\bibinfo {year} {2015}{\natexlab{a}})}\BibitemShut
  {NoStop}%
\bibitem [{\citenamefont {Schmitz}\ \emph {et~al.}(2009)\citenamefont
  {Schmitz}, \citenamefont {Matjeschk}, \citenamefont {Schneider},
  \citenamefont {Glueckert}, \citenamefont {Enderlein}, \citenamefont {Huber},\
  and\ \citenamefont {Schaetz}}]{schmitz2009quantum}%
  \BibitemOpen
  \bibfield  {author} {\bibinfo {author} {\bibfnamefont {H.}~\bibnamefont
  {Schmitz}}, \bibinfo {author} {\bibfnamefont {R.}~\bibnamefont {Matjeschk}},
  \bibinfo {author} {\bibfnamefont {C.}~\bibnamefont {Schneider}}, \bibinfo
  {author} {\bibfnamefont {J.}~\bibnamefont {Glueckert}}, \bibinfo {author}
  {\bibfnamefont {M.}~\bibnamefont {Enderlein}}, \bibinfo {author}
  {\bibfnamefont {T.}~\bibnamefont {Huber}},\ and\ \bibinfo {author}
  {\bibfnamefont {T.}~\bibnamefont {Schaetz}},\ }\bibfield  {title} {\bibinfo
  {title} {Quantum walk of a trapped ion in phase space},\ }\href
  {https://doi.org/10.1103/PhysRevLett.103.090504} {\bibfield  {journal}
  {\bibinfo  {journal} {Phys. Rev. Lett.}\ }\textbf {\bibinfo {volume} {103}},\
  \bibinfo {pages} {090504} (\bibinfo {year} {2009})}\BibitemShut {NoStop}%
\bibitem [{\citenamefont {Z\"ahringer}\ \emph {et~al.}(2010)\citenamefont
  {Z\"ahringer}, \citenamefont {Kirchmair}, \citenamefont {Gerritsma},
  \citenamefont {Solano}, \citenamefont {Blatt},\ and\ \citenamefont
  {Roos}}]{zahringer2010realization}%
  \BibitemOpen
  \bibfield  {author} {\bibinfo {author} {\bibfnamefont {F.}~\bibnamefont
  {Z\"ahringer}}, \bibinfo {author} {\bibfnamefont {G.}~\bibnamefont
  {Kirchmair}}, \bibinfo {author} {\bibfnamefont {R.}~\bibnamefont
  {Gerritsma}}, \bibinfo {author} {\bibfnamefont {E.}~\bibnamefont {Solano}},
  \bibinfo {author} {\bibfnamefont {R.}~\bibnamefont {Blatt}},\ and\ \bibinfo
  {author} {\bibfnamefont {C.~F.}\ \bibnamefont {Roos}},\ }\bibfield  {title}
  {\bibinfo {title} {Realization of a quantum walk with one and two trapped
  ions},\ }\href {https://doi.org/10.1103/PhysRevLett.104.100503} {\bibfield
  {journal} {\bibinfo  {journal} {Phys. Rev. Lett.}\ }\textbf {\bibinfo
  {volume} {104}},\ \bibinfo {pages} {100503} (\bibinfo {year}
  {2010})}\BibitemShut {NoStop}%
\bibitem [{\citenamefont {Genske}\ \emph {et~al.}(2013)\citenamefont {Genske},
  \citenamefont {Alt}, \citenamefont {Steffen}, \citenamefont {Werner},
  \citenamefont {Werner}, \citenamefont {Meschede},\ and\ \citenamefont
  {Alberti}}]{genske2013electric}%
  \BibitemOpen
  \bibfield  {author} {\bibinfo {author} {\bibfnamefont {M.}~\bibnamefont
  {Genske}}, \bibinfo {author} {\bibfnamefont {W.}~\bibnamefont {Alt}},
  \bibinfo {author} {\bibfnamefont {A.}~\bibnamefont {Steffen}}, \bibinfo
  {author} {\bibfnamefont {A.~H.}\ \bibnamefont {Werner}}, \bibinfo {author}
  {\bibfnamefont {R.~F.}\ \bibnamefont {Werner}}, \bibinfo {author}
  {\bibfnamefont {D.}~\bibnamefont {Meschede}},\ and\ \bibinfo {author}
  {\bibfnamefont {A.}~\bibnamefont {Alberti}},\ }\bibfield  {title} {\bibinfo
  {title} {Electric quantum walks with individual atoms},\ }\href
  {https://doi.org/10.1103/PhysRevLett.110.190601} {\bibfield  {journal}
  {\bibinfo  {journal} {Phys. Rev. Lett.}\ }\textbf {\bibinfo {volume} {110}},\
  \bibinfo {pages} {190601} (\bibinfo {year} {2013})}\BibitemShut {NoStop}%
\bibitem [{\citenamefont {Karski}\ \emph {et~al.}(2009)\citenamefont {Karski},
  \citenamefont {F{\"o}rster}, \citenamefont {Choi}, \citenamefont {Steffen},
  \citenamefont {Alt}, \citenamefont {Meschede},\ and\ \citenamefont
  {Widera}}]{karski2009quantum}%
  \BibitemOpen
  \bibfield  {author} {\bibinfo {author} {\bibfnamefont {M.}~\bibnamefont
  {Karski}}, \bibinfo {author} {\bibfnamefont {L.}~\bibnamefont {F{\"o}rster}},
  \bibinfo {author} {\bibfnamefont {J.-M.}\ \bibnamefont {Choi}}, \bibinfo
  {author} {\bibfnamefont {A.}~\bibnamefont {Steffen}}, \bibinfo {author}
  {\bibfnamefont {W.}~\bibnamefont {Alt}}, \bibinfo {author} {\bibfnamefont
  {D.}~\bibnamefont {Meschede}},\ and\ \bibinfo {author} {\bibfnamefont
  {A.}~\bibnamefont {Widera}},\ }\bibfield  {title} {\bibinfo {title} {Quantum
  walk in position space with single optically trapped atoms},\ }\href
  {https://doi.org/10.1126/science.1174436} {\bibfield  {journal} {\bibinfo
  {journal} {Science}\ }\textbf {\bibinfo {volume} {325}},\ \bibinfo {pages}
  {174} (\bibinfo {year} {2009})}\BibitemShut {NoStop}%
\bibitem [{\citenamefont {Preiss}\ \emph {et~al.}(2015)\citenamefont {Preiss},
  \citenamefont {Ma}, \citenamefont {Tai}, \citenamefont {Lukin}, \citenamefont
  {Rispoli}, \citenamefont {Zupancic}, \citenamefont {Lahini}, \citenamefont
  {Islam},\ and\ \citenamefont {Greiner}}]{preiss2015strongly}%
  \BibitemOpen
  \bibfield  {author} {\bibinfo {author} {\bibfnamefont {P.~M.}\ \bibnamefont
  {Preiss}}, \bibinfo {author} {\bibfnamefont {R.}~\bibnamefont {Ma}}, \bibinfo
  {author} {\bibfnamefont {M.~E.}\ \bibnamefont {Tai}}, \bibinfo {author}
  {\bibfnamefont {A.}~\bibnamefont {Lukin}}, \bibinfo {author} {\bibfnamefont
  {M.}~\bibnamefont {Rispoli}}, \bibinfo {author} {\bibfnamefont
  {P.}~\bibnamefont {Zupancic}}, \bibinfo {author} {\bibfnamefont
  {Y.}~\bibnamefont {Lahini}}, \bibinfo {author} {\bibfnamefont
  {R.}~\bibnamefont {Islam}},\ and\ \bibinfo {author} {\bibfnamefont
  {M.}~\bibnamefont {Greiner}},\ }\bibfield  {title} {\bibinfo {title}
  {Strongly correlated quantum walks in optical lattices},\ }\href
  {https://doi.org/10.1126/science.1260364} {\bibfield  {journal} {\bibinfo
  {journal} {Science}\ }\textbf {\bibinfo {volume} {347}},\ \bibinfo {pages}
  {1229} (\bibinfo {year} {2015})}\BibitemShut {NoStop}%
\bibitem [{\citenamefont {Du}\ \emph {et~al.}(2003)\citenamefont {Du},
  \citenamefont {Li}, \citenamefont {Xu}, \citenamefont {Shi}, \citenamefont
  {Wu}, \citenamefont {Zhou},\ and\ \citenamefont {Han}}]{du2003experimental}%
  \BibitemOpen
  \bibfield  {author} {\bibinfo {author} {\bibfnamefont {J.}~\bibnamefont
  {Du}}, \bibinfo {author} {\bibfnamefont {H.}~\bibnamefont {Li}}, \bibinfo
  {author} {\bibfnamefont {X.}~\bibnamefont {Xu}}, \bibinfo {author}
  {\bibfnamefont {M.}~\bibnamefont {Shi}}, \bibinfo {author} {\bibfnamefont
  {J.}~\bibnamefont {Wu}}, \bibinfo {author} {\bibfnamefont {X.}~\bibnamefont
  {Zhou}},\ and\ \bibinfo {author} {\bibfnamefont {R.}~\bibnamefont {Han}},\
  }\bibfield  {title} {\bibinfo {title} {Experimental implementation of the
  quantum random-walk algorithm},\ }\href
  {https://doi.org/10.1103/PhysRevA.67.042316} {\bibfield  {journal} {\bibinfo
  {journal} {Phys. Rev. A}\ }\textbf {\bibinfo {volume} {67}},\ \bibinfo
  {pages} {042316} (\bibinfo {year} {2003})}\BibitemShut {NoStop}%
\bibitem [{\citenamefont {Wang}\ and\ \citenamefont
  {Manouchehri}(2013)}]{wang2013physical}%
  \BibitemOpen
  \bibfield  {author} {\bibinfo {author} {\bibfnamefont {J.}~\bibnamefont
  {Wang}}\ and\ \bibinfo {author} {\bibfnamefont {K.}~\bibnamefont
  {Manouchehri}},\ }\href@noop {} {\emph {\bibinfo {title} {Physical
  implementation of quantum walks}}}\ (\bibinfo  {publisher} {Springer},\
  \bibinfo {year} {2013})\BibitemShut {NoStop}%
\bibitem [{\citenamefont {Schreiber}\ \emph {et~al.}(2011)\citenamefont
  {Schreiber}, \citenamefont {Cassemiro}, \citenamefont
  {Poto\ifmmode~\check{c}\else \v{c}\fi{}ek}, \citenamefont {G\'abris},
  \citenamefont {Jex},\ and\ \citenamefont {Silberhorn}}]{Schreiber:2011}%
  \BibitemOpen
  \bibfield  {author} {\bibinfo {author} {\bibfnamefont {A.}~\bibnamefont
  {Schreiber}}, \bibinfo {author} {\bibfnamefont {K.~N.}\ \bibnamefont
  {Cassemiro}}, \bibinfo {author} {\bibfnamefont {V.}~\bibnamefont
  {Poto\ifmmode~\check{c}\else \v{c}\fi{}ek}}, \bibinfo {author} {\bibfnamefont
  {A.}~\bibnamefont {G\'abris}}, \bibinfo {author} {\bibfnamefont
  {I.}~\bibnamefont {Jex}},\ and\ \bibinfo {author} {\bibfnamefont
  {C.}~\bibnamefont {Silberhorn}},\ }\bibfield  {title} {\bibinfo {title}
  {Decoherence and disorder in quantum walks: From ballistic spread to
  localization},\ }\href {https://doi.org/10.1103/PhysRevLett.106.180403}
  {\bibfield  {journal} {\bibinfo  {journal} {Phys. Rev. Lett.}\ }\textbf
  {\bibinfo {volume} {106}},\ \bibinfo {pages} {180403} (\bibinfo {year}
  {2011})}\BibitemShut {NoStop}%
\bibitem [{\citenamefont {Kitagawa}\ \emph {et~al.}(2012)\citenamefont
  {Kitagawa}, \citenamefont {Broome}, \citenamefont {Fedrizzi}, \citenamefont
  {Rudner}, \citenamefont {Berg}, \citenamefont {Kassal}, \citenamefont
  {Aspuru-Guzik}, \citenamefont {Demler},\ and\ \citenamefont
  {White}}]{Kitagawa:2012}%
  \BibitemOpen
  \bibfield  {author} {\bibinfo {author} {\bibfnamefont {T.}~\bibnamefont
  {Kitagawa}}, \bibinfo {author} {\bibfnamefont {M.~A.}\ \bibnamefont
  {Broome}}, \bibinfo {author} {\bibfnamefont {A.}~\bibnamefont {Fedrizzi}},
  \bibinfo {author} {\bibfnamefont {M.~S.}\ \bibnamefont {Rudner}}, \bibinfo
  {author} {\bibfnamefont {E.}~\bibnamefont {Berg}}, \bibinfo {author}
  {\bibfnamefont {I.}~\bibnamefont {Kassal}}, \bibinfo {author} {\bibfnamefont
  {A.}~\bibnamefont {Aspuru-Guzik}}, \bibinfo {author} {\bibfnamefont
  {E.}~\bibnamefont {Demler}},\ and\ \bibinfo {author} {\bibfnamefont {A.~G.}\
  \bibnamefont {White}},\ }\bibfield  {title} {\bibinfo {title} {Observation of
  topologically protected bound states in photonic quantum walks},\ }\href
  {http://dx.doi.org/10.1038/ncomms1872} {\bibfield  {journal} {\bibinfo
  {journal} {Nature Communications}\ }\textbf {\bibinfo {volume} {3}},\
  \bibinfo {pages} {882 EP } (\bibinfo {year} {2012})},\ \bibinfo {note}
  {article}\BibitemShut {NoStop}%
\bibitem [{\citenamefont {Rechtsman}\ \emph {et~al.}(2013)\citenamefont
  {Rechtsman}, \citenamefont {Zeuner}, \citenamefont {Plotnik}, \citenamefont
  {Lumer}, \citenamefont {Podolsky}, \citenamefont {Dreisow}, \citenamefont
  {Nolte}, \citenamefont {Segev},\ and\ \citenamefont
  {Szameit}}]{rechtsman_photonic_2013}%
  \BibitemOpen
  \bibfield  {author} {\bibinfo {author} {\bibfnamefont {M.~C.}\ \bibnamefont
  {Rechtsman}}, \bibinfo {author} {\bibfnamefont {J.~M.}\ \bibnamefont
  {Zeuner}}, \bibinfo {author} {\bibfnamefont {Y.}~\bibnamefont {Plotnik}},
  \bibinfo {author} {\bibfnamefont {Y.}~\bibnamefont {Lumer}}, \bibinfo
  {author} {\bibfnamefont {D.}~\bibnamefont {Podolsky}}, \bibinfo {author}
  {\bibfnamefont {F.}~\bibnamefont {Dreisow}}, \bibinfo {author} {\bibfnamefont
  {S.}~\bibnamefont {Nolte}}, \bibinfo {author} {\bibfnamefont
  {M.}~\bibnamefont {Segev}},\ and\ \bibinfo {author} {\bibfnamefont
  {A.}~\bibnamefont {Szameit}},\ }\bibfield  {title} {\bibinfo {title}
  {Photonic {Floquet} topological insulators},\ }\href
  {http://dx.doi.org/10.1038/nature12066} {\bibfield  {journal} {\bibinfo
  {journal} {Nature}\ }\textbf {\bibinfo {volume} {496}},\ \bibinfo {pages}
  {196} (\bibinfo {year} {2013})}\BibitemShut {NoStop}%
\bibitem [{\citenamefont {Zeuner}\ \emph {et~al.}(2015)\citenamefont {Zeuner},
  \citenamefont {Rechtsman}, \citenamefont {Plotnik}, \citenamefont {Lumer},
  \citenamefont {Nolte}, \citenamefont {Rudner}, \citenamefont {Segev},\ and\
  \citenamefont {Szameit}}]{zeuner_observation_2015}%
  \BibitemOpen
  \bibfield  {author} {\bibinfo {author} {\bibfnamefont {J.~M.}\ \bibnamefont
  {Zeuner}}, \bibinfo {author} {\bibfnamefont {M.~C.}\ \bibnamefont
  {Rechtsman}}, \bibinfo {author} {\bibfnamefont {Y.}~\bibnamefont {Plotnik}},
  \bibinfo {author} {\bibfnamefont {Y.}~\bibnamefont {Lumer}}, \bibinfo
  {author} {\bibfnamefont {S.}~\bibnamefont {Nolte}}, \bibinfo {author}
  {\bibfnamefont {M.~S.}\ \bibnamefont {Rudner}}, \bibinfo {author}
  {\bibfnamefont {M.}~\bibnamefont {Segev}},\ and\ \bibinfo {author}
  {\bibfnamefont {A.}~\bibnamefont {Szameit}},\ }\bibfield  {title} {\bibinfo
  {title} {Observation of a {Topological} {Transition} in the {Bulk} of a
  {Non}-{Hermitian} {System}},\ }\href
  {https://doi.org/10.1103/PhysRevLett.115.040402} {\bibfield  {journal}
  {\bibinfo  {journal} {Phys. Rev. Lett.}\ }\textbf {\bibinfo {volume} {115}},\
  \bibinfo {pages} {040402} (\bibinfo {year} {2015})}\BibitemShut {NoStop}%
\bibitem [{\citenamefont {Cardano}\ \emph {et~al.}(2016)\citenamefont
  {Cardano}, \citenamefont {Maffei}, \citenamefont {Massa}, \citenamefont
  {Piccirillo}, \citenamefont {de~Lisio}, \citenamefont {De~Filippis},
  \citenamefont {Cataudella}, \citenamefont {Santamato},\ and\ \citenamefont
  {Marrucci}}]{cardano_statistical_2015}%
  \BibitemOpen
  \bibfield  {author} {\bibinfo {author} {\bibfnamefont {F.}~\bibnamefont
  {Cardano}}, \bibinfo {author} {\bibfnamefont {M.}~\bibnamefont {Maffei}},
  \bibinfo {author} {\bibfnamefont {F.}~\bibnamefont {Massa}}, \bibinfo
  {author} {\bibfnamefont {B.}~\bibnamefont {Piccirillo}}, \bibinfo {author}
  {\bibfnamefont {C.}~\bibnamefont {de~Lisio}}, \bibinfo {author}
  {\bibfnamefont {G.}~\bibnamefont {De~Filippis}}, \bibinfo {author}
  {\bibfnamefont {V.}~\bibnamefont {Cataudella}}, \bibinfo {author}
  {\bibfnamefont {E.}~\bibnamefont {Santamato}},\ and\ \bibinfo {author}
  {\bibfnamefont {L.}~\bibnamefont {Marrucci}},\ }\bibfield  {title} {\bibinfo
  {title} {Statistical moments of quantum-walk dynamics reveal topological
  quantum transitions},\ }\href {https://doi.org/10.1038/ncomms11439}
  {\bibfield  {journal} {\bibinfo  {journal} {Nature Communications}\ }\textbf
  {\bibinfo {volume} {7}},\ \bibinfo {pages} {11439} (\bibinfo {year}
  {2016})}\BibitemShut {NoStop}%
\bibitem [{\citenamefont {Xiao}\ \emph {et~al.}(2017)\citenamefont {Xiao},
  \citenamefont {Zhan}, \citenamefont {Bian}, \citenamefont {Wang},
  \citenamefont {Zhang}, \citenamefont {Wang}, \citenamefont {Li},
  \citenamefont {Mochizuki}, \citenamefont {Kim}, \citenamefont {Kawakami}
  \emph {et~al.}}]{xiao2017observation}%
  \BibitemOpen
  \bibfield  {author} {\bibinfo {author} {\bibfnamefont {L.}~\bibnamefont
  {Xiao}}, \bibinfo {author} {\bibfnamefont {X.}~\bibnamefont {Zhan}}, \bibinfo
  {author} {\bibfnamefont {Z.}~\bibnamefont {Bian}}, \bibinfo {author}
  {\bibfnamefont {K.}~\bibnamefont {Wang}}, \bibinfo {author} {\bibfnamefont
  {X.}~\bibnamefont {Zhang}}, \bibinfo {author} {\bibfnamefont
  {X.}~\bibnamefont {Wang}}, \bibinfo {author} {\bibfnamefont {J.}~\bibnamefont
  {Li}}, \bibinfo {author} {\bibfnamefont {K.}~\bibnamefont {Mochizuki}},
  \bibinfo {author} {\bibfnamefont {D.}~\bibnamefont {Kim}}, \bibinfo {author}
  {\bibfnamefont {N.}~\bibnamefont {Kawakami}}, \emph {et~al.},\ }\bibfield
  {title} {\bibinfo {title} {Observation of topological edge states in
  parity--time-symmetric quantum walks},\ }\href
  {https://doi.org/10.1038/nphys4204} {\bibfield  {journal} {\bibinfo
  {journal} {Nature Physics}\ }\textbf {\bibinfo {volume} {13}},\ \bibinfo
  {pages} {1117} (\bibinfo {year} {2017})}\BibitemShut {NoStop}%
\bibitem [{\citenamefont {Cardano}\ \emph {et~al.}(2017)\citenamefont
  {Cardano}, \citenamefont {D’Errico}, \citenamefont {Dauphin}, \citenamefont
  {Maffei}, \citenamefont {Piccirillo}, \citenamefont {de~Lisio}, \citenamefont
  {De~Filippis}, \citenamefont {Cataudella}, \citenamefont {Santamato},
  \citenamefont {Marrucci} \emph {et~al.}}]{cardano2017detection}%
  \BibitemOpen
  \bibfield  {author} {\bibinfo {author} {\bibfnamefont {F.}~\bibnamefont
  {Cardano}}, \bibinfo {author} {\bibfnamefont {A.}~\bibnamefont {D’Errico}},
  \bibinfo {author} {\bibfnamefont {A.}~\bibnamefont {Dauphin}}, \bibinfo
  {author} {\bibfnamefont {M.}~\bibnamefont {Maffei}}, \bibinfo {author}
  {\bibfnamefont {B.}~\bibnamefont {Piccirillo}}, \bibinfo {author}
  {\bibfnamefont {C.}~\bibnamefont {de~Lisio}}, \bibinfo {author}
  {\bibfnamefont {G.}~\bibnamefont {De~Filippis}}, \bibinfo {author}
  {\bibfnamefont {V.}~\bibnamefont {Cataudella}}, \bibinfo {author}
  {\bibfnamefont {E.}~\bibnamefont {Santamato}}, \bibinfo {author}
  {\bibfnamefont {L.}~\bibnamefont {Marrucci}}, \emph {et~al.},\ }\bibfield
  {title} {\bibinfo {title} {Detection of zak phases and topological invariants
  in a chiral quantum walk of twisted photons},\ }\href
  {https://doi.org/10.1038/ncomms15516} {\bibfield  {journal} {\bibinfo
  {journal} {Nature communications}\ }\textbf {\bibinfo {volume} {8}},\
  \bibinfo {pages} {1} (\bibinfo {year} {2017})}\BibitemShut {NoStop}%
\bibitem [{\citenamefont {Barkhofen}\ \emph {et~al.}(2017)\citenamefont
  {Barkhofen}, \citenamefont {Nitsche}, \citenamefont {Elster}, \citenamefont
  {Lorz}, \citenamefont {G\'abris}, \citenamefont {Jex},\ and\ \citenamefont
  {Silberhorn}}]{barkhofen2017measuring}%
  \BibitemOpen
  \bibfield  {author} {\bibinfo {author} {\bibfnamefont {S.}~\bibnamefont
  {Barkhofen}}, \bibinfo {author} {\bibfnamefont {T.}~\bibnamefont {Nitsche}},
  \bibinfo {author} {\bibfnamefont {F.}~\bibnamefont {Elster}}, \bibinfo
  {author} {\bibfnamefont {L.}~\bibnamefont {Lorz}}, \bibinfo {author}
  {\bibfnamefont {A.}~\bibnamefont {G\'abris}}, \bibinfo {author}
  {\bibfnamefont {I.}~\bibnamefont {Jex}},\ and\ \bibinfo {author}
  {\bibfnamefont {C.}~\bibnamefont {Silberhorn}},\ }\bibfield  {title}
  {\bibinfo {title} {Measuring topological invariants in disordered
  discrete-time quantum walks},\ }\href
  {https://doi.org/10.1103/PhysRevA.96.033846} {\bibfield  {journal} {\bibinfo
  {journal} {Phys. Rev. A}\ }\textbf {\bibinfo {volume} {96}},\ \bibinfo
  {pages} {033846} (\bibinfo {year} {2017})}\BibitemShut {NoStop}%
\bibitem [{\citenamefont {Wang}\ \emph {et~al.}(2018)\citenamefont {Wang},
  \citenamefont {Xiao}, \citenamefont {Qiu}, \citenamefont {Wang},
  \citenamefont {Yi},\ and\ \citenamefont {Xue}}]{wang2018detecting}%
  \BibitemOpen
  \bibfield  {author} {\bibinfo {author} {\bibfnamefont {X.}~\bibnamefont
  {Wang}}, \bibinfo {author} {\bibfnamefont {L.}~\bibnamefont {Xiao}}, \bibinfo
  {author} {\bibfnamefont {X.}~\bibnamefont {Qiu}}, \bibinfo {author}
  {\bibfnamefont {K.}~\bibnamefont {Wang}}, \bibinfo {author} {\bibfnamefont
  {W.}~\bibnamefont {Yi}},\ and\ \bibinfo {author} {\bibfnamefont
  {P.}~\bibnamefont {Xue}},\ }\bibfield  {title} {\bibinfo {title} {Detecting
  topological invariants and revealing topological phase transitions in
  discrete-time photonic quantum walks},\ }\href
  {https://doi.org/10.1103/PhysRevA.98.013835} {\bibfield  {journal} {\bibinfo
  {journal} {Phys. Rev. A}\ }\textbf {\bibinfo {volume} {98}},\ \bibinfo
  {pages} {013835} (\bibinfo {year} {2018})}\BibitemShut {NoStop}%
\bibitem [{\citenamefont {Geraldi}\ \emph {et~al.}(2020)\citenamefont
  {Geraldi}, \citenamefont {De}, \citenamefont {Laneve}, \citenamefont
  {Barkhofen}, \citenamefont {Sperling}, \citenamefont {Mataloni},\ and\
  \citenamefont {Silberhorn}}]{geraldi2020subdiffusion}%
  \BibitemOpen
  \bibfield  {author} {\bibinfo {author} {\bibfnamefont {A.}~\bibnamefont
  {Geraldi}}, \bibinfo {author} {\bibfnamefont {S.}~\bibnamefont {De}},
  \bibinfo {author} {\bibfnamefont {A.}~\bibnamefont {Laneve}}, \bibinfo
  {author} {\bibfnamefont {S.}~\bibnamefont {Barkhofen}}, \bibinfo {author}
  {\bibfnamefont {J.}~\bibnamefont {Sperling}}, \bibinfo {author}
  {\bibfnamefont {P.}~\bibnamefont {Mataloni}},\ and\ \bibinfo {author}
  {\bibfnamefont {C.}~\bibnamefont {Silberhorn}},\ }\href
  {https://arxiv.org/abs/2007.12526} {\bibinfo {title} {Subdiffusion via
  disordered quantum walks}} (\bibinfo {year} {2020}),\ \Eprint
  {https://arxiv.org/abs/2007.12526} {arXiv:2007.12526 [quant-ph]} \BibitemShut
  {NoStop}%
\bibitem [{\citenamefont {Schnyder}\ \emph {et~al.}(2008)\citenamefont
  {Schnyder}, \citenamefont {Ryu}, \citenamefont {Furusaki},\ and\
  \citenamefont {Ludwig}}]{Schnyder:2008}%
  \BibitemOpen
  \bibfield  {author} {\bibinfo {author} {\bibfnamefont {A.~P.}\ \bibnamefont
  {Schnyder}}, \bibinfo {author} {\bibfnamefont {S.}~\bibnamefont {Ryu}},
  \bibinfo {author} {\bibfnamefont {A.}~\bibnamefont {Furusaki}},\ and\
  \bibinfo {author} {\bibfnamefont {A.~W.~W.}\ \bibnamefont {Ludwig}},\
  }\bibfield  {title} {\bibinfo {title} {Classification of topological
  insulators and superconductors in three spatial dimensions},\ }\href
  {https://doi.org/10.1103/PhysRevB.78.195125} {\bibfield  {journal} {\bibinfo
  {journal} {Phys. Rev. B}\ }\textbf {\bibinfo {volume} {78}},\ \bibinfo
  {pages} {195125} (\bibinfo {year} {2008})}\BibitemShut {NoStop}%
\bibitem [{\citenamefont {Tarasinski}\ \emph {et~al.}(2014)\citenamefont
  {Tarasinski}, \citenamefont {Asb\'oth},\ and\ \citenamefont
  {Dahlhaus}}]{Tarasinski:2014}%
  \BibitemOpen
  \bibfield  {author} {\bibinfo {author} {\bibfnamefont {B.}~\bibnamefont
  {Tarasinski}}, \bibinfo {author} {\bibfnamefont {J.~K.}\ \bibnamefont
  {Asb\'oth}},\ and\ \bibinfo {author} {\bibfnamefont {J.~P.}\ \bibnamefont
  {Dahlhaus}},\ }\bibfield  {title} {\bibinfo {title} {Scattering theory of
  topological phases in discrete-time quantum walks},\ }\href
  {https://doi.org/10.1103/PhysRevA.89.042327} {\bibfield  {journal} {\bibinfo
  {journal} {Phys. Rev. A}\ }\textbf {\bibinfo {volume} {89}},\ \bibinfo
  {pages} {042327} (\bibinfo {year} {2014})}\BibitemShut {NoStop}%
\bibitem [{\citenamefont {Cedzich}\ \emph {et~al.}(2018)\citenamefont
  {Cedzich}, \citenamefont {Geib}, \citenamefont {Gr\"unbaum}, \citenamefont
  {Stahl}, \citenamefont {Vel\'azquez}, \citenamefont {Werner},\ and\
  \citenamefont {Werner}}]{Cedzich:2018}%
  \BibitemOpen
  \bibfield  {author} {\bibinfo {author} {\bibfnamefont {C.}~\bibnamefont
  {Cedzich}}, \bibinfo {author} {\bibfnamefont {T.}~\bibnamefont {Geib}},
  \bibinfo {author} {\bibfnamefont {F.~A.}\ \bibnamefont {Gr\"unbaum}},
  \bibinfo {author} {\bibfnamefont {C.}~\bibnamefont {Stahl}}, \bibinfo
  {author} {\bibfnamefont {L.}~\bibnamefont {Vel\'azquez}}, \bibinfo {author}
  {\bibfnamefont {A.~H.}\ \bibnamefont {Werner}},\ and\ \bibinfo {author}
  {\bibfnamefont {R.~F.}\ \bibnamefont {Werner}},\ }\bibfield  {title}
  {\bibinfo {title} {The topological classification of one-dimensional
  symmetric quantum walks},\ }\href {https://doi.org/10.1007/s00023-017-0630-x}
  {\bibfield  {journal} {\bibinfo  {journal} {Ann. Henri Poincar\'e}\ }\textbf
  {\bibinfo {volume} {19}},\ \bibinfo {pages} {325} (\bibinfo {year}
  {2018})}\BibitemShut {NoStop}%
\bibitem [{\citenamefont {Asb\'oth}(2012)}]{Asboth:2012}%
  \BibitemOpen
  \bibfield  {author} {\bibinfo {author} {\bibfnamefont {J.~K.}\ \bibnamefont
  {Asb\'oth}},\ }\bibfield  {title} {\bibinfo {title} {Symmetries, topological
  phases, and bound states in the one-dimensional quantum walk},\ }\href
  {https://doi.org/10.1103/PhysRevB.86.195414} {\bibfield  {journal} {\bibinfo
  {journal} {Phys. Rev. B}\ }\textbf {\bibinfo {volume} {86}},\ \bibinfo
  {pages} {195414} (\bibinfo {year} {2012})}\BibitemShut {NoStop}%
\bibitem [{\citenamefont {Anderson}(1958)}]{Anderson:1958}%
  \BibitemOpen
  \bibfield  {author} {\bibinfo {author} {\bibfnamefont {P.~W.}\ \bibnamefont
  {Anderson}},\ }\bibfield  {title} {\bibinfo {title} {Absence of diffusion in
  certain random lattices},\ }\href {https://doi.org/10.1103/PhysRev.109.1492}
  {\bibfield  {journal} {\bibinfo  {journal} {Phys. Rev.}\ }\textbf {\bibinfo
  {volume} {109}},\ \bibinfo {pages} {1492} (\bibinfo {year}
  {1958})}\BibitemShut {NoStop}%
\bibitem [{\citenamefont {Wegner}(1979)}]{wegner1979}%
  \BibitemOpen
  \bibfield  {author} {\bibinfo {author} {\bibfnamefont {F.}~\bibnamefont
  {Wegner}},\ }\bibfield  {title} {\bibinfo {title} {The mobility edge problem:
  Continuous symmetry and a conjecture},\ }\href
  {https://doi.org/10.1007/BF01319839} {\bibfield  {journal} {\bibinfo
  {journal} {Zeitschrift f{\"u}r Physik B: Condensed Matter}\ }\textbf
  {\bibinfo {volume} {35}},\ \bibinfo {pages} {207} (\bibinfo {year}
  {1979})}\BibitemShut {NoStop}%
\bibitem [{\citenamefont {Efetov}\ \emph {et~al.}(1980)\citenamefont {Efetov},
  \citenamefont {Larkin},\ and\ \citenamefont {Khmel'nitskii}}]{efetov1980zh}%
  \BibitemOpen
  \bibfield  {author} {\bibinfo {author} {\bibfnamefont {K.~B.}\ \bibnamefont
  {Efetov}}, \bibinfo {author} {\bibfnamefont {A.~I.}\ \bibnamefont {Larkin}},\
  and\ \bibinfo {author} {\bibfnamefont {D.~E.}\ \bibnamefont
  {Khmel'nitskii}},\ }\bibfield  {title} {\bibinfo {title} {Interaction of
  diffusion modes in the theory of localization},\ }\href
  {http://www.jetp.ac.ru/cgi-bin/e/index/e/52/3/p568?a=list} {\bibfield
  {journal} {\bibinfo  {journal} {JETP}\ }\textbf {\bibinfo {volume} {52}},\
  \bibinfo {pages} {568} (\bibinfo {year} {1980})}\BibitemShut {NoStop}%
\bibitem [{\citenamefont {Pruisken}\ and\ \citenamefont
  {Sch{\"a}fer}(1982)}]{pruisken1982anderson}%
  \BibitemOpen
  \bibfield  {author} {\bibinfo {author} {\bibfnamefont {A.~M.}\ \bibnamefont
  {Pruisken}}\ and\ \bibinfo {author} {\bibfnamefont {L.}~\bibnamefont
  {Sch{\"a}fer}},\ }\bibfield  {title} {\bibinfo {title} {{The Anderson model
  for electron localisation non-linear $\sigma$ model, asymptotic gauge
  invariance}},\ }\href
  {https://doi.org/https://doi.org/10.1016/0550-3213(82)90056-6} {\bibfield
  {journal} {\bibinfo  {journal} {Nuclear Physics B}\ }\textbf {\bibinfo
  {volume} {200}},\ \bibinfo {pages} {20 } (\bibinfo {year}
  {1982})}\BibitemShut {NoStop}%
\bibitem [{\citenamefont {Efetov}\ and\ \citenamefont
  {Larkin}(1983)}]{efetov1983kinetics}%
  \BibitemOpen
  \bibfield  {author} {\bibinfo {author} {\bibfnamefont {K.~B.}\ \bibnamefont
  {Efetov}}\ and\ \bibinfo {author} {\bibfnamefont {A.~I.}\ \bibnamefont
  {Larkin}},\ }\bibfield  {title} {\bibinfo {title} {Kinetics of a quantum
  particle in long metallic wires},\ }\href
  {http://www.jetp.ac.ru/cgi-bin/index/e/58/2/p444?a=list} {\bibfield
  {journal} {\bibinfo  {journal} {Sov. Phys. JETP}\ }\textbf {\bibinfo {volume}
  {58}},\ \bibinfo {pages} {444} (\bibinfo {year} {1983})}\BibitemShut
  {NoStop}%
\bibitem [{\citenamefont {Efetov}(1997)}]{Efetov-book}%
  \BibitemOpen
  \bibfield  {author} {\bibinfo {author} {\bibfnamefont {K.~B.}\ \bibnamefont
  {Efetov}},\ }\href@noop {} {\emph {\bibinfo {title} {{Sypersymmetry in
  Disorder and Chaos}}}}\ (\bibinfo  {publisher} {Cambridge University Press},\
  \bibinfo {address} {Cambridge},\ \bibinfo {year} {1997})\BibitemShut
  {NoStop}%
\bibitem [{\citenamefont {Abrahams}\ \emph {et~al.}(1979)\citenamefont
  {Abrahams}, \citenamefont {Anderson}, \citenamefont {Licciardello},\ and\
  \citenamefont {Ramakrishnan}}]{Abrahams:1979}%
  \BibitemOpen
  \bibfield  {author} {\bibinfo {author} {\bibfnamefont {E.}~\bibnamefont
  {Abrahams}}, \bibinfo {author} {\bibfnamefont {P.~W.}\ \bibnamefont
  {Anderson}}, \bibinfo {author} {\bibfnamefont {D.~C.}\ \bibnamefont
  {Licciardello}},\ and\ \bibinfo {author} {\bibfnamefont {T.~V.}\ \bibnamefont
  {Ramakrishnan}},\ }\bibfield  {title} {\bibinfo {title} {Scaling theory of
  localization: Absence of quantum diffusion in two dimensions},\ }\href
  {https://doi.org/10.1103/PhysRevLett.42.673} {\bibfield  {journal} {\bibinfo
  {journal} {Phys. Rev. Lett.}\ }\textbf {\bibinfo {volume} {42}},\ \bibinfo
  {pages} {673} (\bibinfo {year} {1979})}\BibitemShut {NoStop}%
\bibitem [{\citenamefont {Fulga}\ \emph {et~al.}(2011)\citenamefont {Fulga},
  \citenamefont {Hassler}, \citenamefont {Akhmerov},\ and\ \citenamefont
  {Beenakker}}]{Fulga:2011}%
  \BibitemOpen
  \bibfield  {author} {\bibinfo {author} {\bibfnamefont {I.~C.}\ \bibnamefont
  {Fulga}}, \bibinfo {author} {\bibfnamefont {F.}~\bibnamefont {Hassler}},
  \bibinfo {author} {\bibfnamefont {A.~R.}\ \bibnamefont {Akhmerov}},\ and\
  \bibinfo {author} {\bibfnamefont {C.~W.~J.}\ \bibnamefont {Beenakker}},\
  }\bibfield  {title} {\bibinfo {title} {Scattering formula for the topological
  quantum number of a disordered multimode wire},\ }\href
  {https://doi.org/10.1103/PhysRevB.83.155429} {\bibfield  {journal} {\bibinfo
  {journal} {Phys. Rev. B}\ }\textbf {\bibinfo {volume} {83}},\ \bibinfo
  {pages} {155429} (\bibinfo {year} {2011})}\BibitemShut {NoStop}%
\bibitem [{\citenamefont {Altland}\ \emph {et~al.}(2014)\citenamefont
  {Altland}, \citenamefont {Bagrets}, \citenamefont {Fritz}, \citenamefont
  {Kamenev},\ and\ \citenamefont {Schmiedt}}]{Altland:2014}%
  \BibitemOpen
  \bibfield  {author} {\bibinfo {author} {\bibfnamefont {A.}~\bibnamefont
  {Altland}}, \bibinfo {author} {\bibfnamefont {D.}~\bibnamefont {Bagrets}},
  \bibinfo {author} {\bibfnamefont {L.}~\bibnamefont {Fritz}}, \bibinfo
  {author} {\bibfnamefont {A.}~\bibnamefont {Kamenev}},\ and\ \bibinfo {author}
  {\bibfnamefont {H.}~\bibnamefont {Schmiedt}},\ }\bibfield  {title} {\bibinfo
  {title} {{Quantum criticality of quasi-one-dimensional topological Anderson
  insulators}},\ }\href {https://doi.org/10.1103/PhysRevLett.112.206602}
  {\bibfield  {journal} {\bibinfo  {journal} {Phys. Rev. Lett.}\ }\textbf
  {\bibinfo {volume} {112}},\ \bibinfo {pages} {206602} (\bibinfo {year}
  {2014})}\BibitemShut {NoStop}%
\bibitem [{Note2()}]{Note2}%
  \BibitemOpen
  \bibinfo {note} {Notice that this implies that $\DOTSB \sum@ \slimits@ _{q,
  \sigma '\sigma } P^{\protect \rm chiral}_{\sigma '\sigma }(t,q) =
  1/(4\protect \qopname \relax o{ln}^2 t)$ as shown in Appendix~\ref
  {TransferMatrix}.}\BibitemShut {Stop}%
\bibitem [{\citenamefont {Sinai}(1982)}]{Sinai:1982}%
  \BibitemOpen
  \bibfield  {author} {\bibinfo {author} {\bibfnamefont {Y.~G.}\ \bibnamefont
  {Sinai}},\ }\bibfield  {title} {\bibinfo {title} {The limiting behavior of a
  one-dimensional random walk in a random medium},\ }\href
  {http://epubs.siam.org/doi/abs/10.1137/1127028} {\bibfield  {journal}
  {\bibinfo  {journal} {Theory Probab. Appl.}\ }\textbf {\bibinfo {volume}
  {27}},\ \bibinfo {pages} {256} (\bibinfo {year} {1982})}\BibitemShut
  {NoStop}%
\bibitem [{\citenamefont {Bouchaud}\ \emph {et~al.}(1990)\citenamefont
  {Bouchaud}, \citenamefont {Comtet}, \citenamefont {Georges},\ and\
  \citenamefont {Doussal}}]{Bouchaud:1990}%
  \BibitemOpen
  \bibfield  {author} {\bibinfo {author} {\bibfnamefont {J.}~\bibnamefont
  {Bouchaud}}, \bibinfo {author} {\bibfnamefont {A.}~\bibnamefont {Comtet}},
  \bibinfo {author} {\bibfnamefont {A.}~\bibnamefont {Georges}},\ and\ \bibinfo
  {author} {\bibfnamefont {P.~L.}\ \bibnamefont {Doussal}},\ }\bibfield
  {title} {\bibinfo {title} {Classical diffusion of a particle in a
  one-dimensional random force field},\ }\href
  {https://doi.org/http://dx.doi.org/10.1016/0003-4916(90)90043-N} {\bibfield
  {journal} {\bibinfo  {journal} {Annals of Physics}\ }\textbf {\bibinfo
  {volume} {201}},\ \bibinfo {pages} {285 } (\bibinfo {year}
  {1990})}\BibitemShut {NoStop}%
\bibitem [{\citenamefont {Comtet}\ and\ \citenamefont
  {Dean}(1998)}]{Comtet:1998}%
  \BibitemOpen
  \bibfield  {author} {\bibinfo {author} {\bibfnamefont {A.}~\bibnamefont
  {Comtet}}\ and\ \bibinfo {author} {\bibfnamefont {D.~S.}\ \bibnamefont
  {Dean}},\ }\bibfield  {title} {\bibinfo {title} {{Exact results on Sinai's
  diffusion}},\ }\href {http://stacks.iop.org/0305-4470/31/i=43/a=004}
  {\bibfield  {journal} {\bibinfo  {journal} {Journal of Physics A:
  Mathematical and General}\ }\textbf {\bibinfo {volume} {31}},\ \bibinfo
  {pages} {8595} (\bibinfo {year} {1998})}\BibitemShut {NoStop}%
\bibitem [{Note3()}]{Note3}%
  \BibitemOpen
  \bibinfo {note} {Sinai diffusion has recently been suggested to leave traces
  of unconventional heat propagation in the form of non-monotonically
  propagating thermal current pulses in quasi-one-dimensional topological
  superconducting wires near criticality~\cite {Bagrets:2016}.}\BibitemShut
  {Stop}%
\bibitem [{\citenamefont {Asb\'oth}\ and\ \citenamefont
  {Obuse}(2013)}]{Asboth:2013}%
  \BibitemOpen
  \bibfield  {author} {\bibinfo {author} {\bibfnamefont {J.~K.}\ \bibnamefont
  {Asb\'oth}}\ and\ \bibinfo {author} {\bibfnamefont {H.}~\bibnamefont
  {Obuse}},\ }\bibfield  {title} {\bibinfo {title} {Bulk-boundary
  correspondence for chiral symmetric quantum walks},\ }\href
  {https://doi.org/10.1103/PhysRevB.88.121406} {\bibfield  {journal} {\bibinfo
  {journal} {Phys. Rev. B}\ }\textbf {\bibinfo {volume} {88}},\ \bibinfo
  {pages} {121406(R)} (\bibinfo {year} {2013})}\BibitemShut {NoStop}%
\bibitem [{\citenamefont {Mondragon-Shem}\ \emph {et~al.}(2014)\citenamefont
  {Mondragon-Shem}, \citenamefont {Hughes}, \citenamefont {Song},\ and\
  \citenamefont {Prodan}}]{Mondragon-Shem:2014}%
  \BibitemOpen
  \bibfield  {author} {\bibinfo {author} {\bibfnamefont {I.}~\bibnamefont
  {Mondragon-Shem}}, \bibinfo {author} {\bibfnamefont {T.~L.}\ \bibnamefont
  {Hughes}}, \bibinfo {author} {\bibfnamefont {J.}~\bibnamefont {Song}},\ and\
  \bibinfo {author} {\bibfnamefont {E.}~\bibnamefont {Prodan}},\ }\bibfield
  {title} {\bibinfo {title} {Topological criticality in the chiral-symmetric
  aiii class at strong disorder},\ }\href
  {https://doi.org/10.1103/PhysRevLett.113.046802} {\bibfield  {journal}
  {\bibinfo  {journal} {Phys. Rev. Lett.}\ }\textbf {\bibinfo {volume} {113}},\
  \bibinfo {pages} {046802} (\bibinfo {year} {2014})}\BibitemShut {NoStop}%
\bibitem [{Note4()}]{Note4}%
  \BibitemOpen
  \bibinfo {note} {Topological boundary modes can be introduced, for example,
  by connecting two 1-dimensional quantum walk systems characterized by
  $\protect \hat U(\varphi =0,\theta )$ and $\protect \hat U(\varphi =0,-\theta
  )$, respectively. The spin expectation of boundary modes are eigen values of
  chiral operator $\sigma _2$.}\BibitemShut {Stop}%
\bibitem [{\citenamefont {Zhao}\ and\ \citenamefont
  {Gong}(2015)}]{zhao2015disordered}%
  \BibitemOpen
  \bibfield  {author} {\bibinfo {author} {\bibfnamefont {Q.}~\bibnamefont
  {Zhao}}\ and\ \bibinfo {author} {\bibfnamefont {J.}~\bibnamefont {Gong}},\
  }\bibfield  {title} {\bibinfo {title} {From disordered quantum walk to
  physics of off-diagonal disorder},\ }\href
  {https://doi.org/10.1103/PhysRevB.92.214205} {\bibfield  {journal} {\bibinfo
  {journal} {Phys. Rev. B}\ }\textbf {\bibinfo {volume} {92}},\ \bibinfo
  {pages} {214205} (\bibinfo {year} {2015})}\BibitemShut {NoStop}%
\bibitem [{\citenamefont {Boutari}\ \emph {et~al.}(2016)\citenamefont
  {Boutari}, \citenamefont {Feizpour}, \citenamefont {Barz}, \citenamefont
  {Franco}, \citenamefont {Kim}, \citenamefont {Kolthammer},\ and\
  \citenamefont {Walmsley}}]{boutari2016large}%
  \BibitemOpen
  \bibfield  {author} {\bibinfo {author} {\bibfnamefont {J.}~\bibnamefont
  {Boutari}}, \bibinfo {author} {\bibfnamefont {A.}~\bibnamefont {Feizpour}},
  \bibinfo {author} {\bibfnamefont {S.}~\bibnamefont {Barz}}, \bibinfo {author}
  {\bibfnamefont {C.~D.}\ \bibnamefont {Franco}}, \bibinfo {author}
  {\bibfnamefont {M.~S.}\ \bibnamefont {Kim}}, \bibinfo {author} {\bibfnamefont
  {W.~S.}\ \bibnamefont {Kolthammer}},\ and\ \bibinfo {author} {\bibfnamefont
  {I.~A.}\ \bibnamefont {Walmsley}},\ }\bibfield  {title} {\bibinfo {title}
  {Large scale quantum walks by means of optical fiber cavities},\ }\href
  {https://doi.org/10.1088/2040-8978/18/9/094007} {\bibfield  {journal}
  {\bibinfo  {journal} {Journal of Optics}\ }\textbf {\bibinfo {volume} {18}},\
  \bibinfo {pages} {094007} (\bibinfo {year} {2016})}\BibitemShut {NoStop}%
\bibitem [{\citenamefont {Xu}\ \emph {et~al.}(2018)\citenamefont {Xu},
  \citenamefont {Wang}, \citenamefont {Pan}, \citenamefont {Sun}, \citenamefont
  {Xu}, \citenamefont {Chen}, \citenamefont {Tang}, \citenamefont {Gong},
  \citenamefont {Han}, \citenamefont {Li},\ and\ \citenamefont
  {Guo}}]{xu2018measuring}%
  \BibitemOpen
  \bibfield  {author} {\bibinfo {author} {\bibfnamefont {X.-Y.}\ \bibnamefont
  {Xu}}, \bibinfo {author} {\bibfnamefont {Q.-Q.}\ \bibnamefont {Wang}},
  \bibinfo {author} {\bibfnamefont {W.-W.}\ \bibnamefont {Pan}}, \bibinfo
  {author} {\bibfnamefont {K.}~\bibnamefont {Sun}}, \bibinfo {author}
  {\bibfnamefont {J.-S.}\ \bibnamefont {Xu}}, \bibinfo {author} {\bibfnamefont
  {G.}~\bibnamefont {Chen}}, \bibinfo {author} {\bibfnamefont {J.-S.}\
  \bibnamefont {Tang}}, \bibinfo {author} {\bibfnamefont {M.}~\bibnamefont
  {Gong}}, \bibinfo {author} {\bibfnamefont {Y.-J.}\ \bibnamefont {Han}},
  \bibinfo {author} {\bibfnamefont {C.-F.}\ \bibnamefont {Li}},\ and\ \bibinfo
  {author} {\bibfnamefont {G.-C.}\ \bibnamefont {Guo}},\ }\bibfield  {title}
  {\bibinfo {title} {Measuring the winding number in a large-scale chiral
  quantum walk},\ }\href {https://doi.org/10.1103/PhysRevLett.120.260501}
  {\bibfield  {journal} {\bibinfo  {journal} {Phys. Rev. Lett.}\ }\textbf
  {\bibinfo {volume} {120}},\ \bibinfo {pages} {260501} (\bibinfo {year}
  {2018})}\BibitemShut {NoStop}%
\bibitem [{\citenamefont {Nitsche}\ \emph {et~al.}(2018)\citenamefont
  {Nitsche}, \citenamefont {Barkhofen}, \citenamefont {Kruse}, \citenamefont
  {Sansoni}, \citenamefont {{\v{S}}tefa{\v{n}}{\'a}k}, \citenamefont
  {G{\'a}bris}, \citenamefont {Poto{\v{c}}ek}, \citenamefont {Kiss},
  \citenamefont {Jex},\ and\ \citenamefont {Silberhorn}}]{nitsche2018probing}%
  \BibitemOpen
  \bibfield  {author} {\bibinfo {author} {\bibfnamefont {T.}~\bibnamefont
  {Nitsche}}, \bibinfo {author} {\bibfnamefont {S.}~\bibnamefont {Barkhofen}},
  \bibinfo {author} {\bibfnamefont {R.}~\bibnamefont {Kruse}}, \bibinfo
  {author} {\bibfnamefont {L.}~\bibnamefont {Sansoni}}, \bibinfo {author}
  {\bibfnamefont {M.}~\bibnamefont {{\v{S}}tefa{\v{n}}{\'a}k}}, \bibinfo
  {author} {\bibfnamefont {A.}~\bibnamefont {G{\'a}bris}}, \bibinfo {author}
  {\bibfnamefont {V.}~\bibnamefont {Poto{\v{c}}ek}}, \bibinfo {author}
  {\bibfnamefont {T.}~\bibnamefont {Kiss}}, \bibinfo {author} {\bibfnamefont
  {I.}~\bibnamefont {Jex}},\ and\ \bibinfo {author} {\bibfnamefont
  {C.}~\bibnamefont {Silberhorn}},\ }\bibfield  {title} {\bibinfo {title}
  {Probing measurement-induced effects in quantum walks via recurrence},\
  }\href {https://advances.sciencemag.org/content/4/6/eaar6444} {\bibfield
  {journal} {\bibinfo  {journal} {Science advances}\ }\textbf {\bibinfo
  {volume} {4}},\ \bibinfo {pages} {eaar6444} (\bibinfo {year}
  {2018})}\BibitemShut {NoStop}%
\bibitem [{\citenamefont {Do}\ \emph {et~al.}(2005)\citenamefont {Do},
  \citenamefont {Stohler}, \citenamefont {Balasubramanian}, \citenamefont
  {Elliott}, \citenamefont {Eash}, \citenamefont {Fischbach}, \citenamefont
  {Fischbach}, \citenamefont {Mills},\ and\ \citenamefont
  {Zwickl}}]{do2005experimental}%
  \BibitemOpen
  \bibfield  {author} {\bibinfo {author} {\bibfnamefont {B.}~\bibnamefont
  {Do}}, \bibinfo {author} {\bibfnamefont {M.~L.}\ \bibnamefont {Stohler}},
  \bibinfo {author} {\bibfnamefont {S.}~\bibnamefont {Balasubramanian}},
  \bibinfo {author} {\bibfnamefont {D.~S.}\ \bibnamefont {Elliott}}, \bibinfo
  {author} {\bibfnamefont {C.}~\bibnamefont {Eash}}, \bibinfo {author}
  {\bibfnamefont {E.}~\bibnamefont {Fischbach}}, \bibinfo {author}
  {\bibfnamefont {M.~A.}\ \bibnamefont {Fischbach}}, \bibinfo {author}
  {\bibfnamefont {A.}~\bibnamefont {Mills}},\ and\ \bibinfo {author}
  {\bibfnamefont {B.}~\bibnamefont {Zwickl}},\ }\bibfield  {title} {\bibinfo
  {title} {Experimental realization of a quantum quincunx by use of linear
  optical elements},\ }\href
  {http://josab.osa.org/abstract.cfm?URI=josab-22-2-499} {\bibfield  {journal}
  {\bibinfo  {journal} {J. Opt. Soc. Am. B}\ }\textbf {\bibinfo {volume}
  {22}},\ \bibinfo {pages} {499} (\bibinfo {year} {2005})}\BibitemShut
  {NoStop}%
\bibitem [{\citenamefont {Xue}\ \emph {et~al.}(2015{\natexlab{b}})\citenamefont
  {Xue}, \citenamefont {Zhang}, \citenamefont {Qin}, \citenamefont {Zhan},
  \citenamefont {Bian}, \citenamefont {Li},\ and\ \citenamefont
  {Sanders}}]{xue_experimental_2015}%
  \BibitemOpen
  \bibfield  {author} {\bibinfo {author} {\bibfnamefont {P.}~\bibnamefont
  {Xue}}, \bibinfo {author} {\bibfnamefont {R.}~\bibnamefont {Zhang}}, \bibinfo
  {author} {\bibfnamefont {H.}~\bibnamefont {Qin}}, \bibinfo {author}
  {\bibfnamefont {X.}~\bibnamefont {Zhan}}, \bibinfo {author} {\bibfnamefont
  {Z.~H.}\ \bibnamefont {Bian}}, \bibinfo {author} {\bibfnamefont
  {J.}~\bibnamefont {Li}},\ and\ \bibinfo {author} {\bibfnamefont {B.~C.}\
  \bibnamefont {Sanders}},\ }\bibfield  {title} {\bibinfo {title} {Experimental
  {{Quantum}}-{{Walk Revival}} with a {{Time}}-{{Dependent Coin}}},\ }\href
  {https://doi.org/10.1103/PhysRevLett.114.140502} {\bibfield  {journal}
  {\bibinfo  {journal} {Phys. Rev. Lett.}\ }\textbf {\bibinfo {volume} {114}},\
  \bibinfo {pages} {140502} (\bibinfo {year} {2015}{\natexlab{b}})}\BibitemShut
  {NoStop}%
\bibitem [{\citenamefont {Wang}\ \emph {et~al.}(2019)\citenamefont {Wang},
  \citenamefont {Qin}, \citenamefont {Ding}, \citenamefont {Chen},
  \citenamefont {Chen}, \citenamefont {You}, \citenamefont {He}, \citenamefont
  {Jiang}, \citenamefont {You}, \citenamefont {Wang}, \citenamefont
  {Schneider}, \citenamefont {Renema}, \citenamefont {H\"ofling}, \citenamefont
  {Lu},\ and\ \citenamefont {Pan}}]{wang2019boson}%
  \BibitemOpen
  \bibfield  {author} {\bibinfo {author} {\bibfnamefont {H.}~\bibnamefont
  {Wang}}, \bibinfo {author} {\bibfnamefont {J.}~\bibnamefont {Qin}}, \bibinfo
  {author} {\bibfnamefont {X.}~\bibnamefont {Ding}}, \bibinfo {author}
  {\bibfnamefont {M.-C.}\ \bibnamefont {Chen}}, \bibinfo {author}
  {\bibfnamefont {S.}~\bibnamefont {Chen}}, \bibinfo {author} {\bibfnamefont
  {X.}~\bibnamefont {You}}, \bibinfo {author} {\bibfnamefont {Y.-M.}\
  \bibnamefont {He}}, \bibinfo {author} {\bibfnamefont {X.}~\bibnamefont
  {Jiang}}, \bibinfo {author} {\bibfnamefont {L.}~\bibnamefont {You}}, \bibinfo
  {author} {\bibfnamefont {Z.}~\bibnamefont {Wang}}, \bibinfo {author}
  {\bibfnamefont {C.}~\bibnamefont {Schneider}}, \bibinfo {author}
  {\bibfnamefont {J.~J.}\ \bibnamefont {Renema}}, \bibinfo {author}
  {\bibfnamefont {S.}~\bibnamefont {H\"ofling}}, \bibinfo {author}
  {\bibfnamefont {C.-Y.}\ \bibnamefont {Lu}},\ and\ \bibinfo {author}
  {\bibfnamefont {J.-W.}\ \bibnamefont {Pan}},\ }\bibfield  {title} {\bibinfo
  {title} {{Boson Sampling with 20 Input Photons and a 60-Mode Interferometer
  in a $1{0}^{14}$-Dimensional Hilbert Space}},\ }\href
  {https://doi.org/10.1103/PhysRevLett.123.250503} {\bibfield  {journal}
  {\bibinfo  {journal} {Phys. Rev. Lett.}\ }\textbf {\bibinfo {volume} {123}},\
  \bibinfo {pages} {250503} (\bibinfo {year} {2019})}\BibitemShut {NoStop}%
\bibitem [{\citenamefont {He}\ \emph {et~al.}(2017)\citenamefont {He},
  \citenamefont {Ding}, \citenamefont {Su}, \citenamefont {Huang},
  \citenamefont {Qin}, \citenamefont {Wang}, \citenamefont {Unsleber},
  \citenamefont {Chen}, \citenamefont {Wang}, \citenamefont {He}, \citenamefont
  {Wang}, \citenamefont {Zhang}, \citenamefont {Chen}, \citenamefont
  {Schneider}, \citenamefont {Kamp}, \citenamefont {You}, \citenamefont {Wang},
  \citenamefont {H\"ofling}, \citenamefont {Lu},\ and\ \citenamefont
  {Pan}}]{he2017time}%
  \BibitemOpen
  \bibfield  {author} {\bibinfo {author} {\bibfnamefont {Y.}~\bibnamefont
  {He}}, \bibinfo {author} {\bibfnamefont {X.}~\bibnamefont {Ding}}, \bibinfo
  {author} {\bibfnamefont {Z.-E.}\ \bibnamefont {Su}}, \bibinfo {author}
  {\bibfnamefont {H.-L.}\ \bibnamefont {Huang}}, \bibinfo {author}
  {\bibfnamefont {J.}~\bibnamefont {Qin}}, \bibinfo {author} {\bibfnamefont
  {C.}~\bibnamefont {Wang}}, \bibinfo {author} {\bibfnamefont {S.}~\bibnamefont
  {Unsleber}}, \bibinfo {author} {\bibfnamefont {C.}~\bibnamefont {Chen}},
  \bibinfo {author} {\bibfnamefont {H.}~\bibnamefont {Wang}}, \bibinfo {author}
  {\bibfnamefont {Y.-M.}\ \bibnamefont {He}}, \bibinfo {author} {\bibfnamefont
  {X.-L.}\ \bibnamefont {Wang}}, \bibinfo {author} {\bibfnamefont {W.-J.}\
  \bibnamefont {Zhang}}, \bibinfo {author} {\bibfnamefont {S.-J.}\ \bibnamefont
  {Chen}}, \bibinfo {author} {\bibfnamefont {C.}~\bibnamefont {Schneider}},
  \bibinfo {author} {\bibfnamefont {M.}~\bibnamefont {Kamp}}, \bibinfo {author}
  {\bibfnamefont {L.-X.}\ \bibnamefont {You}}, \bibinfo {author} {\bibfnamefont
  {Z.}~\bibnamefont {Wang}}, \bibinfo {author} {\bibfnamefont {S.}~\bibnamefont
  {H\"ofling}}, \bibinfo {author} {\bibfnamefont {C.-Y.}\ \bibnamefont {Lu}},\
  and\ \bibinfo {author} {\bibfnamefont {J.-W.}\ \bibnamefont {Pan}},\
  }\bibfield  {title} {\bibinfo {title} {Time-bin-encoded boson sampling with a
  single-photon device},\ }\href
  {https://doi.org/10.1103/PhysRevLett.118.190501} {\bibfield  {journal}
  {\bibinfo  {journal} {Phys. Rev. Lett.}\ }\textbf {\bibinfo {volume} {118}},\
  \bibinfo {pages} {190501} (\bibinfo {year} {2017})}\BibitemShut {NoStop}%
\bibitem [{\citenamefont {Yuan}\ \emph {et~al.}(2016)\citenamefont {Yuan},
  \citenamefont {Fr\"ohlich}, \citenamefont {Lucamarini}, \citenamefont
  {Roberts}, \citenamefont {Dynes},\ and\ \citenamefont {Shields}}]{Yuan:2016}%
  \BibitemOpen
  \bibfield  {author} {\bibinfo {author} {\bibfnamefont {Z.~L.}\ \bibnamefont
  {Yuan}}, \bibinfo {author} {\bibfnamefont {B.}~\bibnamefont {Fr\"ohlich}},
  \bibinfo {author} {\bibfnamefont {M.}~\bibnamefont {Lucamarini}}, \bibinfo
  {author} {\bibfnamefont {G.~L.}\ \bibnamefont {Roberts}}, \bibinfo {author}
  {\bibfnamefont {J.~F.}\ \bibnamefont {Dynes}},\ and\ \bibinfo {author}
  {\bibfnamefont {A.~J.}\ \bibnamefont {Shields}},\ }\bibfield  {title}
  {\bibinfo {title} {Directly phase-modulated light source},\ }\href
  {https://doi.org/10.1103/PhysRevX.6.031044} {\bibfield  {journal} {\bibinfo
  {journal} {Phys. Rev. X}\ }\textbf {\bibinfo {volume} {6}},\ \bibinfo {pages}
  {031044} (\bibinfo {year} {2016})}\BibitemShut {NoStop}%
\bibitem [{\citenamefont {Hamilton}\ \emph {et~al.}(2016)\citenamefont
  {Hamilton}, \citenamefont {Barkhofen}, \citenamefont {Sansoni}, \citenamefont
  {Jex},\ and\ \citenamefont {Silberhorn}}]{hamilton2016driven}%
  \BibitemOpen
  \bibfield  {author} {\bibinfo {author} {\bibfnamefont {C.~S.}\ \bibnamefont
  {Hamilton}}, \bibinfo {author} {\bibfnamefont {S.}~\bibnamefont {Barkhofen}},
  \bibinfo {author} {\bibfnamefont {L.}~\bibnamefont {Sansoni}}, \bibinfo
  {author} {\bibfnamefont {I.}~\bibnamefont {Jex}},\ and\ \bibinfo {author}
  {\bibfnamefont {C.}~\bibnamefont {Silberhorn}},\ }\bibfield  {title}
  {\bibinfo {title} {Driven discrete time quantum walks},\ }\href
  {https://doi.org/10.1088/1367-2630/18/7/073008} {\bibfield  {journal}
  {\bibinfo  {journal} {New Journal of Physics}\ }\textbf {\bibinfo {volume}
  {18}},\ \bibinfo {pages} {073008} (\bibinfo {year} {2016})}\BibitemShut
  {NoStop}%
\bibitem [{\citenamefont {Zirnbauer}(1996)}]{Zirnbauer:1996}%
  \BibitemOpen
  \bibfield  {author} {\bibinfo {author} {\bibfnamefont {M.~R.}\ \bibnamefont
  {Zirnbauer}},\ }\bibfield  {title} {\bibinfo {title} {Supersymmetry for
  systems with unitary disorder: circular ensembles},\ }\href
  {http://stacks.iop.org/0305-4470/29/i=22/a=013} {\bibfield  {journal}
  {\bibinfo  {journal} {Journal of Physics A: Mathematical and General}\
  }\textbf {\bibinfo {volume} {29}},\ \bibinfo {pages} {7113} (\bibinfo {year}
  {1996})}\BibitemShut {NoStop}%
\bibitem [{\citenamefont {Altland}\ \emph
  {et~al.}(2015{\natexlab{b}})\citenamefont {Altland}, \citenamefont
  {Gnutzmann}, \citenamefont {Haake},\ and\ \citenamefont
  {Micklitz}}]{Altland:2015a}%
  \BibitemOpen
  \bibfield  {author} {\bibinfo {author} {\bibfnamefont {A.}~\bibnamefont
  {Altland}}, \bibinfo {author} {\bibfnamefont {S.}~\bibnamefont {Gnutzmann}},
  \bibinfo {author} {\bibfnamefont {F.}~\bibnamefont {Haake}},\ and\ \bibinfo
  {author} {\bibfnamefont {T.}~\bibnamefont {Micklitz}},\ }\bibfield  {title}
  {\bibinfo {title} {A review of sigma models for quantum chaotic dynamics},\
  }\href {http://stacks.iop.org/0034-4885/78/i=8/a=086001} {\bibfield
  {journal} {\bibinfo  {journal} {Reports on Progress in Physics}\ }\textbf
  {\bibinfo {volume} {78}},\ \bibinfo {pages} {086001} (\bibinfo {year}
  {2015}{\natexlab{b}})}\BibitemShut {NoStop}%
\bibitem [{\citenamefont {Lamacraft}\ \emph {et~al.}(2004)\citenamefont
  {Lamacraft}, \citenamefont {Simons},\ and\ \citenamefont
  {Zirnbauer}}]{Lamacraft:2004}%
  \BibitemOpen
  \bibfield  {author} {\bibinfo {author} {\bibfnamefont {A.}~\bibnamefont
  {Lamacraft}}, \bibinfo {author} {\bibfnamefont {B.~D.}\ \bibnamefont
  {Simons}},\ and\ \bibinfo {author} {\bibfnamefont {M.~R.}\ \bibnamefont
  {Zirnbauer}},\ }\bibfield  {title} {\bibinfo {title} {Localization from
  $\ensuremath{\sigma}$-model geodesics},\ }\href
  {https://doi.org/10.1103/PhysRevB.70.075412} {\bibfield  {journal} {\bibinfo
  {journal} {Phys. Rev. B}\ }\textbf {\bibinfo {volume} {70}},\ \bibinfo
  {pages} {075412} (\bibinfo {year} {2004})}\BibitemShut {NoStop}%
\bibitem [{\citenamefont {Altland}\ and\ \citenamefont
  {Merkt}(2001)}]{Altland:2001}%
  \BibitemOpen
  \bibfield  {author} {\bibinfo {author} {\bibfnamefont {A.}~\bibnamefont
  {Altland}}\ and\ \bibinfo {author} {\bibfnamefont {R.}~\bibnamefont
  {Merkt}},\ }\bibfield  {title} {\bibinfo {title} {Spectral and transport
  properties of quantum wires with bond disorder},\ }\href
  {https://doi.org/https://doi.org/10.1016/S0550-3213(01)00209-7} {\bibfield
  {journal} {\bibinfo  {journal} {Nuclear Physics B}\ }\textbf {\bibinfo
  {volume} {607}},\ \bibinfo {pages} {511 } (\bibinfo {year}
  {2001})}\BibitemShut {NoStop}%
\bibitem [{\citenamefont {Meier}\ \emph {et~al.}(2018)\citenamefont {Meier},
  \citenamefont {An}, \citenamefont {Dauphin}, \citenamefont {Maffei},
  \citenamefont {Massignan}, \citenamefont {Hughes},\ and\ \citenamefont
  {Gadway}}]{meier2018observation}%
  \BibitemOpen
  \bibfield  {author} {\bibinfo {author} {\bibfnamefont {E.~J.}\ \bibnamefont
  {Meier}}, \bibinfo {author} {\bibfnamefont {F.~A.}\ \bibnamefont {An}},
  \bibinfo {author} {\bibfnamefont {A.}~\bibnamefont {Dauphin}}, \bibinfo
  {author} {\bibfnamefont {M.}~\bibnamefont {Maffei}}, \bibinfo {author}
  {\bibfnamefont {P.}~\bibnamefont {Massignan}}, \bibinfo {author}
  {\bibfnamefont {T.~L.}\ \bibnamefont {Hughes}},\ and\ \bibinfo {author}
  {\bibfnamefont {B.}~\bibnamefont {Gadway}},\ }\bibfield  {title} {\bibinfo
  {title} {Observation of the topological anderson insulator in disordered
  atomic wires},\ }\href {https://doi.org/10.1126/science.aat3406} {\bibfield
  {journal} {\bibinfo  {journal} {Science}\ }\textbf {\bibinfo {volume}
  {362}},\ \bibinfo {pages} {929} (\bibinfo {year} {2018})}\BibitemShut
  {NoStop}%
\end{thebibliography}%

\end{document}